\documentclass[12pt]{article}

\usepackage[margin=1in]{geometry}
\usepackage{float,comment}
\makeatletter
\renewcommand*{\@fnsymbol}[1]{\@alph{#1}}
\makeatother

\usepackage{natbib}
\bibpunct{(}{)}{;}{a}{}{,}

\usepackage{microtype}
\usepackage{graphicx}
\usepackage{subfigure}
\usepackage{booktabs}
\usepackage[dvipsnames]{xcolor}

\usepackage{amsmath,amssymb}

\DeclareMathOperator*{\argmin}{argmin}

\usepackage{amsfonts}
\usepackage{hyperref}
\usepackage{setspace}

\usepackage{thmtools}
\usepackage{thm-restate}
\usepackage{cleveref}

\declaretheorem{assumption}

\usepackage{accents}

\newcommand{\hide}[1]{}

\def\V{\boldsymbol V}
\def\X{\boldsymbol X}
\def\Y{\boldsymbol Y}

\def\u{\boldsymbol u}
\def\U{\boldsymbol U}
\def\W{\boldsymbol W}
\def\w{\boldsymbol w}

\def\h{\boldsymbol h}
\def\E{\mathbb E}
\def\P{\mathbb P}
\def\p{f}

\def\x{\boldsymbol x}

\def\bx{\boldsymbol x}
\def\btheta{\boldsymbol \theta}

\def\H{\boldsymbol H}
\def\hessianV{\boldsymbol M}
\def\sparsityV{K_j^{0}}
\def\truesetV{S_j^0}
\def\loss{\mathcal L}

\def\lipsc{L_1}
\def\thirdder{L_2}
\def\RSC{m}
\def\subexp{M}

\newcommand{\pa}{\textnormal{pa}}

\usepackage{amssymb,tikz}

\usepackage{algorithmic}
\usepackage[ruled, vlined]{algorithm2e}
\usepackage{soul}

\newtheorem{theorem}{Theorem}
\newtheorem{proposition}{Proposition}

\newtheorem{lemma}{Lemma}

\usepackage{multicol}
\usepackage{multirow}

\usepackage{amsmath}
\usepackage{amsfonts}
\usepackage{amssymb}

\def\Z{\boldsymbol Z}
\def\M{\boldsymbol M}
\def\K{\boldsymbol K}

\usepackage[english]{babel}
\DeclareMathOperator*{\Cov}{Cov}

\def\z{\boldsymbol z}

\def\balpha{\boldsymbol \alpha}
\def\bepsilon{\boldsymbol e}
\def\steponesuperset{\overline}

\def\confset{\text{an}}

\def\rootconstone{b_1}
\def\rootconsttwo{b_2}
\def\rootconstthree{b_3}

\def\thmrootone{a_1}
\def\thmroottwo{a_2}

\def\rootconfone{b_4}

\def\lipscthmtwo{L_1}
\def\thirdderthmtwo{D}
\def\cmaxthmtwo{c_0}

\def\hessianTheta{\boldsymbol M}
\def\sumk{ \sum_{k} }
\def\constrainedset{A_j^0}
\def\steponesuperset{\overline}
\def\balpha{\boldsymbol \alpha}

\def\childconstone{b_1}
\def\childconsttwo{b_2}
\def\childconstthree{b_3}

\def\childineqone{\eta_1}

\def\thmchildone{a_1}

\def\thmchilddeltaone{a_3}
\def\thmchilddeltatwo{a_4}

\def\loss{\mathcal L}

\def\lipscthmtwo{L_1}
\def\thirdderthmtwo{D}
\def\cmaxthmtwo{c_0}
\def\estconf{\widehat{\boldsymbol \epsilon}_j}

\def\r{\boldsymbol r}
\def\Re{\boldsymbol R}

\def\supergraph{\mathcal S}

\def\supportTrueMarginal{\widetilde S_j}
\def\supportFidelity{S_j}

\usepackage{tikz} 
\usetikzlibrary{shapes,snakes}
\usetikzlibrary{shapes,decorations,arrows,calc,arrows.meta,fit,positioning}
\tikzset{
    -Latex,auto,node distance =2 cm and 2 cm,semithick,
    state/.style ={ellipse, draw, minimum width = 0.7 cm},
    point/.style = {circle, draw, inner sep=0.04cm,fill,node contents={}},
    bidirected/.style={Latex-Latex,dashed},
    el/.style = {inner sep=2pt, align=left, sloped}
}

\begin{document}


\title{\bf Causal Discovery with Generalized Linear Models through Peeling Algorithms}
\author{Minjie Wang\thanks{School of Statistics, University of Minnesota, Minneapolis, MN},\hspace{.2cm}
Xiaotong Shen\footnotemark[1],\hspace{.2cm}
and Wei Pan\thanks{Division of Biostatistics, University of Minnesota, Minneapolis, MN}}
\date{}
\maketitle

\begin{abstract}
This article presents a novel method for causal discovery with generalized structural equation models suited for analyzing diverse types of outcomes, including discrete, continuous, and mixed data. Causal discovery often faces challenges due to unmeasured confounders that hinder the identification of causal relationships. The proposed approach addresses this issue by developing two peeling algorithms (bottom-up and top-down) to ascertain causal relationships and valid instruments. This approach first reconstructs a super-graph to represent ancestral relationships between variables, using a peeling algorithm based on nodewise GLM regressions that exploit relationships between primary and instrumental variables. 
Then, it estimates parent-child effects from the ancestral relationships using another peeling algorithm while deconfounding a child's model with information borrowed from its parents' models. The article offers a theoretical analysis of the proposed approach, which establishes conditions for model identifiability and provides statistical guarantees for accurately discovering parent-child relationships via the peeling algorithms. Furthermore, the article presents numerical experiments showcasing the effectiveness of our approach in comparison to state-of-the-art structure learning methods without confounders. Lastly, it demonstrates an application to Alzheimer's disease (AD), highlighting the utility of the method in constructing gene-to-gene and gene-to-disease regulatory networks involving Single Nucleotide Polymorphisms (SNPs) for healthy and AD subjects.
\end{abstract}

\noindent%

\newpage

\doublespacing

\section{Introduction}
\label{intro}

Discovering causal relationships among variables is crucial for scientific inquiries in various fields, including genetics, artificial intelligence, and social science. For instance, in genetics, biologists aim to uncover gene-gene regulatory relationships, while neuroscientists focus on causal influences between different regions of interest in a patient's brain. However, unmeasured confounders can arise when randomized experiments are unethical or infeasible, which distort the discovery process and obscure the relationship between exposures and the outcome variable, leading to false discoveries. This article proposes a novel approach to causal discovery using instrumental variables to correct confounding effects, yielding accurate causal discovery, particularly for discrete outcomes such as binary, count-valued, and multinomial.

Causal discovery necessitates estimating parent-child relationships, or equivalently, the graph structure of a directed acyclic graph (DAG). DAGs are an effective tool for describing directional effects in causal discovery, but reconstructing a DAG structure poses computational challenges due to the acyclicity constraint. Two popular approaches for reconstructing a Gaussian DAG structure without confounders are the sequential conditional independence tests, such as the PC algorithm \citep{spirtes2000causation}, and the likelihood-based methods subject to the acyclicity constraint \citep{zheng2018dags,yuan2019constrained}. Recently, \citet{li2023inference} proposed a linear causal discovery method without confounders through interventions. However, causal discovery for discrete outcome data, particularly in the presence of confounders, has received limited attention, and unique challenges arise when handling such data. One challenge is the non-identifiability of the logistic DAG model, even without confounders \citep{park2017learning}. Moreover, in the presence of confounders, unmeasured confounders can distort causal effect estimation, making structural equation models non-identifiable. Another challenge is the typically intractable form of the marginal likelihood, despite an interpretable conditional likelihood and data-specific noise or variance. It also remains unclear how to separate confounders from causal effects in the discovery process. Some recent proposals focus on simple situations, such as the two-stage least squares \citep{theil1992estimation}, an instrumental variable (IV) regression of continuous outcomes given a known causal order, and the two-stage predictor substitution (2SPS, \citet{terza2008two}) and two-stage residual inclusion (2SRI, \citet{hausman1978specification,terza2008two}) for discrete outcome data. However, neither approach applies to causal discovery with an unknown causal order and multiple primary variables.

This article proposes a new approach called GAMPI (Generalized Linear Models with Peeling and Instruments) for causal discovery of multiple primary variables from various data types. GAMPI involves a two-step process. First, we propose a fidelity model as a simple surrogate for the original intractable marginal model, which retains intervention characteristics. Then, we design a bottom-up peeling algorithm to reconstruct the super-graph consisting of ancestral relationships while identifying valid instrumental variables (IVs) for each primary variable by exploiting the connections between the primary and instrumental variables to determine the causal order. For each primary variable, a constrained generalized linear model (GLM, \citet{nelder1972generalized}) subject to the truncated $\ell_1$-penalty constraint (TLP, \citet{shen2012likelihood}) is fit on the instrumental variables to identify nonzero-coefficient IVs, followed by a difference-of-convex (DC) algorithm to solve the corresponding nonconvex minimization. In the second step, given the identified super-graph, we develop a top-down peeling algorithm to estimate the direct causal effects of each primary variable while identifying its parents from ancestors. In this peeling process, we propose a novel deconfounding approach using the estimated confounders from the parents' equation models to correct the confounding effects of a child's equation model. This approach fits a TLP-constrained GLM to each primary variable on its ancestors and residuals from its ancestors' models to identify parents and estimate the direct causal effect of each parent-child relationship.

This article contributes to causal discovery. It introduces a comprehensive approach capable of handling diverse data types with unobserved confounders, ensuring the identification of parent-child relationships through valid instruments for each primary variable. This involves generalized linear models, addressing both discrete and mixed (continuous and discrete) outcomes while considering confounders beyond Gaussian data without confounders by \citet{li2023inference}. In particular,

\begin{enumerate}
	
	\item[(1)] It establishes the identifiability of generalized structural equation models with confounders and instruments, valid and invalid. This result does not require additional assumptions for each primary variable with a nonlinear link, unlike the Gaussian case which requires valid instrumental variables to be the majority of the instrumental variables \citep{kang2016instrumental,windmeijer2019use}.
	
	\item[(2)] It introduces a fidelity model to handle intractable likelihoods and eliminate the confounding effects for identifying ancestral relationships.
	
	\item[(3)] It designs a projection-based difference-convex (DC) algorithm to solve nonconvex minimization for a constrained generalized linear model regression. This algorithm delivers a global minimizer with high probability and a computational complexity of $q^2\max(q,n)\log K^0$, where $q$, $n$, and $K^0$ are the numbers of regressors, the sample size, and the nonzero regression coefficients.
	
	\item[(4)] It develops bottom-up and top-down peeling algorithms to estimate the causal order and the causal effects for primary variables. These algorithms require solving at most $p$ generalized linear model regressions subject to the truncated $\ell_1$-penalty constraint, where $p$ is the number of primary variables.
	
	\item[(5)] It shows that GAMPI yields the correct discovery of all parent-child relationships, providing statistical guarantees for GAMPI.
	
	\item[(6)] It demonstrates the superior performance of GAMPI for logistic and Poisson models over state-of-the-art methods, NOTEARS \citep{zheng2018dags} and a faster
	version of NOTEARS, called DAGMA \citep{bello2022dagma}, especially in the presence of confounders. It suggests that GAMPI corrects the confounding effects without imposing additional noise variance structures to reconstruct a causal graph.

\end{enumerate}

The rest of the article is structured as follows. Section 2 introduces generalized structural equation models with confounders and instruments. Section 3 introduces the fidelity model and three algorithms, one DC and two peeling algorithms, for identifying the ancestral and then parent-child relationships. Section 4 investigates the statistical properties of the proposed approach. Section 5 performs simulation studies, followed by Section 6 with an application to Alzheimer's disease to reconstruct a gene-to-gene and gene-to-disease regulatory network. Section 7 concludes the article. The Appendix contains illustrative examples with technical proofs and additional simulations in the Supplementary Materials.

\section{Generalized Structural Mean Models}

\subsection{Directed Acyclic Graphs, Confounders, and Interventions}

Given a vector of primary variables $\Y = (Y_1,\ldots,Y_p)^{\top}$, the joint probability of a generalized structural equation model (SEM, \citet{pearl2000models}) with confounders $\h=(h_1,\ldots,h_p)$ and instrumental variables $\X = (X_1,\ldots,X_q)^{\top}$ can be factorized as:
\begin{align}
\label{eq:model}
\P \left(\Y | \X,\h \right)=\prod_{j=1}^p \P \left(Y_j | \Y_{\text{pa}(j)}, \X,h_j \right),
\end{align}
where $\P \left(Y_j | \Y_{\text{pa}(j)}, \X,h_j \right)$ denotes the conditional probability of $Y_j$ given $\Y_{\text{pa}(j)}, \X,h_j$, which follows an exponential family distribution. Note that \eqref{eq:model} characterizes a DAG under the acyclicity constraint. Moreover, the conditional distribution of $Y_j$ is characterized by a generalized linear model:
\begin{align}
\psi_j(\E \left[ Y_j | \Y_{\text{pa}(j)}, \X,h_j \right]) 
= \U_{\text{pa}(j),j}^{\top} \Y_{\text{pa}(j)}   + \W_{\text{in}(j),j}^{\top} \X_{\text{in}(j)}  +h_j,     \quad j = 1,\ldots,p, \label{eq:true_model}
\end{align}
where $\psi_j(\cdot)$ is a monotone link function for a GLM (cf. Table \ref{table1}), $\pa(j) \equiv \{k: u_{kj} \neq 0\}=\{ k: Y_k\rightarrow Y_j \}$ denotes a set of parent variables of $Y_j$, defined by the parent-child relationship $Y_k\rightarrow Y_j$, $\text{in}(j) \equiv \{l: w_{lj} \neq 0\}= \{ l: X_l\rightarrow Y_j \}$ denotes a set of the associated instrumental variables of $Y_j$, defined by an intervention from $X_l$ to $Y_j$: $X_l\rightarrow Y_j$, and $\Y_A=(\Y_{k_1},\cdots,\Y_{k_M})^{\top}$, $k_m \in A$, is a sub-vector of $\Y$ indexed by $A$. Here, $\U=(u_{kj})$ and $\W=(w_{lj})$ are the $p \times p$ adjacency and $q \times p$ intervention matrices, and $\U_{\text{pa}(j),j}=(u_{kj})_{k \in \pa(j)}$ and $\W_{\text{in}(j),j}=(w_{lj})_{l \in \text{in}(j)}$ are sub-vectors of the $j$th column vector of $\U$, $\U_{\bullet j}=(u_{kj})$ and the $j$th column vector of $\W$, $\W_{\bullet j}=(w_{lj})$. $^{\top}$ denotes the transpose. Note that the $p$ structural equations can possess different $\psi_j$s, depending on the data type of $Y_j$, reminiscent of the mixed graphical models framework \citep{yang2015graphical}. We refer the reader to Section~\ref{case_study} for an illustrative example.

The adjacency matrix $\U$ specifies a directed acyclic graph (DAG) with each primary variable as a node, and its non-zero elements represent directed edges between nodes. To prevent directed cycles, $\U$ is subject to the acyclicity constraint \citep{zheng2018dags, yuan2019constrained}.

\begin{table}[ht]
	\footnotesize
	\centering
	\caption{Examples of distributions in generalized linear models}
	\label{table1}
	\begin{tabular}{l l l c c c }
		\hline \multicolumn{1}{l}{Distribution} & \multicolumn{1}{l}{Support} & \multicolumn{1}{c}{Link} & \multicolumn{1}{c}{Density} \\
		\hline
		Bernoulli, $Bern(\mu)$  & Integer: $\{0,1\}$ & $\psi_j(\mu) = \ln(\frac{\mu}{1-\mu})$ 
		& $\mu^{y} (1-\mu)^{1-y}$ \\
		Binomial, $Bin(N,\mu)$  & Integer: $0,\ldots, N$ & $\psi_j(\mu) = \ln(\frac{\mu}{1-\mu})$ 
		& ${N\choose y} \mu^{y}(1-\mu)^{N-y}$ \\ 
		Gaussian, $N(\mu,\sigma^2)$ & Real: $(-\infty,\infty)$ & $\psi_j(\mu) = \mu$  & 
		$\frac{1}{\sqrt{2 \pi \sigma^2} } \exp(-\frac{(y-\mu)^2}{2 \sigma^2})$ \\ 
		Poisson,  $Poisson(\mu)$ &  Integer: $0,1,\ldots$    &  $\psi_j(\mu) = \ln \mu$ &  
		$\frac{\mu^{y} \exp(-\mu)}{ y!}$ \\
		Multinomial,  $Multi(\mu_1,\ldots,\mu_K)$ &  $K$-vector of integer: $[0,\ldots,N]$   
		&  $\psi_j(\mu) = \ln (\frac{\mu}{1-\mu})$ &  $\frac{n!}{y_1 !\ldots y_K !}
		\prod_{k=1}^K \mu_k^{y_k}$ \\
		\hline
	\end{tabular}
\end{table}

\subsection{Identifiability}

Model \eqref{eq:true_model} encodes a DAG model describing multiple parent-child relationships, which however, is generally not identifiable in the presence of unmeasured confounders $\h$. Note that \eqref{eq:true_model} may not be identifiable even in the absence of confounders $\h$,  for instance, a logistic model without instrumental variables and confounders \citep{park2017learning}. However, as suggested by Proposition \ref{identifiability_prop}, with suitable instruments, \eqref{eq:true_model} is identifiable.

To proceed, we first categorize instrumental variables (IVs) into valid IVs and non-valid IVs (covariates). A valid instrument $X_l$ for primary variable $Y_j$ satisfies:

(i) Relevance: it intervenes on $Y_j$;

(ii) Exclusion: it does not intervene on other primary variables.

Next, we make some assumptions on instruments for model \eqref{eq:true_model}. 
\begin{assumption}\label{identifiability_assumption} Assume that for $j=1,\ldots, p$, model~\eqref{eq:true_model} satisfies:

(A) (Local faithfulness) $\Cov(Y_j,X_l|\X_{\{1,\cdots,q\} \backslash \{l \}}) \neq 0$ when $X_l$ intervenes on an immediate parent of $Y_j$, where $\Cov$ denotes the covariance.

(B) (Instrumental sufficiency) Each primary variable is intervened by at least one valid IV. If $\psi_k$ is linear, then the number of valid IVs for $Y_j$ that is a child of $Y_k$ is required to exceed 50\% of its total number of IVs, known as the majority rule. Otherwise, the majority rule is not required for a specific nonlinear $\psi_j$.

(C) (Validity) Confounders $\h=(h_1,\ldots,h_p)$ and instrumental variables $\X = (X_1,\ldots,X_q)^{\top}$ are independent. That is, for each pair of $(l,j)$, $X_l$ and $h_j$ are independent.

\end{assumption}

Assumption~\ref{identifiability_assumption}(A) guarantees that other interventions don't offset an intervention from $X_l$ to $Y_j$, while Assumption~\ref{identifiability_assumption}(B) ensures that each primary variable has at least one valid IV. Both are necessary for the identifiability of a Gaussian structural model \citep{li2023inference}. The second condition in Assumption~\ref{identifiability_assumption}(B) requires the majority rule for a linear link, which amounts to the so-called majority requirement for Gaussian data \citep{kang2016instrumental,windmeijer2019use}. However, such a majority condition is not required for a nonlinear link function. We provide an illustrative example of the majority rule in Appendix~\ref{majority}. 
Assumption~\ref{identifiability_assumption}(C) is also required by the two-stage least squares methods for the IVs \citep{terza2008two,johnston2008use}, known as the instrumental validity assumption.

\begin{proposition}[Identifiability]\label{identifiability_prop}
Under Assumption~\ref{identifiability_assumption}, model~\eqref{eq:true_model} is identifiable for model parameters $(\U,\W)$.
\end{proposition}

Proposition \ref{identifiability_prop} suggests that a nonlinear link function permits the identification of the parents of a primary variable, which is unlike the linear link for Gaussian data. This new result highlights the importance of a link function concerning the model
identifiability of causal effects.

\section{Method}

This section estimates $(\U,\W)$ to identify parent-child relationships and the corresponding interventions in~\eqref{eq:true_model}. Due to the model identifiability issue of \eqref{eq:true_model}, direct estimation of $\U$ is impossible without the help of instrumental variables $\X$. To estimate parent sets $\pa(j)$, $j=1,\ldots,p$, and thus $\U$, we first need to determine the causal order, which amounts to determining ancestral relationships, including all parent-child relationships. Here, $Y_k$ is an ancestor of $Y_j$, or $Y_j$ is an offspring of $Y_k$, denoted by $Y_k \rightsquigarrow Y_j$, if there exists a directed pathway $Y_k \to Y_{k_1} \to \ldots \to Y_{k_m} \to Y_j$, where $Y_k \to Y_{k_1}$ is a parent-child relationship defined by $\U$. Subsequently, $\text{an}(j)$ denotes a set of ancestors of $Y_j$. Once $\text{an}(j)$ is identified, we then pinpoint $\pa(j)$, $j=1,\ldots,p$, through a deconfounding approach in Section 3.3.

\subsection{Fidelity Models}

This subsection introduces a working model termed as the ``fidelity model", to identify all ancestral relationships. 
The term ``fidelity model" is named as it yields the same support as the marginal distribution of the original model. Towards this end, we exploit the connections between a primary variable and the associated instrumental variables, described by the conditional distribution of $Y_j$ given $\X$ from \eqref{eq:true_model}, $\P(Y_j|\X)$, to identify the causal orders among primary variables. However, $\P(Y_j|\X)$ is generally intractable even given an analytic expression of $\P(Y_j|\Y_{\text{pa(j)}},\X,h_j)$ in \eqref{eq:true_model}. To overcome this difficulty, we introduce the fidelity model that is also a GLM:
\begin{align}
\psi_j(\E(Y_j | \X)) = \V_{\bullet j}^{\top} \X , \quad j = 1,\cdots,p. \label{eq:fidelity_model}
\end{align}
Here, $\V_{\bullet j}=(V_{1j},\ldots,V_{qj})$ is the $j$th column vector of a $q \times p$ matrix $\V=(\V_{\bullet 1},\ldots,\V_{\bullet p})$.
This model \eqref{eq:fidelity_model} is motivated by the observation that the conditional distribution of $Y_j$ given $\X$, denoted by $\P^*(Y_j|\X)$ and defined by \eqref{eq:fidelity_model}, satisfies $\frac{\partial \P^*(Y_j|\X)}{\partial X_m} \neq 0$ if and only if $\frac{\partial \P(Y_j|\X)}{\partial X_m} \neq 0$ based on \eqref{eq:true_model} due to the properties of GLMs, as shown in Proposition~\ref{fidelity_model_support}, where $\frac{\partial}{\partial X_m}$ denotes the partial derivative with respect to $X_m$.

The conditional distribution $\P^*(Y_j \mid \X)$ defined by the fidelity model \eqref{eq:fidelity_model} not only provides a simple form to work with, but also has the same support as the intractable marginal distribution $\P(Y_j|\X)$ under \eqref{eq:true_model}, although with different intervention magnitudes. 
In particular, a nonzero $l$-th element of $\V_{\bullet j}$ indicates that $Y_k$ is an ancestor of $Y_j$ if $X_l$ is a valid IV of $Y_k$. This property permits the identification of the super-graph characterizing all the ancestral relationships, as shown in Proposition \ref{prop_peeling}.

We define the index set of $X_1,\cdots,X_q$ with nonzero coefficients in the fidelity model \eqref{eq:fidelity_model} and in the true model $\mathbb P \left(Y_j | \X \right)$ marginalized from \eqref{eq:true_model} as $\supportFidelity=\{m: V_{mj} \neq 0\}$ and $\supportTrueMarginal=\{m: \frac{\partial \P(Y_j| \X)}{\partial X_m}\neq 0\}$, respectively, for $j=1,\ldots,p$.

\begin{proposition}[Support preservation]\label{fidelity_model_support}
	Assume that Assumption~\ref{identifiability_assumption} is satisfied and the link function $\psi_j$s in \eqref{eq:true_model} are differentiable. Then, $\mathbb P^*(Y_j|\X)$ defined by the fidelity model \eqref{eq:fidelity_model} has the same support as $\mathbb P \left(Y_j | \X \right)$ under the full model \eqref{eq:true_model}, that is, $\supportFidelity = \supportTrueMarginal$, $j=1,\ldots,p$.
\end{proposition}

Proposition~\ref{fidelity_model_support} suggests that the fidelity model \eqref{eq:fidelity_model} retains the intervention structure of $\P(Y_j \mid \X)$ in the original model concerning the presence or absence of a specific intervention. It is worth mentioning that the fidelity model \eqref{eq:fidelity_model} eliminates the confounding effects when identifying the support of $\P(Y_j \mid \X)$ and hence the ancestral relationships or the causal order among $Y_1,\ldots,Y_p$. This property is due to Assumption~\ref{identifiability_assumption}(C) that $\X$ are independent of confounders $\h$. Consequently, the confounders are marginalized for $\X$ and thus have no impact on the support of $\mathbb{P}(Y_j \mid \X)$. We include an illustrative example of the fidelity model in Appendix~\ref{appen_fidelity}.

\subsection{Identifying Ancestral Relationships via Peeling}

This subsection proposes nodewise constrained GLM regressions subject to the $\ell_0$-constraint based on the fidelity model to estimate nonzero elements of $\V$ in \eqref{eq:fidelity_model}.

\begin{sloppypar}
Consider the data matrix $(\X_{n\times q},\Y_{n\times p})$ where $\X_{i\bullet}$ and $\Y_{i\bullet}$ refer to the $i$th row of $\X$ and $\Y$. Given independent observations $(\X_{i\bullet},\Y_{i\bullet})_{i=1}^n$, let $\loss(\V_{\bullet j})=n^{-1} \sum_{i=1}^n \ell(Y_{ij},\V^{\top}_{\bullet j} \X_{i\bullet} )$ denote the negative log-likelihood for a GLM, where $\ell(Y_{ij},\V^{\top}_{\bullet j} \X_{i\bullet})$ is the negative log-likelihood for $Y_{ij}$ given $\X_{i\bullet}$; refer to Table \ref{table1} and \eqref{simple-like} for details.
For example, ${\ell(Y_{ij},\V^{\top}_{\bullet j} \X_{i\bullet})= \left( -Y_{i j} \left( \V^{\top}_{\bullet j} \X_{i\bullet}   \right) + \log( 1 + \exp (\V^{\top}_{\bullet j} \X_{i\bullet}  )) \right)}$ for a logistic model.
\end{sloppypar}

For $j=1,\ldots,p$, the nodewise constrained GLM regression solves the following minimization with a nonconvex constraint: 
\begin{align}
\widehat{\V}_{\bullet j}  =\arg \min_{\V_{\bullet j}} \hspace{2mm} \loss(\V_{\bullet j})  \hspace{2mm}
\text {  subject to } \quad   \sum_{l=1}^{q} I\left(V_{l j} \neq 0\right) \leq K_{j}, \label{eq:constrained}
\end{align}
where $1 \leq K_j \leq q$ is an integer-valued tuning parameter. Note that $K_j \geq 1$ ensures that each variable $Y_j$ receives at least one valid IV, as required by Assumption \ref{identifiability_assumption}(B). Here, we impose the $\ell_0$-constraint to obtain the exact number of non-zeros as opposed to the $\ell_1$ version.

To solve the nonconvex minimization \eqref{eq:constrained}, we propose a projection-based difference-convex (DC) algorithm for efficient computation. 
The constrained problem is equivalent to solving a penalized version of \eqref{eq:constrained} by adding a penalty term to the objective function. Specifically, we minimize $\loss(\V_{\bullet j})+\lambda_j \sum_{l=1}^{q} I\left(V_{l j} \neq 0\right)$, where $\lambda_j > 0$ is a computational parameter corresponding to the constrained parameter $K_j$ in \eqref{eq:constrained}. Next, we replace the $\ell_0$-indicator function with its computational surrogate, the truncated $\ell_1$-function (TLP) denoted by $J_{\tau}(\cdot)$, where $J_{\tau}(z) = \min(|z|/ \tau,1)$, as suggested by \citep{shen2012likelihood}. 
We decompose $J_{\tau}$ into a difference of two convex functions: $J_{\tau}(z)= S_1(z) - S_2(z)  \equiv |z|/\tau - \max(|z|/\tau-1, 0)$, to construct an upper approximation of the cost function iteratively. At the $t$-th iteration, we approximate $J_{\tau}$ by $S_1(z) - S_2(z^{[t-1]}) -  \nabla S_2(z^{[t-1]})^{\top} (z-z^{[t-1]})= \frac{|z|}{\tau} \cdot I\left(|z^{[t-1]}| \leq \tau \right)  +  1 - I\left(|z^{[t-1]}| \leq \tau \right)$ based on the DC decomposition.
Then, we solve the unconstrained minimization problem:
\begin{align}
               \widetilde{\V}_{\bullet j}^{[t]}=\arg \min _{V_{l j}} \hspace{2mm}
                     \loss(\V_{\bullet j}) +\gamma_j \tau_j \sum_{l=1}^{q} I\left(\left|\widetilde{V}_{l j}^{[t-1]}\right| \leq \tau_{j}\right)\left|V_{l j}\right|,
\label{eq:unconstrained}
\end{align}
where $\gamma_j = \lambda_j/\tau_j^2$. The DC algorithm iterates until a stopping criterion is met. Finally, the estimated solution $\widehat{\V}_{\bullet j}$ is computed by projecting the penalized solution onto the constraint set $\left\{\left\|\V_{\bullet j}\right\|_{0} \leq K_{j}\right\}$. In this paper, $\|\cdot\|_q$ denotes the $\ell_q$-norm of a vector and $\| \bx \|_0 = \sum_j I(x_j \neq 0)$.
In practice, we use either 5-fold cross-validation or the extended Bayesian information criterion (EBIC, \citet{chen2008extended}) to choose $(\tau_j,K_j)$.
We recommend EBIC due to its computational efficiency and strong empirical performance.

Algorithm \ref{alg:dc_alg} summarizes the DC algorithm for solving nonconvex minimization \eqref{eq:constrained}.

\begin{algorithm}[H]
	\caption{DC algorithm for nonconvex minimization \eqref{eq:constrained}}
	\label{alg:dc_alg}
	\begin{algorithmic}
		\STATE {1.} \textbf{(Initialization)}   
		Specify tuning parameters $(\tau_j,K_j)$.
		Initialize $\| \widetilde \V_{\bullet j}^{[0]} \|_0 \leq K_j$, and choose a sequence of $\gamma_j$ so that $|C_j| \geq K_j$ in Step 4.
		
		\STATE {2.} \textbf{(Relaxation) }   Compute the penalized solution $\widetilde \V_{\bullet j}^{[t]}$ of \eqref{eq:unconstrained}.
		
		\STATE {3.} \textbf{(Termination)}
		Repeat Step 2 until a termination criterion is met. Compute $\widetilde \V$: $\widetilde{\V}_{\bullet j}=\argmin _{V_{\bullet j}} \loss(\V_{\bullet j})$ 
		with $\V_{\bullet j} \in\left(\widetilde{\V}_{\bullet j}^{[t]}\right)_{t=1}^{T}$, where $T$ is the iteration index at termination.

		\STATE {4.} \textbf{(Projection)} 
		Let $C_j= \{l: |\widetilde{V}_{l j}|> |\widetilde{V}_{\bullet j}|_{(K_{j}+1)}\}$, where $|\widetilde{V}_{\bullet j}|_{(K_{j}+1)}$ is the $\left(K_{j}+1\right)$th largest absolute value of the coefficients. 
		Set $\widehat{\V}_{\bullet j}  =  \argmin_{\V_{\bullet j}} 
		\loss(\V_{\bullet j})$ subject to $V_{l j}=0$ for $l \notin C_j$.

	\end{algorithmic}
\end{algorithm}

{Remark: Computing $\widehat{\V}=(\widehat{\V}_{\bullet 1}, \ldots,\widehat{\V}_{\bullet p})$ amounts to applying Algorithm \ref{alg:dc_alg} $p$ times. The computational complexity of Algorithm \ref{alg:dc_alg} to solve one $\ell_0$-constrained regression in \eqref{eq:constrained} is the number of DC iterations multiplied by that of solving a weighted Lasso regression for a GLM, which is $q^2 \max(q,n) \log K^0_j$ \citep{efron2004least}.}

We now introduce a bottom-up peeling algorithm to estimate ancestral relationships through the nonzero elements of $\widehat{\V}$ using Proposition~\ref{prop_peeling}. This algorithm constructs a hierarchy of different layers of primary variables, defined by the causal ordering of the variables. The algorithm begins with leaf variables at the bottom, and proceeds by recursively identifying and peeling off one leaf layer of primary variables along with the associated instrumental variables in the graph. Specifically, at iteration $h$, based on Proposition~\ref{prop_peeling} (b), the algorithm first identifies all leaf nodes $Y_k$ in the subgraph with $\widehat V_{lk}^{[h]} \neq 0$ and instrumental variables $X_l$ such that $\|\widehat V_{l \bullet}^{[h]}\|_0 = 1$. In practice, the condition $\|\widehat V_{l \bullet}^{[h]}\|_0 = 1$ may not hold due to estimation error. To address this issue, we identify the rows of $\widehat \V^{[h]}$ with the smallest positive $\ell_0$-norm, that is, $\left\{l^{*}: l^{*}=\arg \min_{l=1}^q \|\widehat V_{l \bullet}^{[h]}\|_{0}, \; \text{s.t} \; \|\widehat V_{l \bullet}^{[h]}\|_0 \geq  1 \right\}$, followed by identifying the largest absolute value element index $k^{*}=\arg \max_{k=1}^p \left|\widehat V_{l^{*} k}^{[h]}\right|$ of the $l^*$th row for each $l^*$. By Proposition~\ref{prop_peeling} (b), $X_{l^*} \to Y_{k^*}$. Moreover, the algorithm identifies the ancestral relationship $Y_{k^*} \rightsquigarrow Y_{j}$ if an instrument $X_{l^*}$ for the primary variable $Y_{k^*}$ also satisfies $\widehat V_{l^* j} \neq 0$ for a previously peeled off $Y_j$, according to Proposition~\ref{prop_peeling} (c). The algorithm continues by peeling off all the current leaf-instrument $X_l \to Y_k$ pairs (i.e., removing the $l$th row and $k$th column from the current $\widehat \V^{[h]}$) to focus on the subgraph. This peeling process repeats until all primary variables are removed. The super-graph $\hat \supergraph$ contains all the ancestral relationships identified during this process. Lastly, the algorithm computes the causal ordering from the super-graph $\hat \supergraph$, which is defined as a linear ordering of the nodes where each node appears before all nodes to which it has edges.

\begin{proposition}[Identification of ancestral relationships via $\V$]\label{prop_peeling} Assume that Assumption~\ref{identifiability_assumption} is met. Then,
	
	(a) If $V_{lj} \neq 0$, then $X_l$ intervenes on $Y_j$ or an ancestor of $Y_j$.
	
	(b) $Y_j$ is a leaf variable with no children if and only if there exists an instrument $X_l$ such that $V_{lj} \neq 0$ and $\| V_{l \bullet}\|_0 = 1$.
	
	(c) If $V_{lj} \neq 0$ and $X_l$ is an instrument for $Y_k$, then $Y_k$ is an ancestor of $Y_j$, that is, $Y_k \rightsquigarrow Y_j$.
\end{proposition}

Algorithm \ref{alg:peeling} summarizes the peeling process for identifying all ancestral relationships or the causal order among primary variables. We include an illustrative example of the peeling algorithm in Appendix~\ref{appen_peeling}. In Step 3, the peeling algorithm identifies all ancestral relationships via Proposition~\ref{prop_peeling}, reconstructing a superset that includes all parent-child relationships. Given the superset, we propose a deconfounding approach to identify parent-child relationships.

\begin{algorithm}[H]
	\caption{Peeling algorithm for identifying all ancestral relationships}
	\label{alg:peeling}
	\begin{algorithmic}
		\STATE {1.} \textbf{(Initialization)}  $\widehat{\V}^{[1]}=\widehat{\V}$ and $\hat \supergraph = \emptyset$.  
		
		Begin iteration $h=1, \cdots$:  at iteration $h$,
		\STATE {2.} \textbf{(Leaf-IV pairs)}
		\begin{itemize}
			\item[(a)] Identify rows of $\widehat \V^{[h]}$ with the smallest positive $\ell_0$-norm. Store
			indices of all IVs associated with leaf variables in $A^{[h]}=\left\{l^{*}: l^{*}=\arg \min \|\widehat V_{l \bullet}^{[h]}\|_{0}\right\}$.
			
			\item[(b)] Identify the largest absolute value element index of the $l^*$th row for each $l^* \in A^{[h]}$: $B_{l^{*}}^{[h]}=\left\{k^{*}: k^{*}=\arg \max \left|\widehat V_{l^{*} k}^{[h]}\right|\right\}$. Identify all leaf-IV pairs: $X_{l^{*}} \rightarrow Y_{k^{*}}$. Let $B^{[h]}=\bigcup_{l^{*}} B_{l^{*}}^{[h]}$.
		\end{itemize}
		\STATE {3.} \textbf{(Ancestral relationships)}
		Identify ancestral relationships $Y_{k^{*}} \rightsquigarrow Y_{j}$ if i) $X_{l^{*}} \rightarrow Y_{k^{*}}$ for $l^{*} \in A^{[h]}$ and ii) $\widehat V_{l^{*} j} \neq 0$ where $Y_{j}$ has been previously removed.
			Update $\hat \supergraph$ = $\hat \supergraph \cup \{(k^*,j)\}$.

			\STATE {4.} \textbf{(Peeling)}  Remove leaf variables and associated IVs. Let $\widehat \V^{[h+1]}=\widehat \V_{\backslash\left(A^{[h]} , B^{[h]}\right)}^{[h]}$ where ${\widehat \V}_{\backslash\left(A^{[h]}, B^{[h]}\right)}^{[h]}$ is a submatrix by removing the rows and columns indexed by $A^{[h]}$ and $B^{[h]}$ from $\widehat \V^{[h]}$.
			
			\STATE {5.} \textbf{(Termination)} Let $h\rightarrow h+1$ and repeat steps 2-4 until all $Y_j$'s are removed. Update $\hat \supergraph = \hat \supergraph \cup \{(k,j): Y_k \to \cdots \to Y_j$ in $\hat \supergraph \}$. Compute the causal ordering $\hat \pi = (\hat \pi_1,\cdots,\hat \pi_p)$ from $\hat \supergraph$. Return the ancestors and IVs identified for each $Y_j$, $(\steponesuperset{\text{an}}(j),\steponesuperset{\text{in}}(j))$.
		\end{algorithmic}
	\end{algorithm}

\subsection{Identifying Parent-Child Relationships via Deconfounding}

This subsection identifies parent-child relationships given the estimated ancestral relationships from Algorithm \ref{alg:peeling}.

\subsubsection{Deconfounding}

Given estimated ancestral relationships from the first stage, we develop a novel deconfounding approach based on residual inclusion, called DRI, to estimate parent-child relationships in the presence of confounders. From \eqref{eq:true_model},
\begin{align}
\label{confounding}
\psi_j\left(\E ( Y_j | \Y_{\text{pa}(j)}, \X,h_j) \right)
= \U_{\text{pa}(j),j}^{\top} \Y_{\text{pa}(j)}  + \W_{\text{in}(j),j}^{\top} \X_{\text{in}(j)} +h_j, \quad j = 1,\cdots,p,
\end{align}
where $h_1,\ldots,h_p$ may be correlated. When there is no confounder, we could identify parents by fitting a constrained GLM regression of $Y_j$ on its ancestors $\Y_{\text{an}(j)}$ and instruments $\X_{\text{in}(j)}$. However, in the presence of confounders, unobserved confounders $h_j$ and $\h_{\text{pa}(j)}$ can be correlated. Thus $Y_j$'s parent variables $\Y_{\text{pa}(j)}$ depends on $h_j$ through $\h_{\text{pa}(j)}$, which biases the estimation of $\U_{\text{pa}(j), j}$ as $h_j$ is one resource of the model error for the regression of $Y_j$.

To address the confounding issue, we propose a novel deconfounding approach, DRI, to correct the confounding effects in the child structural equations by treating the residuals from its parent GLM regression as predictors. In this way, this approach utilizes the connections between confounders in a parent and its child equations. To facilitate DRI, we make the practically sensible assumption that the confounders $h_1,\cdots,h_p$ are jointly normal. Assumption \ref{confounder_assumption1} simplifies the implementation of DRI and makes it computationally efficient.

\begin{assumption}
	\label{confounder_assumption1}
	The confounders $h_1,\cdots,h_p$ are jointly normal with an unknown mean and an unknown covariance. 
\end{assumption}

{Remark: Assumption \ref{confounder_assumption1} can be relaxed to the assumption that each confounder can be represented as a linear function of other confounders. In the literature, most assume one common underlying confounding (i.e., one $\h$ across all equations) while we here consider a more general case of $h_1,\cdots,h_p$. For complex problems, Assumption \ref{confounder_assumption1} is sensible as the confounder is in fact an ensemble of many confounding effects.}

To implement DRI, we estimate the confounding effect $h_j$ using the parent equations for each $Y_j$ based on Assumption \ref{confounder_assumption1}, that is, $h_j |\{h_k, k \in \confset(j)\} \sim N(\sum_{k \in \confset(j)} \alpha_{kj} h_k,\sigma^2)$, or $h_j = \sum_{k \in \confset(j)} \alpha_{kj} h_k + \bepsilon_j$, where $\bepsilon_j \sim N(0,\sigma^2)$ is the unobserved error orthogonal to the projection space spanned by $\{h_k: k \in \confset(j)\}$, and uncorrelated with and thus independent of $\{h_k: k \in \confset(j)\}$ and $\Y_{\text{pa}(j)}$. By Assumption~\ref{identifiability_assumption}(C), $\bepsilon_j$ is also independent of $\X$.
Then,
\begin{align}
\psi_j(\E \left[ Y_j | \Y_{\text{pa}(j)}, \X,h_j \right]) 
= \U_{\text{pa}(j),j}^{\top}  \Y_{\text{pa}(j)}  + \W_{\text{in}(j),j}^{\top} \X_{\text{in}(j)}  + \sum_{k \in \confset(j)} \alpha_{kj} h_k+ \bepsilon_j,    
\label{child}
\end{align}
where DRI replaces $h_k$ with the residuals $\widehat h_k$ estimated from the parent equations of $Y_j$. As a result, $\bepsilon_j$ is independent of $\Y_{\text{pa}(j)},\X_{\text{in}(j)}, \sum_{k \in \confset(j)} \alpha_{kj} h_k$ in \eqref{child}, resolving the dependence issue of $\Y_{\text{pa}(j)}$ on $h_j$ in \eqref{confounding} due to confounding.

We propose a top-down algorithm to estimate parent-child relationships through deconfounding, given the causal ordering of the primary variables $\hat \pi$, and $(\steponesuperset{\text{an}}(j),\steponesuperset{\text{in}}(j))$, $j=1,\ldots,p$, identified by Algorithm~\ref{alg:peeling}. Note that the causal ordering represents the direction of edges in a DAG in that for every directed edge $(k,j)$, i.e., $Y_k \to Y_j$, $k$ appears before $j$ in the ordering.
The algorithm proceeds from the top to the bottom of a hierarchy defined by the causal order while identifying the parents for each primary variable and iterates this process until the last element of the ordering.

The algorithm starts from a root variable $Y_k$ that has no parents. First, a GLM regression of $Y_k$ is fit on its valid IVs $\X_{\steponesuperset{\text{in}}(k)}$ via the model: $\psi_k(\E \left[ Y_k | \X \right]) = \W_{\steponesuperset{\text{in}}(k), k}^{\top} \X_{\steponesuperset{\text{in}}(k)} $. Then, we compute the residuals $Y_{ik} - \varphi_k(\widehat  \W_{\steponesuperset{\text{in}}(k), k}^{\top} \X_{i,\steponesuperset{\text{in}}(k)} )$ to estimate the confounding effect $h_{ik}$, where $\varphi_k(\cdot)$ is the inverse link function for the $k$-th GLM model.
It is important to note that the confounders do not bias the estimation of residuals in root equations by the independence assumption of the IVs and confounders.
Our simulations and theory suggest that this approach works well, as in the IV regression \citep{johnston2008use}.
Alternatively, we can also fit a generalized linear mixed-effect model for root equations when the data has repeated measurements. Details are given in Algorithm~\ref{alg:peeling_confounder_mix_effect} of the Appendix.

The algorithm then moves to a non-root variable $Y_j$ and considers the GLM regression on its ancestors $\Y_{\steponesuperset{\text{an}}(j)}$, its IVs $\X_{\steponesuperset{\text{in}}(j)}$, and the estimated confounders $\widehat h_k$ from the ancestor equations via the model:
$\psi_j(\E \left[ Y_j | \Y_{\text{pa}(j)}, \X,h_j \right]) = \U_{\steponesuperset{\text{an}}(j), j}^{\top} \Y_{\steponesuperset{\text{an}}(j)}  + \W_{\steponesuperset{\text{in}}(j), j}^{\top} \X_{\steponesuperset{\text{in}}(j)}  + \sum_{k \in \steponesuperset{\text{an}}(j)} \alpha_{kj} \widehat h_k + \bepsilon_j$, where $(\text{pa}(j),h_k)$ in \eqref{child} is replaced by $(\steponesuperset{\text{an}}(j),\widehat h_k)$. Specifically, it fits TLP-constrained GLM regressions:
\begin{align}
&   (\widehat{\W}_{{\steponesuperset{\text{in}}(j)}, j},\widehat{\U}_{\steponesuperset{\text{an}}(j), j},\widehat{\balpha}_{\steponesuperset{\text{an}}(j), j})  \nonumber \\ & =\argmin_{\W_{{\steponesuperset{\text{in}}(j)}, j},\U_{\steponesuperset{\text{an}}(j), j},\balpha_{\steponesuperset{\text{an}}(j), j}}
\loss(\W_{{\steponesuperset{\text{in}}(j)}, j},\U_{\steponesuperset{\text{an}}(j), j},\balpha_{\steponesuperset{\text{an}}(j), j} | \X_{\steponesuperset{\text{in}}(j)} ,\Y_{\steponesuperset{\text{an}}(j)},\widehat \h_{\steponesuperset{\text{an}}(j)}) \nonumber \\
&\text {  subject to } \quad     \sum_{k \in \steponesuperset{\text{an}}(j)} I(U_{kj} \neq 0) \leq K_{j}, \quad \sum_{k \in \steponesuperset{\text{an}}(j)} I(\alpha_{kj} \neq 0) \leq K'_{j}, 
\quad j=1, \ldots, p, 
\label{full-like}
\end{align}
where $0 \leq K_j \leq  |\steponesuperset{\text{an}}(j)|$ and $0 \leq K'_j \leq  |\steponesuperset{\text{an}}(j)|$
can be tuned as in \eqref{eq:constrained}, with $|\cdot|$ denoting the size of a set; $\W_{\steponesuperset{\text{in}}(j), j}$ is unconstrained so that Assumption \eqref{identifiability_assumption}(B) continues to satisfy; $\loss(\W_{{\steponesuperset{\text{in}}(j)}, j},\U_{\steponesuperset{\text{an}}(j), j},\balpha_{\steponesuperset{\text{an}}(j), j} | \X_{\steponesuperset{\text{in}}(j)} ,\Y_{\steponesuperset{\text{an}}(j)},\widehat \h_{\steponesuperset{\text{an}}(j)}) = n^{-1} \sum_{i=1}^n \ell(Y_{ij}, \W^{\top}_{\steponesuperset{\text{in}}(j), j} \X_{i,\steponesuperset{\text{in}}(j)}  +  \U^{\top}_{\steponesuperset{\text{an}}(j), j} \Y_{i,\steponesuperset{\text{an}}(j)}  +  \balpha^{\top}_{\steponesuperset{\text{an}}(j), j} \widehat \h_{i,\steponesuperset{\text{an}}(j)} )$; $\h_{i,\steponesuperset{\text{an}}(j)}$ denotes a column vector consisting of $\{h_{ik}: k \in \steponesuperset{\text{an}}(j)\}$ and $\balpha^{\top}_{\steponesuperset{\text{an}}(j), j} \widehat \h_{i,\steponesuperset{\text{an}}(j)} = \sum_{k \in \widehat{\text{an}}(j)} \widehat \alpha_{kj} \widehat h_{ik}$. 
From \eqref{full-like}, we obtain the estimated set $\widehat{\text{pa}}(j)=\{k \in \steponesuperset{\text{an}}(j): \widehat U_{k j} \neq 0\} \subset \steponesuperset{\text{an}}(j)$, and $\widehat{\text{in}}(j)= \steponesuperset{\text{in}}(j)$.
Finally, we compute the residuals 
\begin{align}
\label{residual}
\widehat h_{ij} = Y_{ij} -\varphi_j(\widehat \U_{\widehat{\text{pa}}(j), j}^{\top} \Y_{i,\widehat{\text{pa}}(j)}  + \widehat \W_{\widehat{\text{in}}(j), j}^{\top} \X_{i,\widehat{\text{in}}(j)}  + \sum_{k \in \widehat{\text{an}}(j)} \widehat \alpha_{kj} \widehat h_{ik}). 
\end{align}

Algorithm \ref{alg:peeling_confounder} summarizes the peeling process for identifying parent-child relationships using the proposed deconfounders.

\begin{algorithm}
	\caption{Peeling algorithm for estimating parent-child relationships via DRI}
	\label{alg:peeling_confounder}
	\begin{algorithmic}
		\STATE {1.} Input $(\steponesuperset{\text{an}}(j), \steponesuperset{\text{in}}(j))_{j=1}^p$ and $\hat \pi$ from Algorithm~\ref{alg:peeling}. Input data matrix $(Y_{ij},X_{ij})_{n \times (p+q)}=(\Y_{i\bullet},\X_{i\bullet})_{i=1}^n$ of primary variables $\Y_{n \times p}$ and instruments $\X_{n \times q}$.
		
		Begin Iteration: for $d=1\cdots,p$,
		
		\STATE {2.} \textbf{(Estimating the confounding effects via IV regression)} If $\hat \pi_d$ is a root variable indexed by $Y_k$, compute $\widehat{\W}_{{\steponesuperset{\text{in}}(k)}, k}$ by fitting a GLM regression of $Y_k$ on $\X$: $\E[  Y_k | \X ] = \varphi_k(  \W_{\steponesuperset{\text{in}}(k), k}^{\top} \X_{\steponesuperset{\text{in}}(k)}  )$.
		Compute the residuals: $\widehat h_{ik} =  Y_{ik} - \varphi_k(  \widehat \W_{\steponesuperset{\text{in}}(k), k}^{\top} \X_{i,\steponesuperset{\text{in}}(k)}  )$.

		\STATE {3.}  \textbf{(Deconfounding)} If $\hat \pi_d$ is a non-root variable indexed by $Y_j$, compute $(\widehat{\W}_{{\steponesuperset{\text{in}}(j)}, j},\widehat{\U}_{\steponesuperset{\text{an}}(j), j},\widehat{\balpha}_{\steponesuperset{\text{an}}(j), j})$ by fitting a TLP-constrained GLM regression of $Y_j$ in \eqref{full-like}.
		Compute the residuals $\widehat h_{ij}$  in \eqref{residual}.

	\end{algorithmic}
\end{algorithm}
{Remark:
The computational complexity of Algorithm \ref{alg:peeling_confounder} amounts to solving at most
$p$ TLP-constrained regressions in \eqref{eq:constrained} of size $|\steponesuperset{\text{in}}(j)|+
2 |\steponesuperset{\text{an}}(j)|$ via Algorithm \ref{alg:dc_alg},  
which is of order $p (p+q)^2 \max(n,(p+q)) \log K^0_j$.
}

\subsubsection{Connections with 2SRI and 2SPS}

DRI is reminiscent of, but fundamentally different from the two-stage predictor substitution (2SPS, \citep{terza2008two}) and two-stage residual inclusion (2SRI, \citep{hausman1978specification,terza2008two}), both of which require a known causal order between two primary variables. In 2SRI, the residuals obtained at the first stage serve as an additional predictor as opposed to replacing the endogenous variables with their predicted values in 2SPS, which is also known as two-stage least squares for Gaussian data. However, neither applies to our situation of multiple primary variables with an unknown causal order and confounders.

For our problem, we also include a version of predictor substitution, referred to as DPS, to compare with DRI in the Appendix. In practice, we recommend DRI for causal discovery due to its superior performance and theoretical guarantees, and therefore integrate it with our top-down peeling algorithm for implementation. DRI explores the connection between parent and child equations to eliminate the confounding effect in a child equation through the residuals, whereas DPS cannot capture this aspect. This recommendation is consistent with the observation that 2SRI suits more than 2SPS for binary or discrete outcomes \citep{terza2008two,ying2019two}.

\section{Theory}

This section presents a novel theoretical analysis of the proposed approach, offering theoretical guarantees even in the presence of confounders. First, we demonstrate in Theorem~\ref{thm_anc_sele_cons} that the proposed DC algorithm, Algorithm~\ref{alg:dc_alg}, successfully recovers the true support of $\V^0$, terminates within a finite number of steps, and achieves a global minimizer for the nonconvex minimization \eqref{eq:constrained}, with probability approaching one. Based on this, our bottom-up peeling algorithm, Algorithm~\ref{alg:peeling}, retrieves the true super-graph $\supergraph$. Secondly, we prove in Theorem~\ref{thm_confounder_consistency} that our top-down peeling algorithm, Algorithm~\ref{alg:peeling_confounder}, accurately reconstructs the true causal graph, thereby identifying all parent-child relationships.

Consider a generalized linear model with the canonical link, where the negative log-likelihood of $Y_{ij}$ given $\X_{i\bullet}$ based on independent observations $(Y_{ij},\X_{i\bullet})_{i=1}^n$ can be expressed as:
\begin{align}
\label{GLM-like}
\ell(Y_{ij},\theta(\X_{i\bullet}))=-Y_{i j} \theta(\X_{i\bullet}) + A_j(\theta(\X_{i\bullet})), \quad i=1,\ldots,n.
\end{align}
Here, $A_j(\theta)$ represents the cumulant function of an exponential family distribution, with $\theta$ denoting the regression function. For instance, in the case of the logistic regression, $A_j(\theta) = \log(1 + \exp(\theta))$. Given the canonical link, $A_j^{'}(\theta) = E(Y_j | \cdot ) = \psi_j^{-1}(\theta) = \varphi_j(\theta) \vert_{\theta = \V^{\top}_{\bullet j} \X_{i\bullet}}$. Hence, the log-likelihood of $Y_{ij}$ given $\X_{i\bullet}$ for the fidelity model \eqref{eq:fidelity_model} can be written as:
\begin{align}
\label{simple-like}
\ell(Y_{ij},\V^{\top}_{\bullet j} \X_{i\bullet})=-Y_{i j} \big( \V^{\top}_{\bullet j} \X_{i\bullet}   \big) + A_j(\V^{\top}_{\bullet j} \X_{i\bullet}), \quad i=1,\ldots,n. 
\end{align}

Subsequently, we denote $^0$ as the true parameter; for example, $\V^0$ means the true parameter values of $\V$. Denote $S_j^0 = \{l: V_{lj}^{0} \neq 0\}$. Let $K_j^{0} = \| \V^0_{\bullet j}\|_0 = |S_j^0|$ and $K_{\max }^{0}  = \max_{1\leq j \leq p} K_j^{0}$. The following technical conditions are assumed for the fidelity model \eqref{eq:fidelity_model}.

\begin{assumption}[GLM residuals]
	\label{GLM-residual}
	Assume that for some positive constants $\lipsc$ and $\thirdder$, $|A_j^{''}(\theta)| \leq \lipsc$, $|A_j^{'''}(\theta)| \leq 
	\thirdder$, $j=1,\ldots,p$, where $^{''}$ and $^{'''}$ denote the second and third derivatives. 
	Moreover, $\{\xi_{ij}\}_{i=1}^n$ with $\xi_{ij}= Y_{ij} - \varphi_j({{\V}_{\bullet j}^{0}}^{\top} \X_{i\bullet})$ is sub-exponential with mean zero, so that for any real $t>0$,
	\begin{align*}
	& \P\big( \big| n^{-1} \sum_{i=1}^n \xi_{ij}  \big|  \geq t  \big)  \leq 2 \exp \big( -  \min\big( \frac{t^2}{2 \subexp^2} , \frac{t}{2 \subexp} \big) n   \big), \quad  j=1,\ldots,p.
	\end{align*}
\end{assumption}

\begin{assumption}[Restricted strong convexity]
	\label{REcondition}
	For a constant $m >0$, 
	\begin{eqnarray}
	\Lambda_{\min }=\min _{A:|A| \leq 2 K_{\max }^{0}} \min _{\left\{(\Delta,
		\V_{\bullet j}):\left\|\Delta_{A^c} \right\|_{1} \leq 3\left\|\Delta_{A}\right\|_{1}, \V_{\bullet j} \in 
		(\V_{\bullet j}^0-\Delta, \V_{\bullet j}^0+\Delta)
		\right\}} \frac{ \Delta^{\top} \nabla^2 \loss(\V_{\bullet j}) \Delta }{\|\Delta\|_{2}^{2}} \geq m.
	\label{eigen}
	\end{eqnarray}
\end{assumption}
Note that \eqref{eigen} is the restricted strong convexity (eigenvalue) condition and requires the log-likelihood $\loss(\V_{\bullet j})$ to be strongly convex in a neighborhood of $\V_{\bullet j}^0$, where $\nabla^2 \loss(\V_{\bullet j}^0) = \X^{\top} \hessianV^{j} \X$ and $\hessianV^{j}$ is a diagonal matrix with $\hessianV_{ii}^{j} =  A_j''({\V_{\bullet j}^0}^{\top} \X_{i\bullet})$ depending on $\X$ and $\V^0$ only. This condition has been commonly used for the analysis of the error bound of parameter estimation and the convergence analysis of optimization algorithms \citep{lee2015model,negahban2012unified, hastie2015statistical,zhang2017restricted}. Note that Assumption~\ref{REcondition} permits correlated designs $\X$ and 
is a weaker condition than the irrepresentable condition required by the Lasso \citep{10.1214/09-EJS506}.

\begin{assumption}[Bounded domain for interventions]
	\label{interv_boundedness}
For some constants
$c_0$-$c_2$ and $C_1 > 0$,
	\begin{align*}
	&\| \X \|_{\infty} \leq c_1, \quad  \| \V_{\bullet j}^0 \|_2 \leq C_1, \quad \|(\X_{S_j^0}^{\top} \hessianV^{j} \X_{S_j^0} / n)^{-1} \X_{S_j^0}^{\top}\|_{\infty} \leq c_2, \quad  \Omega_{\max}(\X_{S_j^0}^{\top} \X_{S_j^0} / n) \leq c_0,   
	\end{align*}
where $\Omega_{\max}(\cdot)$ refers to the maximum eigenvalue of a matrix. 	
\end{assumption}

\begin{assumption}[Minimum signal strength]
	\label{SNRcondition}
\begin{align*}
\min_{V_{lj}^0 \neq 0} | V_{lj}^0 |  \geq    100 \subexp c_{2}   \sqrt{\frac{\log q + \log n}{n}} .
\end{align*}
\end{assumption}
Assumption~\ref{SNRcondition} specifies the minimal signal strength over candidate interventions. Such an assumption 
has been used for establishing selection consistency in high-dimensional variable selection 
\citep{zhao2018pathwise}.

\begin{theorem}[Reconstruction of super-graph via Algorithm~\ref{alg:dc_alg}] \label{thm_anc_sele_cons}
	Under Assumptions~\ref{GLM-residual}-\ref{SNRcondition}, for $j=1, \ldots, p$, if the tuning parameters $(\tau_j,K_j)$ of Algorithm~\ref{alg:dc_alg} satisfy:
	\begin{enumerate}

		\item[(1)] (Computation) $\gamma_j \in [8 \tau_{j}^{-1} \cdot  \subexp c_{1}  \sqrt{   (\log q + \log n) / n}  , m / 6]$,
		\item[(2)] (Tuning parameters) $8 \subexp c_{2}  \sqrt{\frac{\log q + \log n}{n}} \leq \tau_{j} \leq 0.4 \min_{V_{lj}^0 \neq 0} | V_{lj}^0 |$, $K_{j}=K_{j}^{0}$,
	\end{enumerate}
	then Algorithm~\ref{alg:dc_alg} terminates in at most $1+\left\lceil\log(2 K_j^{0})/ \log 4\right\rceil$ iterations for \eqref{eq:constrained}, where $\lceil\cdot \rceil$ is the ceiling function. Moreover, for $1 \leq j \leq p$, 
	\begin{align*}
	\mathbb{P}\left(\widetilde{\V}_{\bullet j} \text { is not a global minimizer of }~\eqref{eq:constrained}\right) \leq 8 q \exp (-2(\log (q)+\log (n))) = 8 q^{-1} n^{-2}. 
	\end{align*}
	As a result, Algorithm~\ref{alg:dc_alg} yields  a global minimizer of~\eqref{eq:constrained},
$\widetilde{\V}_{\bullet j}$, with probability tending to 1 as $n \to \infty$.
Importantly, Algorithm~\ref{alg:dc_alg}, together with 
Algorithm~\ref{alg:peeling}, recovers the true super-graph $\supergraph^0$ containing ancestral relations with probability
\begin{eqnarray*}
	\P(\widehat \supergraph \neq \supergraph^0) \leq  8 p q^{-1} n^{-2},
\end{eqnarray*}
where ${\widehat \supergraph}$ is obtained from Algorithm~\ref{alg:peeling}
and $\supergraph^0 \equiv \{ (k,j): k \in \text{an}(j) \}$. Under Assumption~\ref{identifiability_assumption}(C) (i.e., $p \leq q$), with probability tending to one, $\widehat \supergraph$ correctly reconstructs the true super-graph $\supergraph^0$ and thus the causal order of $Y_1,\ldots,Y_p$ as $n\to \infty$.
\end{theorem}
Theorem~\ref{thm_anc_sele_cons} ensures the consistent reconstruction of the super-graph $\supergraph^0$ by Algorithm~\ref{alg:dc_alg} and Algorithm~\ref{alg:peeling}, which characterizes ancestral relationships and determines the causal order of primary variables.
Also, it says that Algorithm \ref{alg:dc_alg} (DC algorithm) attains a global minimizer almost surely as $n \rightarrow \infty$ under the data generating distribution. This result is in contrast to the strong hardness result of \citet{chen2019approximation} that there does not exist a polynomial-time algorithm achieving the globality of the $\ell_0$-constrained optimization \eqref{eq:constrained} in the worst-case scenario. We here show that with probability tending to one, this problem can be solved. In other words, the probability of the worst-case scenario tends to zero.
Note that Algorithm \ref{alg:dc_alg} is indeed a polynomial-time algorithm with time complexity $O(q^2 \max(q,n) \log K^0_j)$ for solving one $\ell_0$-constrained regression in \eqref{eq:constrained}.

Next, we establish causal graph selection consistency of the estimated causal graph based on the estimates $\widehat \U_{\bullet j}$ by Algorithm~\ref{alg:peeling_confounder}. 
On this ground, we ensure that all parent-child relationships are correctly identified.
Let $s = \max_{1 \leq j \leq p} |\text{an}(j)|$, $\tilde s = \max_{1 \leq j \leq p} \| \W_{\bullet j}^0 \|_0$, and $\widetilde \Z = [\X_{\text{in}(j)},\Y_{\text{pa}(j)},\widehat \h_{\text{an}(j)}]$. Under Assumption~\ref{interv_boundedness} with $\widetilde \Z$, $\| \widetilde \Z^{\top} \|_{\infty} \leq \childconstone$, $\| (\widetilde \Z^{\top} \hessianTheta \widetilde \Z/n )^{-1} \widetilde \Z^{\top} \|_{\infty} \leq \childconsttwo$, and 
$\Omega_{\max}({\widetilde \Z}^{\top} {\widetilde \Z} / n) \leq b_0$.

\begin{theorem}[Reconstruction of causal graph via Algorithm~\ref{alg:peeling_confounder}] \label{thm_confounder_consistency}
Under Assumptions~\ref{REcondition}-\ref{interv_boundedness} with 
	$\widetilde \Z = [\X_{\text{in}(j)},\Y_{\text{pa}(j)},\widehat \h_{\text{an}(j)}]$ in the GLM regression \eqref{full-like}, if tuning parameters of Algorithm~\ref{alg:peeling_confounder} satisfy: 
	\begin{enumerate} 
		\item[(1)] (Computation) {\scriptsize{$\gamma_j \in [\tau_{j}^{-1} \cdot   8 \subexp \childconstone  \sqrt{   (\log (2s + \widetilde s) + \log n) / n}  , m / 6]$}},
		\item[(2)] (Tuning parameters) $C  \sqrt{\frac{\log (2s + \widetilde s) + \log n}{n}} \leq \tau_{j} \leq 0.4 \min_{U_{kj}^0 \neq 0} | U_{kj}^0 |$, $K_{j}=K_{j}^{0}$,
	\end{enumerate}
	where $C$ is a constant depending on $ \childconstone$, $ \childconsttwo$ and $b_0$, then, Algorithm~\ref{alg:peeling_confounder} 
reconstructs the causal graph consistently with probability tending to one, or 
	\begin{eqnarray*}
		P(\widehat E = E^0)  \to 1, \quad \quad \text{ as } n \to \infty,
	\end{eqnarray*}	
where $\widehat E = \{(k,j): \widehat U_{kj} \neq 0 \}$ and $E^0 = \{(k,j): U^0_{kj} \neq 0 \}$.
\end{theorem}

Theorem~\ref{thm_confounder_consistency} suggests that Algorithm \ref{alg:peeling_confounder} recovers the true causal graph and thus causal relationships with probability tending to one as the sample size is sufficiently large. In the Supplementary Materials, we prove this by establishing the error bounds of the estimates $\widehat \U_{\bullet j}$, $\widehat \W_{\bullet j}$ for estimating $\U$ and $\W$ by Algorithm~\ref{alg:peeling_confounder}.

\section{Simulations}\label{simulation}

This section investigates the empirical performance of the proposed method. We assess the performance of GAMPI and compare it against the structure learning method NOTEARS \citep{zheng2018dags}, under various graph structures (hub, chain, and random graphs) and types of outcome variables. Further, we compare GAMPI with a recently proposed structure learning method DAGMA \citep{bello2022dagma} based on a log-det constraint. Note that DAGMA is designed exclusively for the Gaussian and logistic outcomes.

\subsection{Simulation Setting}\label{simulation_setting}

The data simulation process is as follows. Firstly, we generate an adjacency matrix $\U$ based on the graph structure and construct an intervention matrix $\W$ with $W_{jj}= 1$, for $j= 1,\cdots,p$ and $W_{lj}= 0$, for $1 \leq l \neq j \leq q$. For the hub graph, $U_{1j} = 1$, $j = 2,\cdots,p$, and 0 otherwise. The random graph is simulated similarly as \citet{li2023inference}. Secondly, we generate Gaussian instrumental variables $\X= (\X_1,\ldots, \X_q) \sim N(0,1)$. Note that our approach also allows for correlated $\X$ satisfying Assumption~\ref{REcondition}. For the confounders, we simulate $\h \sim N(\textbf{0},\boldsymbol \Sigma)$, where $\Sigma_{ij} = 0.95$. In the Supplementary Materials, we explore the simulation setup where the data is generated without confounders, i.e., $\h = \textbf{0}$. Given $\X,\U$,$\W$, and $\h$, we generate random samples $\Y$ according to \eqref{eq:true_model}. In this section, we consider two data types for the outcome variable $Y$: binary and count outcomes. In the binary case,
$Y_j$ is generated from the Bernoulli distribution with $P(Y_j= 1)$ equal to $\frac{\exp(\alpha_0 \w_j^{\top} X_{\text{in}(j)} + h_j)}{(1+\exp(\alpha_0 \w_j^{\top} X_{\text{in}(j)}+ h_j))}$ if $Y_j$ is a root variable, and $\frac{\exp(\beta_1 \u_j^{\top} Y_{\text{pa}(j)} + \alpha_1 \w_j^{\top} X_{\text{in}(j)} + h_j )} {1+\exp(\beta_1 \u_j^{\top} Y_{\text{pa}(j)} + \alpha_1 \w_j^{\top} X_{\text{in}(j)}+ h_j)}$ otherwise.
For the hub graph, we set $\alpha_0 = 5$, $\beta_1 = 2.5$, and $\alpha_1 = 2$. For the chain graph, we set $\alpha_0 = 5$, $\beta_1 = 2.5$, and $\alpha_1 = 3$. For the random graph, we set $\alpha_0 = 5$, $\beta_1 = 3$, and $\alpha_1 = 3$.

For the count outcome, to avoid extreme values, we employ standard copula transforms to simulate $Y$, as described by \citet{yang2015graphical,nelsen2007introduction}. Specifically, we first generate data using $\tilde Y_j = \beta_1 \u_j^{\top} \Y_{\text{pa}(j)} + \alpha_1 \w_j^{\top} \X_{\text{in}(j)}+ h_j + \epsilon_j$, where $\epsilon_j$ are i.i.d. Gaussian errors. We then use a standard copula transform to ensure that the marginals of the generated data $Y_j$ are approximately Poisson. For the hub graph, we set $\alpha_0 = 5$, $\beta_1 = 0.5$, and $\alpha_1 = 2$. For the chain graph, we set $\alpha_0 = 5$, $\beta_1 = 0.5$, and $\alpha_1 = 3$. For the random graph, we set $\alpha_0 = 4$, $\beta_1 = 1$, and $\alpha_1 = 2$. We consider three different graph structures: the hub, chain (of length 4), and random graphs. In addition, we fix the sample size $n=500$ while varying the number of variables from 100 to 300.

To evaluate the accuracy of estimating the directed edges of a graph, we consider five evaluation metrics: the false positive rate (FPR), the false discovery rate (FDR), the F-score, the Matthews correlation coefficient (MCC), and the structural Hamming distance (SHD). The Matthews correlation coefficient is a binary classification metric defined as
\begin{eqnarray*}
	\frac{\mathrm{TP} \times \mathrm{TN}-\mathrm{FP} \times \mathrm{FN}}{\{(\mathrm{TP}+\mathrm{FP})(\mathrm{TP}+\mathrm{FN})(\mathrm{TN}+\mathrm{FP})(\mathrm{TN}+\mathrm{FN})\}^{1/2}},
\end{eqnarray*}
where $\mathrm{TP}$, $\mathrm{FP}$, $\mathrm{TN}$ and $\mathrm{FN}$ denote the true positive, false positive, true negative, and false negative rates for edge selection. A large MCC value close to 1 indicates that the estimated edge set is close to the true edge set. In addition, the structural Hamming distance measures edge directionality between two directed graphs, which is the number of edge insertions, deletions, or flips needed to transform one graph to another graph \citep{tsamardinos2006max}. A small structural Hamming distance between two graphs of the same size indicates their closeness.

\subsection{Results}

This subsection reports the simulation results in a situation where we simulate the data in the presence of confounders. Table~\ref{sim_conf_ebic} suggests that GAMPI outperforms NOTEARS across all setups in terms of causal graph recovery, as measured by five metrics: FPR, FDR, F-score, MCC, and SHD. Table~\ref{sim_conf_ebic} shows that NOTEARS can yield an empty graph with no edges selected when ``NA" occurs. Table~\ref{sim_conf_ebic_DAGMA} in the Supplementary Materials suggests that GAMPI outperforms DAGMA significantly in most scenarios, except for the simple case of the hub graph, where both methods perform equally well. Further, note that unlike GAMPI, NOTEARS and DAGMA do not guarantee acyclicity or estimate the parameters of causal effects.

\begin{table}[ht]
	\begin{center}
		\centering Binary
		\vskip 0.05in
		\resizebox{\linewidth}{!}{
			\begin{tabular}{lccccccccccc}
				\toprule
				\multirow{2}{*}{Graph} &
				\multirow{2}{*}{$(p,q,n)$} &
				\multicolumn{2}{c}{FPR} &
				\multicolumn{2}{c}{FDR} & 
				\multicolumn{2}{c}{F-score} &
				\multicolumn{2}{c}{MCC}  &
				\multicolumn{2}{c}{SHD}  \\
				&  & {No-tears} & {GAMPI} & {No-tears} & {GAMPI} & {No-tears} & {GAMPI}  & {No-tears} & {GAMPI}  & {No-tears} & {GAMPI}\\
				\midrule
				Hub          &  (100,100,500) &    0.00 (0.00) &   0.00 (0.00) &   0.01 (0.01) &   0.05 (0.01) &   0.16 (0.01) &   0.96 (0.01) &   0.29 (0.01) &   0.96 (0.01) &  90.30 (0.78) &   8.10 (1.46)  \\ 
				&  (200,200,500) &   0.00 (0.00) &    0.00 (0.00) &    0.01 (0.01) &    0.04 (0.01) &    0.13 (0.02) &    0.95 (0.01) &    0.25 (0.03) &    0.95 (0.01) &  184.90 (2.37) &   20.40 (3.95)  \\ 
				&  (300,300,500) & 0.00 (0.00) &    0.00 (0.00) &    0.00 (0.00) &    0.04 (0.01) &    0.21 (0.02) &    0.95 (0.01) &    0.34 (0.02) &    0.95 (0.01) &  263.50 (3.87) &   28.20 (7.61)  \\
				Chain        &  (100,100,500)   &  0.00 (0.00) &   0.00 (0.00) &    NA (NA) &   0.16 (0.02) &    NA (NA) &   0.87 (0.01) &   0.00 (0.00) &   0.87 (0.01) &  75.00 (0.00) &  21.00 (2.72)  \\   
				&  (200,200,500) &  0.00 (0.00) &    0.00 (0.00) &     NA (NA) &    0.21 (0.01) &     NA (NA) &    0.84 (0.01) &    0.02 (0.01) &    0.84 (0.01) &  149.80 (0.13) &   52.30 (2.31) \\   
				&  (300,300,500) &  0.00 (0.00) &    0.00 (0.00) &     NA (NA) &    0.22 (0.01) &     NA (NA) &    0.83 (0.01) &    0.01 (0.01) &    0.83 (0.01) &  224.80 (0.13) &   84.30 (5.17)   \\     
				Random       &  (100,100,500) & 0.00 (0.00) &   0.00 (0.00) &    NA (NA) &   0.14 (0.01) &   NA (NA) &   0.74 (0.02) &   0.08 (0.02) &   0.74 (0.02) &  73.00 (1.56) &  33.90 (1.98)\\ 
				&  (200,200,500) & 0.00 (0.00) &    0.00 (0.00) &    0.52 (0.08) &    0.17 (0.01) &    0.03 (0.01) &    0.69 (0.01) &    0.09 (0.01) &    0.70 (0.01) &  147.90 (5.32) &   78.40 (3.25) \\   
				&  (300,300,500) &   0.00 (0.00) &    0.00 (0.00) &    0.61 (0.04) &    0.26 (0.01) &    0.03 (0.00) &    0.64 (0.00) &    0.07 (0.01) &    0.65 (0.00) &  224.30 (6.86) &  144.00 (3.69)   \\   
				\bottomrule
			\end{tabular}
		}
	\end{center}
	\vskip -0.1in
	\begin{center}
		\centering Count
		\vskip 0.05in
		\resizebox{\linewidth}{!}{
			\begin{tabular}{lccccccccccc}
				\toprule
				\multirow{2}{*}{Graph} &
				\multirow{2}{*}{$(p,q,n)$} &
				\multicolumn{2}{c}{FPR} &
				\multicolumn{2}{c}{FDR} & 
				\multicolumn{2}{c}{F-score} &
				\multicolumn{2}{c}{MCC}  &
				\multicolumn{2}{c}{SHD}  \\
				&    & {No-tears} & {GAMPI} & {No-tears} & {GAMPI} & {No-tears} & {GAMPI}  & {No-tears} & {GAMPI}  & {No-tears} & {GAMPI}\\
				\midrule
				Hub          &  (100,100,500) &  0.00 (0.00) &   0.00 (0.00) &    NA (NA) &   0.00 (0.00) &    NA (NA) &   1.00 (0.00) &   0.00 (0.00) &   1.00 (0.00) &  99.00 (0.00) &   0.30 (0.30)  \\ 
				&  (200,200,500) &  0.00 (0.00) &    0.00 (0.00) &     NA (NA) &    0.00 (0.00) &     NA (NA) &    1.00 (0.00) &    0.00 (0.00) &    1.00 (0.00) &  199.00 (0.00) &    1.40 (1.19) \\ 
				&  (300,300,500) &  0.00 (0.00) &    0.00 (0.00) &     NA (NA) &    0.00 (0.00) &     NA (NA) &    1.00 (0.00) &    0.00 (0.00) &    1.00 (0.00) &  299.00 (0.00) &    2.50 (0.79)\\
				Chain        &  (100,100,500)   &  0.00 (0.00) &   0.00 (0.00) &    NA (NA) &   0.00 (0.00) &    NA (NA) &   0.95 (0.00) &   0.00 (0.00) &   0.95 (0.00) &  75.00 (0.00) &   7.30 (0.58) \\   
				&  (200,200,500) &  0.00 (0.00) &    0.00 (0.00) &     NA (NA) &    0.00 (0.00) &     NA (NA) &    0.94 (0.01) &    0.00 (0.00) &    0.94 (0.01) &  150.00 (0.00) &   17.80 (1.58)    \\   
				&  (300,300,500) &  0.00 (0.00) &    0.00 (0.00) &     NA (NA) &    0.00 (0.00) &     NA (NA) &    0.92 (0.00) &    0.00 (0.00) &    0.92 (0.00) &  225.00 (0.00) &   32.60 (1.45)  \\           
				Random        &  (100,100,500)   &  0.00 (0.00) &   0.00 (0.00) &    NA (NA) &   0.00 (0.00) &    NA (NA) &   0.92 (0.01) &   0.00 (0.00) &   0.92 (0.01) &  73.00 (2.93) &  11.10 (1.28)  \\   
				&  (200,200,500) &  0.00 (0.00) &    0.00 (0.00) &     NA (NA) &    0.01 (0.00) &     NA (NA) &    0.89 (0.01) &    0.00 (0.00) &    0.89 (0.01) &  154.60 (3.88) &   31.40 (2.79) \\   
				&  (300,300,500) &  0.00 (0.00) &    0.00 (0.00) &     NA (NA) &    0.00 (0.00) &     NA (NA) &    0.88 (0.01) &    0.00 (0.00) &    0.89 (0.01) &  228.60 (2.93) &   49.20 (2.45)    \\                   
				\bottomrule
			\end{tabular}
		}
	\end{center}
	\vskip -0.15in
	\caption{Comparison of causal graph reconstruction accuracy of GAMPI and NOTEARS in the presence of confounders, with GAMPI employing EBIC for tuning parameter selection and NOTEARS applying the default value of 0.1. Metrics include FPR, FDR, F-score, MCC, and SHD. NA indicates that the method returns an empty graph with no edges selected.}
	\label{sim_conf_ebic}
\end{table}

In the Supplementary Materials, we further compare our deconfounding approach via DRI with employing the standard GLM in Algorithm~\ref{alg:peeling_confounder} for binary outcomes. For the chain graph, the standard logistic regression without adjusting for confounders does not perform well in terms of causal discovery. This is because the unobserved confounders induce false positive edges between the node and its ancestors. By contrast, the deconfounding approach corrects the bias of the confounders and recovers the true graph structure. For the hub graph, though both two approaches recover the true causal graph, the confounding approach still outperforms the standard logistic regression in terms of parameter estimation. Note that our peeling algorithm in the first stage identifies the correct ancestral relationships (super-graph) as the confounders are independent of the instrumental variables by assumption.

In addition, in the Supplementary Materials, we consider the special case when the data is simulated without confounders. The result suggests that our deconfounding approach performs well even when the data is simulated without confounders. Last, we consider the simulation setup where the data has repeated measurements. Still, our deconfounding approach using a mixed effect model outperforms the standard GLM approach. To summarize, our deconfounding approach demonstrates strong empirical performance and outperforms the existing methods in most cases.

\section{Mixed DAG Networks: Direct Effect to AD}\label{case_study}

This section applies GAMPI to a publicly available Alzheimer's Disease Neuroimaging Initiative (ADNI) dataset. Our goal is to estimate a regulatory gene expression network of a subset of genes related to Alzheimer's disease (AD) and identify which of the genes have a direct causal effect on AD through gene-to-gene and gene-to-AD regulatory networks.

First, we download the raw data from the ANDI website (\url{https://adni.loni.usc.edu}), containing gene expression, DNA sequencing, and phenotypic data. Then, for preprocessing, we clean and merge these data to obtain 712 subjects with complete records. In addition, from the KEGG database \citep{kanehisa2002kegg}, we extract the AD reference pathway (hsa05010, \url{https://www.genome.jp/pathway/hsa05010}) and therefore obtain 146 genes from the ANDI data. Meanwhile, the subjects are categorized into four groups: Cognitive Normal (CN), Early Mild Cognitive Impairment (EMCI), Late Mild Cognitive Impairment (LMCI), and Alzheimer's Disease (AD). We treat the 247 CN individuals as the control group and the remaining 465 AD and MCI individuals as the case group. We then include the disease status, a binary outcome with $0/1$ indicating normal/AD, as an additional variable (node) to identify which genes are directly related to AD.

To perform data analysis, we first regress the gene expressions on the additional covariates, including age, gender, education, handedness, and intracranial volume. Next, for each SNP from a gene, we perform significance tests with the gene and disease status marginally and select the genes which have at least one SNP whose i) significance level with the gene is less than $0.05$ and ii) significance level with the disease status is less than $0.02$, rendering $p = 39$ primary variables.
For these genes, we extract their two most correlated SNPs with the disease status based on the p-values given the significance level with the gene less than $0.05$, yielding $q = 39 \times 2 = 78$ instrumental variables. Removing duplicate SNPs and the gene that has the same SNPs as other genes results in $p=38$ and $q=76$. To summarize, we use the gene expressions along with the disease status as primary variables and SNPs as instrumental variables to reconstruct a causal network for gene-to-gene and gene-to-disease regulatory relationships.

\begin{figure}
	\centering
    \includegraphics[scale = 0.55]{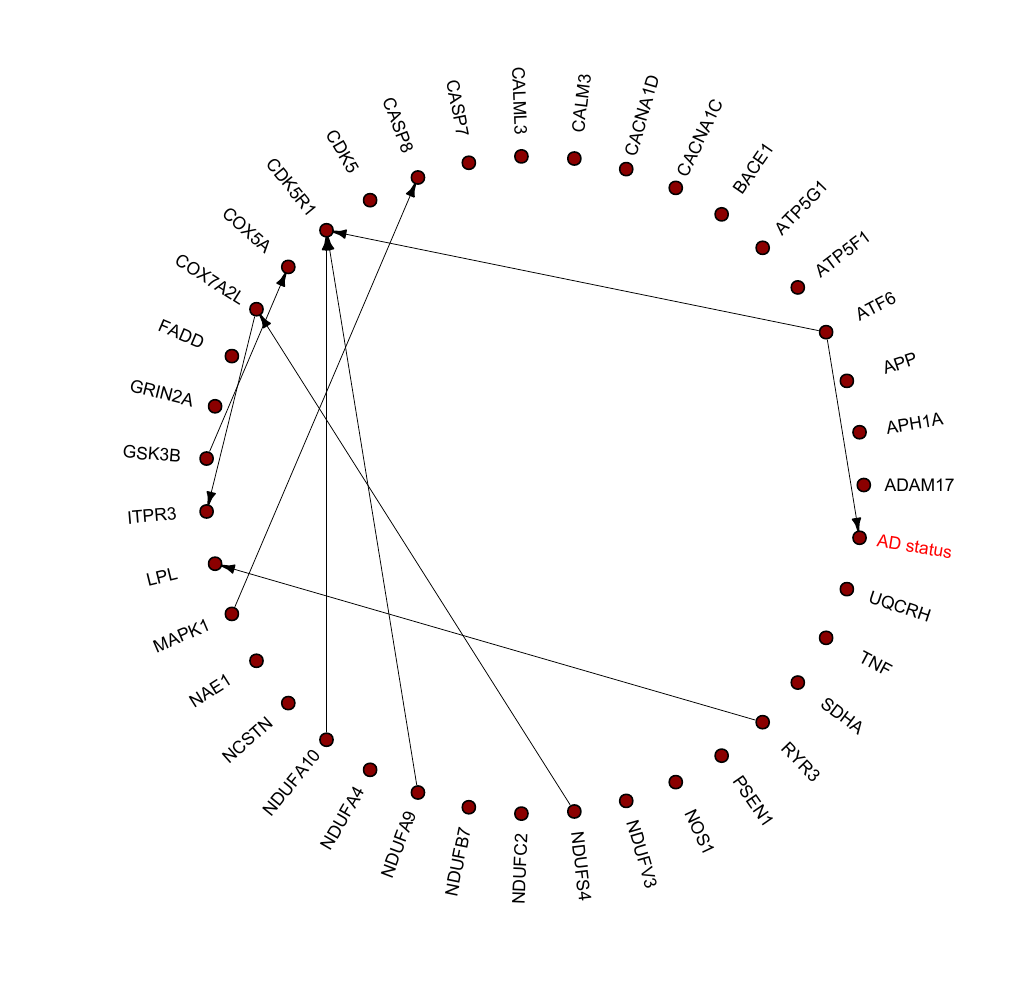}
    \vspace{-0.5cm}
	\caption{Reconstructed gene-to-gene and gene-to-AD regulatory network. The ``AD status" is a binary outcome with $0/1$ indicating normal/AD. Directed edges indicate causal relationships identified by the proposed GAMPI.
}
	\label{fig:case_study_disease_status}
\end{figure}

As shown in Figure~\ref{fig:case_study_disease_status}, GAMPI identifies a direct causal effect of  gene ATF6 on the AD status. In the literature, ATF6 is a transcription factor that acts during endoplasmic reticulum (ER) stress by activating UPR target genes, and ER stress is known to be closely associated with AD. Furthermore, \citet{du2020activating} suggested that ATF6 could be a potential hub for targeting the treatment of AD, which protects the retention of spatial memory in AD model mice. \citet{zhang2022activating} found that the expression of both ATF6 and CTH are decreased in AD patients and ATF6 positively regulates the expression of CTH so that the addition of CTH reduces the loss of spatial learning and memory ability in mice caused by ATF6 reduction. 
In addition, GAMPI uncovers some known regulatory relationships related to AD in the literature for both
the AD and control groups. For example, for the directed connection {MAPK1 $\to$ CASP8}, it has been shown that phosphorylation of p38 MAPK induced by oxidative stress is associated with the activation of caspase-8-mediated apoptotic pathways in dopaminergic neurons \citep{choi2004phosphorylation}. The connection {ATF6 $\to$ CDK5R1} is in the AD KEGG pathway \url{https://www.genome.jp/pathway/hsa05010}. Furthermore, the approach also identifies some potential gene regulatory relationships for future biological investigations. For example, the two genes in the connection {RYR3 $\to$ LPL} are among the 13 genes directly associated with AD in the DEX DFC geneset analysis \citep{sharma2021common}, while the two genes in the connection {GSK3B $\to$ COX5A} are in the same AD-related protein association network in AD-iPS5 neurons \citep{hossini2015induced}.

\section{Discussion}

The article introduces a new causal discovery approach, GAMPI, which reconstructs a directed acyclic graph using instruments in the presence of unmeasured confounders. GAMPI involves generalized structural equation models that are identifiable with the help of instruments under certain conditions.
GAMPI involves two steps. First, we proposed a fidelity model that is also a generalized linear model, having the same support as the marginal model regarding
instrumental interventions. On this ground, we designed a bottom-up peeling algorithm to identify ancestral relationships and valid instruments by exploiting the connection between primary and instrumental variables. In the second step, we proposed a deconfounding approach to further select parent-child relationships from the identified ancestral relationships. This approach estimates the confounding effects from the parent's equations and uses them in subsequent child equations to correct the confounding effects.
The theoretical properties of GAMPI are also analyzed, including the globality of the DC solution for nonconvex minimization, estimation accuracy, and causal graph selection consistency.

Overall, GAMPI provides a promising approach to causal discovery, with potential applications in various fields beyond Alzheimer's disease. For instance, the method can be used to explore causal relationships in complex systems with unmeasured confounders, such as in economics or public health. Furthermore, GAMPI's flexibility to adapt to different distributions of confounders and link functions makes it suitable for a wide range of scenarios. For instance, it can handle directed graphical models with mixed variables \citep{chowdhury2022dagbagm}.  In conclusion, GAMPI offers a valuable contribution to causal inference by providing a practical method for identifying causal relationships under challenging situations.

\section*{Acknowledgements}

The research is supported in part by NSF grant DMS-1952539, NIH grants R01GM113250, R01GM126002, R01AG065636, R01AG074858, R01AG069895, U01AG073079.

\newpage
\begin{appendix}

\begin{center}
{\bf \LARGE Causal Discovery with Generalized Linear Models through Peeling Algorithms: Supplementary Materials} 
\bigskip

{\large Minjie Wang, Xiaotong Shen, and Wei Pan}
\end{center}

\section{Illustrative Examples}

In this section, we delve into detailed examples that elucidate the fidelity models, the peeling algorithm, and the majority rule for a linear link as outlined in Assumption~\ref{identifiability_assumption}(B).

\subsection{Fidelity Model}\label{appen_fidelity}

\begin{figure}[H]
    \centering
    \begin{tikzpicture}[scale=0.8,transform shape ,main/.style = {draw, circle}] 
\node[main] (1) {$Y_1$}; 
\node[main] (2) [right of =1] {$Y_2$};
\node[main] (3) [right of =2] {$Y_3$};
\node[main] (4) [right of =3] {$Y_4$};
\node[main] (5) [right of =4] {$Y_5$};
\node[main] (6) [above of =1,node distance=1.75cm] {$X_1$}; 
\node[main] (7) [above of =2,node distance=1.75cm] {$X_2$}; 
\node[main] (8) [above of =3,node distance=1.75cm] {$X_3$}; 
\node[main] (9) [above of =4,node distance=1.75cm] {$X_4$}; 
\node[main] (10) [above of =5,node distance=1.75cm] {$X_5$};
\node[main] (h1) [below of =1,node distance=1.75cm] {$h_1$};
\node[main] (h2) [below of =2,node distance=1.75cm] {$h_2$}; 
\node[main] (h3) [below of =3,node distance=1.75cm] {$h_3$}; 
\node[main] (h4) [below of =4,node distance=1.75cm] {$h_4$}; 
\node[main] (h5) [below of =5,node distance=1.75cm] {$h_5$}; 
\path (1) edge (2);
\path (2) edge (3);
\path (3) edge (4);
\path (6) edge (1);
\path (7) edge (2);
\path (8) edge (3);
\path (9) edge (4);
\path (10) edge (5);
\path[bend right=30] (1) edge (4);
\path (h1) edge (1);
\path (h2) edge (2);
\path (h3) edge (3);
\path (h4) edge (4);
\path (h5) edge (5);
\end{tikzpicture}
    \caption{Example DAG defined by model~\eqref{eq:example_smm}.}
    \label{fig:example_smm}
\end{figure}
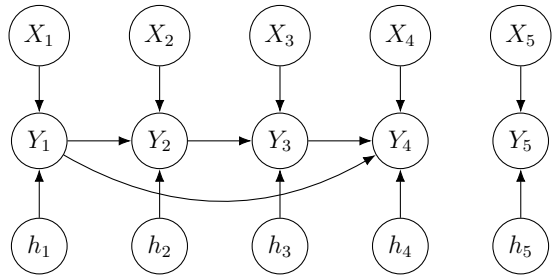

Consider an example of a generalized structural equation model for binary outcomes with $p = q = 5$:
\begin{align}
&\psi(\E[Y_1|X_1,h_1]) = 2X_1 + h_1, && \psi(\E[Y_2|Y_1,X_2,h_2]) = 1.5 Y_1 + 2X_2 + h_2,  \nonumber \\
&\psi(\E[Y_3|Y_2,X_3,h_3]) = 1.5 Y_2 + 2X_3 + h_3, && \psi(\E[Y_4|Y_3,Y_1,X_4,h_4]) = -1.5 Y_1 + 1.5 Y_3  + 2X_4 + h_4, \nonumber\\
& \psi(\E[Y_5|X_5,h_5]) = 2X_5 + h_5, \label{eq:example_smm}
\end{align}
where $\psi_1 = \cdots = \psi_5 = \psi$ is the logit link function. Here, \eqref{eq:example_smm} defines a DAG demonstrated in Figure~\ref{fig:example_smm}. In addition, marginalizing each equation in~\eqref{eq:example_smm} over $\Y$ does not lead to closed-form expressions.

The proposed fidelity model that has the same support as the marginal model of \eqref{eq:example_smm} is:
\begin{align*}
&\psi(\E[Y_1|X_1]) = V_{11} X_1, && \psi(\E[Y_3|X_1,X_2,X_3]) = V_{13} X_1 + V_{23} X_2 + V_{33} X_3 + h_3, \\
&\psi(\E[Y_2|X_1,X_2]) = V_{12} X_1 + V_{22} X_2, && \psi(\E[Y_4|X_1,X_2,X_3,X_4]) = V_{14} X_1 + V_{24} X_2 + V_{34} X_3  + V_{44} X_4  + h_4, \\
& \psi(\E[Y_5|X_5]) = V_{55} X_5 .
\end{align*}

Note that the fidelity model has the same support as the true marginal model and ancestral relationships can be identified via $\V$ by Proposition~\ref{prop_peeling} using the peeling algorithm.

\subsection{Peeling Algorithm}\label{appen_peeling}

We now illustrate the peeling algorithm (Algorithm~\ref{alg:peeling}) with the above motivating example. From~\eqref{eq:example_smm}, we generate the data of sample size $n = 500$ and compute $\widehat \V$ using Algorithm~\ref{alg:dc_alg}. The estimated $\widehat \V$ is:

\begin{align*}
\widehat \V_{q \times p} = \begin{pmatrix}
2.06 & 0.35 & 0.00 & -0.35 & 0.00 \\
0.00 & 1.84 & 0.46 & 0.00 & 0.00 \\
0.00 & 0.00 & 1.77 & 0.39 & 0.00 \\
0.00 & 0.00 & 0.00 & 1.76 & 0.00 \\
0.00 & 0.00 & 0.00 & 0.00 & 1.96
\end{pmatrix}.
\end{align*}

Algorithm~\ref{alg:peeling} proceeds as follows.

\begin{itemize}
    \item \textit{Iteration 1:} $X_4$ is identified as an instrument of leaf node $Y_4$ ($X_4  \rightarrow Y_4$) as row 4 has the smallest row-wise $\ell_0$-norm and $\widehat V_{44}$ is the only nonzero item in row 4.

    $X_5$ is identified as an instrument of leaf node $Y_5$ ($X_5  \rightarrow Y_5$) as row 5 has the smallest row-wise $\ell_0$-norm and $\widehat V_{55}$ is the only nonzero item in row 5.

    $Y_4$, $Y_5$, $X_4$, and $X_5$ are removed.

    \item \textit{Iteration 2:} $X_3$ is identified as an instrument of leaf node $Y_3$ ($X_3  \rightarrow Y_3$) in the subgraph for $Y_1$, $Y_2$ and $Y_3$ as row 3 has the smallest row-wise $\ell_0$-norm of the submatrix for $Y_1$, $Y_2$ and $Y_3$, with $\widehat V_{33}$ the only nonzero item in row 3. Moreover, since $\widehat V_{34} \neq 0$ and $Y_4$ is removed in the previous iteration, $Y_3 \rightsquigarrow Y_4$.

    $Y_3$ and $X_3$ are removed.

    \item \textit{Iteration 3:} Similarly, $X_2$ is identified as an instrument of leaf node $Y_2$ ($X_2  \rightarrow Y_2$) in the subgraph for $Y_1$, $Y_2$. Moreover, since $\widehat V_{23} \neq 0$ and $Y_3$ is removed in the previous iteration, $Y_2 \rightsquigarrow Y_3$.

    $Y_2$ and $X_2$ are removed.

    \item \textit{Iteration 4:} Similarly, $X_1$ is identified as an instrument of leaf node $Y_1$ ($X_1  \rightarrow Y_1$). Moreover, since $\widehat V_{12}$, $\widehat V_{14} \neq 0$ and $Y_2$, $Y_4$ are removed in the previous iterations, $Y_1 \rightsquigarrow Y_2$ and $Y_1 \rightsquigarrow Y_4$.

     $Y_1$ and $X_1$ are removed and the peeling process is terminated.

\end{itemize}

Finally, step 5 adds ancestral relations: $Y_1 \rightsquigarrow Y_3$ and $Y_2 \rightsquigarrow Y_4$. 
To conclude, Algorithm~\ref{alg:peeling} identifies ancestral relationships: $Y_1 \rightsquigarrow Y_2$, $Y_1 \rightsquigarrow Y_3$, $Y_1 \rightsquigarrow Y_4$, $Y_2 \rightsquigarrow Y_3$, $Y_2 \rightsquigarrow Y_4$ and $Y_3 \rightsquigarrow Y_4$.

\subsection{Majority Rule}\label{majority}

Consider the following example of a generalized structural equation model:
\begin{align}
&\psi_1(\E[Y_1|X_1,X_2,X_3,h_1]) = W_{11} X_1 + W_{21} X_2 + W_{31} X_3 + h_1, \nonumber \\
&\psi_2(\E[Y_2|Y_1,X_2,X_3,X_4,h_2]) = U_{12} Y_1 + W_{22} X_2  + W_{32} X_3 +  W_{42} X_4 + h_2,   \label{eq:example_majority}
\end{align}
where $X_1$ is an valid IV of $Y_1$, $X_4$ is an valid IV of $Y_2$, and $X_2$, $X_3$ are invalid IVs.
Note if $\psi_1$ is linear, then~\eqref{eq:example_majority} is not identifiable as the majority rule is not satisfied \citep{kang2016instrumental,windmeijer2019use}. 
If $\psi_1$ is non-linear, then the linear effect of the instruments in the first equation cannot be represented by the one in the second equation. Hence, identifiability is achieved through non-linearity and the majority rule is not required (details are given in the proof of Proposition~\ref{identifiability_prop}).

\section{General Form of Deconfounding Algorithm}

Algorithm~\ref{alg:peeling_confounder_base} serves as a general version of Algorithm~\ref{alg:peeling_confounder} in the main paper for estimating parent-child relationships in the presence of confounders. In Algorithm~\ref{alg:peeling_confounder} of the main paper, we utilize residuals from a GLM to impute confounders. The underlying intuition of this deconfounding approach is to achieve accurate parameter estimates and construct a consistent estimate of unmeasured confounders via residuals through the root equations. In this regard, different models can be employed to estimate confounding effects in the root equations, as described in Algorithm~\ref{alg:peeling_confounder_base}. For instance, the marginal likelihood that integrates the confounding effect $h_k$ from the complete likelihood can be applied, in addition to the Markov Chain Monte Carlo approach \citep{knudson2021likelihood}.

\begin{algorithm}
	\caption{General peeling algorithm for estimating parent-child relationships via DRI}
	\label{alg:peeling_confounder_base}
	\begin{algorithmic}
		
		\STATE {1.} Input $(\steponesuperset{\text{an}}(j), \steponesuperset{\text{in}}(j))_{j=1}^p$ and $\hat \pi$ from Algorithm~\ref{alg:peeling}. Input data matrix $(Y_{ij},X_{ij})_{n \times (p+q)}=(\Y_{i\bullet},\X_{i\bullet})_{i=1}^n$ of primary variables $\Y_{n \times p}$ and instruments $\X_{n \times q}$.
		
		Begin Iteration: for $d=1\cdots,p$,
		
		\STATE {2.} \textbf{(Estimation of confounding effects)} If $\hat \pi_d$ is a root variable indexed by $Y_k$, obtain an estimate of the confounding effect $\widehat h_k$.
		
		\STATE {3.}  \textbf{(Deconfounding)} If $\hat \pi_d$ is a non-root variable indexed by $Y_j$, compute $(\widehat{\W}_{{\steponesuperset{\text{in}}(j)}, j},\widehat{\U}_{\steponesuperset{\text{an}}(j), j},\widehat{\balpha}_{\steponesuperset{\text{an}}(j), j})$ by fitting a TLP-constrained GLM regression of $Y_j$ in \eqref{full-like}.
		Compute the residuals $\widehat h_j$  in \eqref{residual}.

	\end{algorithmic}
\end{algorithm}

Specifically, when the data has repeated measurements, we propose to use a generalized linear mixed model (GLMM) for root equations in Algorithm~\ref{alg:peeling_confounder_mix_effect} as an alternative to Algorithm~\ref{alg:peeling_confounder} in the main paper. Consider the structural equation model:
\begin{align}
\psi_j(\E \left( \Y_{ij} | \Y_{i,\text{pa}(j)}, \X_{i\bullet},h_{ij} \right)) 
= \u_{\text{pa}(j)}^{\top} \Y_{i,\text{pa}(j)}+\w_{\text{in}(j)}^{\top} \X_{i,\text{in}(j)}+ h_{ij} \mathbf{1}_{n_i},     \quad j = 1,\cdots,p. \label{eq:true_mixed_effect_model}
\end{align}
Here, $i = 1,\cdots,N$ represents a group index with $k = 1,\cdots, n_i$ observations within a group. For each group, we observe an $n_i \times 1$ vector of responses, $\Y_{ij}$. Let $\X_{i\bullet}$ be an $n_i \times q$ fixed-effects design matrix, $\w_{\text{in}(j)}$ a $q \times 1$, and $\u_{\text{pa}(j)}$ a $p \times 1$ vector of fixed regression coefficients. Further, $h_{ij}$ denotes a group-specific vector of random intercepts.

Similarly, we adopt a two-stage deconfounding procedure to estimate parent-child relationships, with a GLMM in the root equation for improved confounder estimation. Specifically, we fit a GLMM on $Y_k$ using instrumental variables $X_{\text{in}(k)}$ and estimate the confounding effect $\widehat h_k$ via the estimated random effect. In the child equation, we impute unmeasured confounders using estimated values from the parent equation and fit a TLP-constrained GLM, similar to the previous approach.

\begin{algorithm}
	\caption{Peeling algorithm for estimating parent-child relationships in the presence of confounders using GLMM and DRI}
	\label{alg:peeling_confounder_mix_effect}
	\begin{algorithmic}
		
		\STATE {1.} Input $(\steponesuperset{\text{an}}(j), \steponesuperset{\text{in}}(j))_{j=1}^p$ and $\hat \pi$ from Algorithm~\ref{alg:peeling}. Input data matrix $(Y_{ij},X_{ij})_{n \times (p+q)}=(\Y_{i\bullet},\X_{i\bullet})_{i=1}^n$ of primary variables $\Y_{n \times p}$ and instruments $\X_{n \times q}$.
		
		Begin Iteration: for $d=1\cdots,p$,
		
		\STATE {2.} \textbf{(Estimation of confounding effects using GLMM for root equations)} If $\hat \pi_d$ is a root variable indexed by $Y_k$, estimate the confounding effect $h_k$ by fitting a GLMM on $Y_k$:
		\begin{align*}
		\E[  Y_k | \X_{\text{in}(k)}, h_k ] = \varphi_k(  \X_{\steponesuperset{\text{in}}(k)} \W_{\steponesuperset{\text{in}}(k), k}   + h_k ),
		\end{align*}
		where $h_k$ denotes 
		the random effects. Obtain the estimated confounding effect $\widehat h_k$.
		
		\STATE {3.}  \textbf{(Deconfounding)} If $\hat \pi_d$ is a non-root variable indexed by $Y_j$, compute $(\widehat{\W}_{{\steponesuperset{\text{in}}(j)}, j},\widehat{\U}_{\steponesuperset{\text{an}}(j), j},\widehat{\balpha}_{\steponesuperset{\text{an}}(j), j})$ by fitting a TLP-constrained GLM regression of $Y_j$ in \eqref{full-like}.
		Compute the residuals $\widehat h_j$  in \eqref{residual}.

	\end{algorithmic}
\end{algorithm}

Besides the residual inclusion approach proposed in the main paper for deconfounding, we also include a version of incorporating predictor substitution approach in GAMPI, referred to as DPS, in Algorithm~\ref{alg:peeling_confounder_DPS}.
Here, $\loss(\W_{{\steponesuperset{\text{in}}(j)}, j},\U_{\steponesuperset{\text{an}}(j), j}| \X_{\steponesuperset{\text{in}}(j)} ,\widehat \Y_{\steponesuperset{\text{an}}(j)}) = n^{-1} \sum_{i=1}^n \ell(Y_{ij}, \W^{\top}_{\steponesuperset{\text{in}}(j), j} \X_{i,\steponesuperset{\text{in}}(j)}  +  \U^{\top}_{\steponesuperset{\text{an}}(j), j} \widehat \Y_{i,\steponesuperset{\text{an}}(j)} )$, which indicates the endogenous variables are replaced by their predicted values.

\begin{algorithm}
	\caption{Peeling algorithm for estimating parent-child relationships via DPS}
	\label{alg:peeling_confounder_DPS}
	\begin{algorithmic}
		\STATE {1.} Input $(\steponesuperset{\text{an}}(j), \steponesuperset{\text{in}}(j))_{j=1}^p$ and $\hat \pi$ from Algorithm~\ref{alg:peeling}. Input data matrix $(Y_{ij},X_{ij})_{n \times (p+q)}=(\Y_{i\bullet},\X_{i\bullet})_{i=1}^n$ of primary variables $\Y_{n \times p}$ and instruments $\X_{n \times q}$.
		
		Begin Iteration: for $d=1\cdots,p$,
		
		\STATE {2.} \textbf{(Predictor substitution for root equation)} If $\hat \pi_d$ is a root variable indexed by $Y_k$, compute $\widehat{\W}_{{\steponesuperset{\text{in}}(k)}, k}$ by fitting a GLM regression of $Y_k$ on $\X$: $\E[  Y_k | \X ] = \varphi_k(  \X_{\steponesuperset{\text{in}}(k)} \W_{\steponesuperset{\text{in}}(k), k} )$.
		Impute the predictor: $\widehat Y_k =  \varphi_k( \X_{\steponesuperset{\text{in}}(k)}  \widehat \W_{\steponesuperset{\text{in}}(k), k} )$.

		\STATE {3.}  \textbf{(Predictor substitution for child equation)} If $\hat \pi_d$ is a non-root variable indexed by $Y_j$, compute $(\widehat{\W}_{{\steponesuperset{\text{in}}(j)}, j},\widehat{\U}_{\steponesuperset{\text{an}}(j), j})$ by fitting a TLP-constrained GLM regression of $Y_j$:
		\begin{align*}
		&   (\widehat{\W}_{{\steponesuperset{\text{in}}(j)}, j},\widehat{\U}_{\steponesuperset{\text{an}}(j), j})  \nonumber  =\argmin_{\W_{{\steponesuperset{\text{in}}(j)}, j},\U_{\steponesuperset{\text{an}}(j), j}}
		\loss(\W_{{\steponesuperset{\text{in}}(j)}, j},\U_{\steponesuperset{\text{an}}(j), j} | \X_{\steponesuperset{\text{in}}(j)} ,\widehat \Y_{\steponesuperset{\text{an}}(j)}) \nonumber \\
		&\text {  subject to } \quad     \sum_{k \in \steponesuperset{\text{an}}(j)} I(U_{kj} \neq 0)  \leq K_{j}, \quad j=1, \ldots, p.   
		\end{align*}
		
		Impute the predictor: $\widehat Y_j = \varphi_j(\Y_{\widehat{\text{pa}}(j)} \widehat \U_{\widehat{\text{pa}}(j), j} + \X_{\widehat{\text{in}}(j)} \widehat \W_{\widehat{\text{in}}(j), j})$.

	\end{algorithmic}
\end{algorithm}

\section{Additional Simulations}\label{additional_simulation}

This section provides additional simulations in the paper to demonstrate the necessity of deconfounding in GAMPI.

Ideally, one might suggest estimating the causal relationships directly using the nodewise GLM regression subject to the $\ell_0$-constraint in Algorithm~\ref{alg:peeling_confounder} of the main paper, without employing the deconfounding approach or adjusting for confounders. That is,
\begin{align}
&   (\widehat{\W}_{{\steponesuperset{\text{in}}(j)}, j},\widehat{\U}_{\steponesuperset{\text{an}}(j), j})  \nonumber  =\argmin_{\W_{{\steponesuperset{\text{in}}(j)}, j},\U_{\steponesuperset{\text{an}}(j), j}}
\loss(\W_{\steponesuperset{\text{in}}(j), j}, \U_{\steponesuperset{\text{an}}(j), j} |  \X_{\steponesuperset{\text{in}}(j)} ,\Y_{\steponesuperset{\text{an}}(j)}  ) \nonumber \\
&\text {  subject to } \quad     \sum_{k \in \steponesuperset{\text{an}}(j)} I(U_{kj} \neq 0) \leq K_{j}, 
\quad j=1, \ldots, p. 
\label{full-like-noconf}
\end{align}
Similarly, to select parent-child relationships from the ancestral relationships identified in the first stage, we penalize the number of nonzero elements of $\U$. That is, if $\widehat U_{kj} \neq 0$, then $Y_k$ is a parent of $Y_j$, or $Y_k \to Y_j$.

We show in Section~\ref{section_sim_no_conf} and~\ref{sim_conf_DPS} that GAMPI adjusting for confounders, performs as well as the above approach~\eqref{full-like-noconf} when the data is simulated without confounders and outperforms it in the presence of confounders.

\subsection{Absence of Confounders}\label{section_sim_no_conf}

This subsection considers the special case when the data is simulated without confounders for binary outcomes. Recall that the binary data is simulated from the Bernoulli distribution in Section~\ref{simulation_setting} of the main paper with $\h = 0$.

\begin{table}[ht]
	\begin{center}
		\resizebox{\linewidth}{!}{
			\begin{tabular}{lccccccccccc}
				\toprule
				\multirow{2}{*}{Graph} &
				\multirow{2}{*}{$(p,q,n)$} &
				\multicolumn{2}{c}{FPR} &
				\multicolumn{2}{c}{FDR} & 
				\multicolumn{2}{c}{F-score} &
				\multicolumn{2}{c}{MCC}  &
				\multicolumn{2}{c}{SHD}  \\
				&    & {GAMPI-no deconf} & {GAMPI} & {GAMPI-no deconf} & {GAMPI} & {GAMPI-no deconf} & {GAMPI}  & {GAMPI-no deconf} & {GAMPI}  & {GAMPI-no deconf} & {GAMPI} \\
				\midrule
				Hub         &  (100,100,300) &   0.00 (0.00) &  0.00 (0.00) &  0.03 (0.00) &  0.04 (0.01) &  0.98 (0.00) &  0.98 (0.00) &  0.98 (0.00) &  0.98 (0.00) &  3.70 (0.52) &  4.80 (0.71)   \\ 
				&  (100,100,400) &  0.00 (0.00) &  0.00 (0.00) &  0.02 (0.01) &  0.02 (0.00) &  0.99 (0.00) &  0.99 (0.00) &  0.99 (0.00) &  0.99 (0.00) &  2.20 (0.55) &  2.30 (0.52) \\ 
				&  (100,100,500) &  0.00 (0.00) &  0.00 (0.00) &  0.02 (0.00) &  0.02 (0.00) &  0.99 (0.00) &  0.99 (0.00) &  0.99 (0.00) &  0.99 (0.00) &  2.10 (0.38) &  1.80 (0.29)   \\
				Chain       &  (100,100,300) &  0.00 (0.00) &   0.00 (0.00) &   0.08 (0.01) &   0.07 (0.01) &   0.82 (0.01) &   0.82 (0.01) &   0.83 (0.01) &   0.83 (0.01) &  23.90 (1.45) &  23.80 (1.23)  \\ 
				&  (100,100,400) &  0.00 (0.00) &   0.00 (0.00) &   0.06 (0.00) &   0.06 (0.00) &   0.91 (0.01) &   0.91 (0.01) &   0.91 (0.01) &   0.91 (0.01) &  13.50 (0.95) &  13.20 (0.95) \\ 
				&  (100,100,500) &  0.00 (0.00) &  0.00 (0.00) &  0.05 (0.01) &  0.04 (0.01) &  0.94 (0.00) &  0.95 (0.00) &  0.94 (0.00) &  0.95 (0.00) &  8.30 (0.40) &  7.70 (0.45)  \\               
				Random      &  (100,100,300) &  0.00 (0.00) &   0.00 (0.00) &   0.10 (0.01) &   0.09 (0.01) &   0.85 (0.01) &   0.85 (0.01) &   0.85 (0.01) &   0.85 (0.01) &  20.10 (1.75) &  20.10 (1.67)  \\ 
				&  (100,100,400) &  0.00 (0.00) &   0.00 (0.00) &   0.07 (0.01) &   0.07 (0.01) &   0.93 (0.01) &   0.93 (0.01) &   0.93 (0.01) &   0.93 (0.01) &  10.80 (1.75) &  10.60 (1.61) \\   
				&  (100,100,500) &  0.00 (0.00) &  0.00 (0.00) &  0.04 (0.01) &  0.04 (0.00) &  0.97 (0.00) &  0.97 (0.01) &  0.97 (0.00) &  0.97 (0.01) &  4.80 (0.59) &  4.50 (0.64)    \\   
				\bottomrule
			\end{tabular}
		}
		\caption{Evaluating GAMPI's reconstruction accuracy for binary outcomes without confounders, utilizing the extended BIC (EBIC) for tuning parameter selection. Evaluation metrics include false positive rate (FPR), false discovery rate (FDR), F-score, Matthews correlation coefficient (MCC), and structural Hamming distance (SHD). ``GAMPI-no deconf" method refers to employing nodewise GLM regression approach without adjusting for confounders based on \eqref{full-like-noconf}.}
		\label{sim_no_conf}
	\end{center}
	\vskip -0.1in
\end{table}
Table~\ref{sim_no_conf} suggests that our deconfounding approach, GAMPI, performs well even when the data is simulated without confounders.

\subsection{Presence of Confounders}\label{sim_conf_DPS}

This subsection compares the DRI approach with that without adjusting for confounders under the simulation setting in the presence of confounders. Furthermore, we compare our deconfounding approach, DRI in Algorithm~\ref{alg:peeling_confounder} of the main paper, with that via predictor substitution (DPS) in Algorithm~\ref{alg:peeling_confounder_DPS}. In addition to the five metrics in the paper, we compute the estimation error $\|\widehat{\U} - \U^0\|_F^2$ to evaluate the accuracy of parameter estimation, where $\|\cdot\|_F$ denotes the Frobenius-norm.

\begin{table}[ht]
	\begin{center}
		\resizebox{\linewidth}{!}{
			\begin{tabular}{lccccccc}
				\toprule
				\multirow{2}{*}{Graph} &
				\multirow{2}{*}{$(p,q,n)$} &
				\multicolumn{3}{c}{F-score} &
				\multicolumn{3}{c}{$\| \widehat \U - \U^* \|_F^2$}  \\
				&     & {GAMPI-no deconf} & {GAMPI-DRI} & {GAMPI-DPS} & {GAMPI-no deconf} & {GAMPI-DRI} & {GAMPI-DPS}\\
				\midrule
				Hub    &  (100,100,500) &   0.96 (0.01) &   0.96 (0.01) &   0.91 (0.01) &  77.45 (5.28) &  58.11 (6.66) &  108.04 (12.11) \\
				&  (200,200,500) &   0.95 (0.01) &   0.95 (0.01) &   0.89 (0.01) &  185.12 (25.01) &  140.73 (25.02) &  255.65 (24.69)\\
				&  (300,300,500) &   0.95 (0.01) &   0.95 (0.01) &   0.90 (0.01) &  279.17 (35.82) &  200.55 (45.36) &  373.97 (47.27)\\
				Chain  &  (100,100,500) &   0.74 (0.01) &   0.87 (0.01) &   0.87 (0.01) &  118.06 (6.37) &  74.67 (4.74) &  93.06 (5.2) \\
				&  (200,200,500) &   0.71 (0.01) &    0.84 (0.01) &    0.84 (0.01) &  281.96 (12.01) &  189.05 (9.82) &  217.29 (10.96)\\
				&  (300,300,500) &   0.71 (0.01) &    0.83 (0.01) &    0.84 (0.01) &  415.11 (20.26) &  294.81 (12.57) &  332.47 (13.82) \\
				Random &  (100,100,500) &   0.71 (0.02) &   0.74 (0.02) &   0.74 (0.02) &  282.58 (15.98) &  278.67 (15.95) &  321.49 (15.86) \\
				&  (200,200,500) &   0.64 (0.01) &   0.69 (0.01) &   0.69 (0.01) &  656.68 (30.57) &  633.23 (31.49) &  711.17 (31.42)\\
				&  (300,300,500) &   0.59 (0.00) &    0.64 (0.00) &  0.62 (0.01) &  1111.06 (37.05) &  1053.11 (32.15) &  1174.76 (37.16)\\
				\bottomrule
			\end{tabular}
		}
		\caption{Assessing GAMPI's reconstruction accuracy for binary outcomes with confounders, employing the extended BIC (EBIC) for tuning parameter selection. Evaluation metrics include F-score and parameter estimation error in Frobenius norm. ``GAMPI-DPS" employs the predictor substitution (DPS) approach for deconfounding as proposed in Algorithm~\ref{alg:peeling_confounder_DPS}. ``GAMPI-DRI" uses the residual inclusion approach proposed in Algorithm~\ref{alg:peeling_confounder} of the main paper. In other tables, ``GAMPI" refers to the recommended ``GAMPI-DRI" approach.}
		\label{sim_conf_ebic_DPS}
	\end{center}
	\vskip -0.1in
\end{table}

Table~\ref{sim_conf_ebic_DPS} suggests that our deconfounding approach outperforms the standard GLM approach~\eqref{full-like-noconf} without adjusting for confounders in the presence of confounders. Moreover, deconfounding via DRI proposed in Algorithm~\ref{alg:peeling_confounder} of the main paper outperforms that using predictor substitution (DPS) in Algorithm~\ref{alg:peeling_confounder_DPS} in terms of parameter estimation. Our simulation result indicates that DRI is more suited than DPS for binary or count outcomes, which is concordant with the observation of \citet{terza2008two}.

\subsection{Presence of Confounders with Replicates}\label{sim_conf_mixed_effect}

In this subsection, we evaluate the performance of GAMPI using the generalized linear mixed model (GLMM) for root equations proposed in Algorithm~\ref{alg:peeling_confounder_mix_effect} under the simulation setting with repeated measurements.

Table~\ref{sim_conf_mixed_effect_ebic} suggests that our deconfounding approach using a GLMM outperforms the standard GLM approach, as it better estimates the confounders.

\begin{table}[ht]
	\begin{center}
		\resizebox{\linewidth}{!}{
			\begin{tabular}{lccccccc}
				\toprule
				\multirow{2}{*}{Graph} &
				\multirow{2}{*}{$(p,q,n)$} &
				\multicolumn{3}{c}{F-score} &
				\multicolumn{3}{c}{$\| \widehat \U - \U^* \|_F^2$}  \\
				&    & {GAMPI-no deconf} & {GAMPI} & {GAMPI-GLMM} & {GAMPI-no deconf} & {GAMPI} & {GAMPI-GLMM} \\
				\midrule
				Hub    &  (100,100,500) &  0.94 (0.02) &   0.94 (0.02) &   0.95 (0.02) &  102.96 (24.02) &  79.71 (23.14) &  61.6 (21.58) \\
				&  (200,200,500) &  0.92 (0.01) &   0.92 (0.01) &   0.93 (0.01) &  259.2 (24.82) &  209.5 (25.42) &  160.8 (21.94) \\
				&  (300,300,500) &  0.91 (0.02) &   0.91 (0.02) &   0.93 (0.02) &  357.33 (74.04) &  321.79 (70.8) &  251.28 (68.07)  \\
				Chain  &  (100,100,500) &  0.75 (0.01) &   0.87 (0.01) &   0.93 (0.01) &  125.64 (6.81) &  81.8 (5.5) &  65.99 (6.38)  \\  
				&  (200,200,500) &  0.71 (0.01) &    0.83 (0.01) &    0.91 (0.00) &  285.35 (12.08) &  201.56 (4.04) &  155.85 (3.66)  \\
				&  (300,300,500) &  0.69 (0.01) &    0.80 (0.01) &    0.89 (0.01) &  500.19 (20.73) &  366.98 (16.24) &  287.8 (11.69)  \\

				\bottomrule
			\end{tabular}
		}
	\caption{Evaluating GAMPI's reconstruction accuracy for binary outcomes in repeated measurements, using the extended BIC (EBIC) for tuning. Evaluation metrics include F-score and parameter estimation error in Frobenius norm. ``GAMPI-GLMM" refers to GAMPI using GLMM for root equations proposed in Algorithm~\ref{alg:peeling_confounder_mix_effect}.}
\label{sim_conf_mixed_effect_ebic}
	\end{center}
	\vskip -0.1in
\end{table}

\subsection{Comparison with DAGMA}\label{sim_DAGMA}

This subsection compares GAMPI with a recently proposed structure learning method, called DAGMA. DAGMA is designed only for Gaussian or logistic outcomes. Thus, we compare GAMPI with DAGMA for the binary outcomes case. Table~\ref{sim_conf_ebic_DAGMA} suggests that GAMPI continues to outperform DAGMA in most scenarios. Specifically, DAGMA performs equally well in the easy case, namely, the hub graph. However, in challenging situations like the chain and random graphs, GAMPI outperforms DAGMA significantly.

\begin{table}[ht]
	\begin{center}
		\resizebox{\linewidth}{!}{
			\begin{tabular}{lccccccccccc}
				\toprule
				\multicolumn{1}{l}{Binary} &
				\multirow{2}{*}{$(p,q,n)$} &
				\multicolumn{2}{c}{FPR} &
				\multicolumn{2}{c}{FDR} & 
				\multicolumn{2}{c}{F-score} &
				\multicolumn{2}{c}{MCC}  &
				\multicolumn{2}{c}{SHD}  \\
				Graph   & & {DAGMA} & {GAMPI} & {DAGMA} & {GAMPI} & {DAGMA} & {GAMPI}  & {DAGMA} & {GAMPI}  & {DAGMA} & {GAMPI}\\
				\midrule
				Hub          &  (100,100,500) & 0.00 (0.00) &  0.00 (0.00) &  0.04 (0.00) &  0.05 (0.01) &  0.98 (0.00) &  0.96 (0.01) &  0.98 (0.00) &  0.96 (0.01) &  4.20 (0.57) &  8.10 (1.46) \\ 
				&  (200,200,500) &  0.00 (0.00) &   0.00 (0.00) &   0.05 (0.01) &   0.04 (0.01) &   0.97 (0.00) &   0.95 (0.01) &   0.97 (0.00) &   0.95 (0.01) &  11.60 (1.60) &  20.40 (3.95) \\ 
				&  (300,300,500) & 0.00 (0.00) &   0.00 (0.00) &   0.07 (0.01) &   0.04 (0.01) &   0.96 (0.00) &   0.95 (0.01) &   0.96 (0.00) &   0.95 (0.01) &  24.40 (1.84) &  28.20 (7.61) \\
				Chain        &  (100,100,500)   &   0.01 (0.00) &   0.00 (0.00) &   0.43 (0.01) &   0.16 (0.02) &   0.68 (0.01) &   0.87 (0.01) &   0.70 (0.01) &   0.87 (0.01) &  51.70 (2.63) &  21.00 (2.72) \\   
				&  (200,200,500) &  0.00 (0.00) &    0.00 (0.00) &    0.53 (0.01) &    0.21 (0.01) &    0.60 (0.01) &    0.84 (0.01) &    0.62 (0.01) &    0.84 (0.01) &  145.60 (2.84) &   52.30 (2.31)  \\   
				&  (300,300,500) &   0.00 (0.00) &    0.00 (0.00) &    0.57 (0.01) &    0.22 (0.01) &    0.56 (0.01) &    0.83 (0.01) &    0.59 (0.01) &    0.83 (0.01) &  247.60 (4.02) &   84.30 (5.17)  \\                
				Random       &  (100,100,500) &  0.01 (0.00) &    0.00 (0.00) &    0.80 (0.01) &    0.14 (0.01) &    0.29 (0.01) &    0.74 (0.02) &    0.30 (0.02) &    0.74 (0.02) &  141.60 (3.54) &   33.90 (1.98)  \\ 
				&  (200,200,500) &  0.01 (0.00) &    0.00 (0.00) &    0.82 (0.01) &    0.17 (0.01) &    0.26 (0.01) &    0.69 (0.01) &    0.29 (0.01) &    0.70 (0.01) &  342.90 (5.44) &   78.40 (3.25) \\   
				&  (300,300,500) &   0.01 (0.00) &    0.00 (0.00) &    0.84 (0.01) &    0.26 (0.01) &    0.24 (0.01) &    0.64 (0.00) &    0.27 (0.01) &    0.65 (0.00) &  573.20 (6.88) &  144.00 (3.69)   \\   
				\bottomrule
			\end{tabular}
		}
		\caption{
Comparing reconstruction accuracy of GAMPI and DAGMA for binary outcomes with confounders, where GAMPI employs the extended BIC (EBIC) for tuning and DAGMA uses the default setting with a tuning parameter value of 0.02. Evaluation metrics include false positive rate (FPR), false discovery rate (FDR), F-score, Matthews correlation coefficient (MCC), and structural Hamming distance (SHD).}
		\label{sim_conf_ebic_DAGMA}
	\end{center}
	\vskip -0.1in
\end{table}

\subsection{Tuning Parameter Selection}\label{sim_tuning}

This section examines the performance of two tuning parameter selection approaches for GAMPI. We use either 5-fold cross-validation or the extended Bayesian information criterion (EBIC) to choose $(\tau_j,K_j)$ by minimizing the predictive likelihood or the EBIC criterion.
For cross-validation, we adopt the one-standard error rule which is commonly used for the high-dimensional data. We consider the base simulation in the presence of confounders. Table~\ref{sim_conf_tuning} suggests that the EBIC approach outperforms cross-validation in all settings.

\begin{table}[ht]
	\begin{center}
        \scalebox{0.8}{
			\begin{tabular}{lccccc}
				\toprule
				\toprule
				\multirow{2}{*}{Graph} &
				\multirow{2}{*}{$(p,q,n)$} &
				\multicolumn{2}{c}{F-score} &
				\multicolumn{2}{c}{SHD}  \\
				&     & {CV} & {EBIC} & {CV} & {EBIC} \\
				\midrule
				Hub    &  (100,100,500) &   0.70 (0.03) &   0.96 (0.01) &  44.30 (3.97) &   8.10 (1.46) \\
				&  (200,200,500) &   0.70 (0.04) &   0.95 (0.01) &  89.20 (10.25) &  20.40 (3.95) \\
				&  (300,300,500) &    0.74 (0.05) &    0.95 (0.01) &  120.30 (15.74) &   28.20 (7.61) \\
				Chain  &  (100,100,500) &  0.56 (0.03) &   0.87 (0.01) &  46.20 (1.96) &  21.00 (2.72)  \\
				&  (200,200,500) &  0.59 (0.01) &   0.84 (0.01) &  88.20 (2.10) &  52.30 (2.31) \\
				&  (300,300,500) &   0.56 (0.01) &    0.83 (0.01) &  137.50 (2.93) &   84.30 (5.17)\\
				Random &  (100,100,500) &   0.55 (0.02) &   0.74 (0.02) &  46.20 (2.03) &  33.90 (1.98) \\
				&  (200,200,500) &   0.48 (0.02) &    0.69 (0.01) &  102.50 (4.47) &   78.40 (3.25)  \\
				&  (300,300,500) &   0.47 (0.01) &    0.64 (0.00) &  158.90 (6.32) &  144.00 (3.69) \\
				\bottomrule
			\end{tabular}
   }
		\caption{Reconstruction accuracy of causal graph of GAMPI for binary outcomes in the presence of confounders, where GAMPI uses cross validation (CV) or the extended BIC (EBIC) for tuning parameter selection. Evaluation metrics include F-score and structural Hamming distance (SHD).}
		\label{sim_conf_tuning}
	\end{center}
	\vskip -0.1in
\end{table}

\section{Technical Proofs}

\subsection{Proof of Proposition~\ref{identifiability_prop}}
Assume that two structural equation models as in equation~\eqref{eq:true_model}, defined by $\boldsymbol{\theta} = (\U,\W)$ and $\tilde{\boldsymbol{\theta}} = (\widetilde \U,\widetilde \W)$, induce the same distribution of $(\Y,\X)$. We will show that $\boldsymbol{\theta} = \tilde{\boldsymbol{\theta}}$.

Let $G(\boldsymbol{\theta})$ and $G(\tilde{\boldsymbol{\theta}})$ be the DAGs corresponding to  $\boldsymbol{\theta}$ and $\tilde{\boldsymbol{\theta}}$. First, we show that the topological order of $Y_{1}, \cdots, Y_{p}$ is identifiable if two DAGs $G(\boldsymbol{\theta})$ and $G(\tilde{\boldsymbol{\theta}})$ have the same topological depth $d_{G}(j)=d(j)$ of each variable $Y_j$, $j=1, \cdots, p$. 
For $G(\boldsymbol{\theta})$, assume, without loss of generality, that $Y_{1}$ is a leaf node in $G(\boldsymbol{\theta})$. By Assumption~\ref{identifiability_assumption}(B), there exists a valid instrument, say $X_{1}$, that intervenes on $Y_{1}$. By Assumptions~\ref{identifiability_assumption}(A) and (ii),
\begin{align}
    \Cov\left(Y_{1}, X_{1} \mid \boldsymbol{Y}_{S}, \boldsymbol{X}_{\{2, \ldots, q\}}\right) \neq 0,  \quad \quad &\text {for any } S \subseteq\{2, \ldots, p\} ,  \label{eq:eq_identi} \\
    \Cov \left(Y_{j}, X_{1} \mid \boldsymbol{X}_{\{2, \ldots, q\}}\right)=0,   \quad \quad &j=2, \ldots, p. \label{eq:eq_identi2}
\end{align}
Hence, \eqref{eq:eq_identi} implies that $X_1 \to Y_1$ in $G(\tilde{\boldsymbol{\theta}})$.
Now suppose $Y_1$ is not a leaf node in $G(\tilde{\boldsymbol{\theta}})$ and there exists $Y_2$ such that $Y_1 \to Y_2$. Then, $\Cov \left(Y_2, X_{1} \mid \boldsymbol{X}_{\{2, \ldots, q\}}\right)=0$ by~\eqref{eq:eq_identi2} but $X_1 \to Y_1$ and $Y_1 \to Y_2$, which contradicts Assumption~\ref{identifiability_assumption}(A). Therefore, if $Y_1$ is a leaf node in $G(\boldsymbol{\theta})$, then $Y_1$ must also be a leaf node in $G(\tilde{\boldsymbol{\theta}})$.

Second, we show that $\boldsymbol{\theta} = \tilde{\boldsymbol{\theta}}$. Recall that for the $j$th equation,
$\psi_j(\E \left( Y_j | \Y_{\text{pa}(j)}, \X,h_j \right)) 
= \u_j^{\top} \Y_{\text{pa}(j)}+\w_j^{\top} \X_{\text{in}(j)}+h_j$, $j = 1,\cdots,p$.
Let $Y_k$ be a parent of $Y_j$. We can rewrite the above equation as
$\psi_j(\E \left( Y_j | \Y_{\text{pa}(j)}, \X,h_j \right)) 
= U_{kj} \Y_k       +  \U_{\text{pa}(j)\backslash k,j} \Y_{\text{pa}(j)\backslash k}     +\w_j^{\top} \X_{\text{in}(j)}+h_j$.
Similarly, from the $k$th equation,
$\E \left( \Y_k | \Y_{\text{pa}(k)}, \X,h_k \right) 
= \psi_k^{-1}(\u_k^{\top} \Y_{\text{pa}(k)}+\w_k^{\top} \X_{\text{in}(k)}+h_k)$.   
Therefore,
\begin{align}
&\E \left( \psi_j(\E \left( Y_j | \Y_{\text{pa}(j)}, \X,h_j \right)) -     \U_{\text{pa}(j)\backslash k,j} \Y_{\text{pa}(j)\backslash k} | \Y_{\text{pa}(k)}, \X,h_k \right) \nonumber
 \\
&  = U_{kj} \psi_k^{-1}(\u_k^{\top} \Y_{\text{pa}(k)}+\w_k^{\top} \X_{\text{in}(k)}+h_k)   +  \W_{{\text{in}(j)},j}^{\top} \X_{\text{in}(j)} + \E \left( h_j | \Y_{\text{pa}(k)}, \X,h_k \right) .   \label{eq:eq_identi3}
\end{align}
Note that the left-hand side is not equal to $\psi_j(\E \left( Y_j | \Y_{\text{pa}(k)}, \X,h_j \right))$ but
characterizes a proper conditional distribution.
Suppose $\u_k = \U_{\bullet k}$ is identified. We will show that $U_{kj}$ is identifiable and therefore $\boldsymbol{\theta}$ is also identifiable by induction on the topological depth.
If there exists $\tilde U_{kj} \neq U_{kj}$ and $\text{in}(j) \neq \widetilde{\text{in}}(j)$, which renders
the same conditional distribution~\eqref{eq:eq_identi3} in that
$U_{kj} \psi_k^{-1}(\u_k^{\top} \Y_{\text{pa}(k)}+\w_k^{\top} \X_{\text{in}(k)}+h_k) + \W_{{\text{in}(j)},j}^{\top} \X_{\text{in}(j)}+ \E \left( h_j | \Y_{\text{pa}(k)}, \X,h_k \right)= \widetilde{U}_{kj} \psi_k^{-1}(\u_k^{\top} \Y_{\text{pa}(k)}+\w_k^{\top} \X_{\text{in}(k)}+h_k) + \widetilde{\W}_{{\widetilde{\text{in}}(j)},j}^{\top} \X_{\widetilde{\text{in}}(j)}+ \E \left( h_j | \Y_{\text{pa}(k)}, \X,h_k \right)$.
Rearranging terms yields that 
\begin{align*}
    &U_{kj} \psi_k^{-1}(\u_k^{\top} \Y_{\text{pa}(k)}+\w_k^{\top} \X_{\text{in}(k)}+h_k) - \widetilde{U}_{kj} \psi_k^{-1}(\u_k^{\top} \Y_{\text{pa}(k)}+\w_k^{\top} \X_{\text{in}(k)}+h_k) \\
    &= \widetilde{\W}_{{\widetilde{\text{in}}(j)},j}^{\top} \X_{\widetilde{\text{in}}(j)} - \W_{{\text{in}(j)},j}^{\top} \X_{\text{in}(j)}  .
\end{align*}
If $\psi_k^{-1}(\cdot)$ is a non-linear function, then the left-hand side cannot be linearly represented by $\X_{\text{in}(j) \cup \widetilde{\text{in}}(j)}$. This implies $\widetilde U_{kj} = U_{kj}$, $\widetilde{\W}_{{\widetilde{\text{in}}(j)},j} = \W_{{\text{in}(j)},j}$ and  $\text{in}(j) = \widetilde{\text{in}}(j)$. If $\psi_k^{-1}(\cdot)$ is a linear function, then the same conclusion holds
under the majority rule that the number of valid IVs for $Y_j$ exceeds 50\% of its total number of IVs. To see this, denoting the set of valid IVs by $\text{in}_{*}$, by the majority rule, $|\text{in}_{*}(j)| > |\text{in}(j)|/2$, hence there must exist some valid IV, $l \in \text{in}_{*}(j) \cap \widetilde{\text{in}}_{*}(j)$, such that $X_l$ cannot be linearly represented by $\X_{\text{in}(k)}$. Again, we have $\widetilde U_{kj} = U_{kj}$, $\widetilde{\W}_{{\widetilde{\text{in}}(j)},j} = \W_{{\text{in}(j)},j}$ and  $\text{in}(j) = \widetilde{\text{in}}(j)$.
This completes the proof.

Before proving Proposition~\ref{fidelity_model_support} and Proposition~\ref{prop_peeling}, we first introduce Lemma~\ref{lemma_peeling} which investigates the marginal distribution defined by the true model $\P(Y_j | \X)$ and instruments.

\begin{lemma}\label{lemma_peeling} 
	If the marginal distribution under the true model $\P(Y_j | \X)$ satisfies: $\frac{\partial}{\partial X_l} \P(Y_j | \X) \neq 0$, then $X_l$ intervenes on $Y_j$ or an ancestor of $Y_j$. 
\end{lemma}

\noindent {\bf Proof of Lemma~\ref{lemma_peeling}}.
Note that the marginal distribution can be written as
\begin{align}
\p(Y_j|\X) =  \iint  \p(Y_j | \Y_{\text{pa}(j)}, \X_{\text{in}(j)},h_{j} )  \p(\Y_{\text{pa}(j)} |  \X )  f(h_j) \,d \Y_{\text{pa}(j)}  \,d h_j. \label{eq:product}
\end{align}
If $\frac{\partial}{\partial X_l} \P(Y_j | \X) \neq 0$, or equivalently, $\frac{\partial}{\partial X_l} \p(Y_j | \X) \neq 0$, then, by the product rule
and Assumption~\ref{identifiability_assumption}(C), i) $\frac{\partial}{\partial X_l} \p(Y_j | \Y_{\text{pa}(j)}, \X_{\text{in}(j)},h_{j} ) \neq 0$, or ii) $\frac{\partial}{\partial X_l} \p(\Y_{\text{pa}(j)} |  \X )  \neq 0$.
Note $\p(Y_j | \Y_{\text{pa}(j)}, \X_{\text{in}(j)},h_j) = \exp(  Y_{j} \big( \U_{\text{pa}(j),j}^{\top} \Y_{\text{pa}(j)} + \W_{\text{in}(j),j}^{\top} \X_{\text{in}(j)} + h_{j}  \big) - A_j( \U_{\text{pa}(j),j}^{\top} \Y_{\text{pa}(j)} + \W_{\text{in}(j),j}^{\top} \X_{\text{in}(j)} + h_{j} ))$ under \eqref{eq:true_model}. By the chain rule, 
$ \frac{\partial \p(Y_j | \Y_{\text{pa}(j)}, \X_{\text{in}(j)},h_j)}{\partial X_l} = \p(Y_j | \Y_{\text{pa}(j)}, \X_{\text{in}(j)},h_j) (Y_j - \varphi_j(\U_{\text{pa}(j),j}^{\top} \Y_{\text{pa}(j)}   + \W_{\text{in}(j),j}^{\top} \X_{\text{in}(j)} +h_j)) W_{lj}$.
Therefore, condition i) implies that $W_{lj} \neq 0$ and  $l \in \text{in}(j)$. Condition ii) implies that there exists an $m \in \text{pa}(j)$ such that $\frac{\partial}{\partial X_l} \p( Y_m |  \X )  \neq 0$. Similarly, this implies $l \in \text{in}(m), m \in \text{pa}(j)$, or there exists an $r \in \text{pa}(m)$ such that $\frac{\partial}{\partial X_l} \p( Y_r |  \X )  \neq 0$. By induction, we conclude that if $V_{lj} \neq 0$, then (i) there exists an $l \in \text{in}(j)$ such that $W_{lj} \neq 0$, or (ii) there exists an $k \in \text{an}(j)$ and $l \in \text{in}(k)$ such that $W_{lk} \neq 0$. Hence, $X_l$ intervenes on $Y_j$ or an ancestor of $Y_j$.

\subsection{Proof of Proposition~\ref{fidelity_model_support}}

Recall that $\supportFidelity=\{l: V_{lj} \neq 0\}$ for the fidelity model \eqref{eq:fidelity_model} and $\supportTrueMarginal=\{l: \frac{\partial \mathbb P(Y_j| \X)}{\partial X_l}\neq 0\}$ for the true model $\mathbb P \left(Y_j | \X \right)$. Next, we will show that $\{l: V_{lj} \neq 0\} = \supportTrueMarginal$, implying that the fidelity model $\mathbb P^*(Y_j|\X)$ and the marginal model $\P(Y_j|\X)$ have the same support.

For any $l \in \supportTrueMarginal$, $\frac{\partial \mathbb P(Y_j| \X)}{\partial X_l}\neq 0$. By Lemma~\ref{lemma_peeling}, $X_l$ intervenes on $Y_j$ or an ancestor of $Y_j$, that is, $l \in \{\text{in}(j)\} \cup \text{in}({\text{an}(j)})$.
Hence, there exists a path in the graph from $X_l$ to $Y_j$: $X_l \to Y_k \to \cdots \to Y_j$.
By the local faithfulness in Assumption 1(A), $\Cov(X_l,Y_j) \neq 0$. This implies that $V_{lj} \neq 0$ in the fidelity model. Otherwise, suppose $V_{lj} = 0$. By \eqref{eq:fidelity_model}, $\E( Y_j | X_l, \X_{-l}) = \E(Y_j |  \X_{-l})$, implying that $\P(Y_j | X_l, \X_{-l}) = 
\P(Y_j |  \X_{-l})$ and thus $\Cov(X_l,Y_j) = 0$ by the definition of conditional independence, which contradicts $\Cov(X_l,Y_j) \neq 0$. Hence, $V_{lj} \neq 0$ in the fidelity model or $l \in \supportFidelity$, implying $\supportTrueMarginal \subset \supportFidelity$.

On the other hand, for any $l \in \supportFidelity$,  $V_{lj} \neq 0$. Then, $\E[ Y_j | X_l, \X_{-l} ] \neq \E[ Y_j |  \X_{-l} ]$. Now, suppose $\frac{\partial \P(Y_j|\X)}{\partial X_l} = 0$. Then, as in \eqref{eq:product}, there does not exist a path in the graph from $X_l$ to $Y_j$. Thus, $\Cov(X_l,Y_j) = 0$, which contradicts $\E[ Y_j | X_l, \X_{-l} ] \neq \E[ Y_j |  \X_{-l} ]$. Hence, $l \in \supportTrueMarginal$ and thus $\supportFidelity \subset \supportTrueMarginal$. This establishes that $\supportFidelity = \supportTrueMarginal$.

\subsection{Proof of Proposition~\ref{prop_peeling}}

If $V_{lj} \neq 0$, then $\frac{\partial}{\partial X_l} \P(Y_j | \X) \neq 0$ by Proposition~\ref{fidelity_model_support}. By Lemma~\ref{lemma_peeling}, 
	$X_l$ intervenes on $Y_j$ or an ancestor of $Y_j$.

Moreover, for a leaf node $Y_j$, there exists an instrument $X_l \to Y_j$ by Assumption~\ref{identifiability_assumption} (B). If there exists $j' \neq j$ such that $V_{lj'} \neq 0$, then $Y_j$ must be an ancestor of $Y_{j'}$, which contradicts the fact that $Y_j$ is a leaf node. 
On the other hand, suppose $V_{lj} \neq 0$ and $V_{lj'} = 0$, $\forall j' \neq j$. If $Y_j$ is not a leaf node, then there exists a $Y_{j'}$ such that $Y_j$ is a parent of $Y_{j'}$. This implies $\frac{\partial}{\partial X_l} \p(Y_{j'} | \X) \neq 0$ and thus $V_{lj'} \neq 0$, which contradicts $\| V_{l \bullet}\|_0 = 1$.

\subsection{Proof of Theorem~\ref{thm_anc_sele_cons}}
Let $\truesetV = \{l: V_{lj}^0 \neq 0\}$ and $S_{j}^{[t]} = \{l: |\widetilde V_{lj}^{[t]} | \geq \tau_j \}$ be the indices of the true and estimated non-zero elements of the $j$th columns $\widehat{\V}_{\bullet j}^{0}$ and $\widetilde{\V}_{\bullet j}^{[t]}$ at the $t$-th iteration of Algorithm 1, respectively. Let the corresponding false negative and positive sets be $\text{FN}_j^{[t]} = \truesetV \backslash S_{j}^{[t]}$ and $\text{FP}_j^{[t]} = S_{j}^{[t]} \backslash \truesetV$ at iteration $t$. Let an event $\mathcal{E}_{j}=\big\{\|\X^{\top} \widehat{\boldsymbol \xi}_{j} / n\|_{\infty} \leq 0.5 \gamma_j \tau_j \big\}  \cap  \big\{\|\widehat{\V}_{\bullet j}^{0}-\V_{\bullet j}^{0}\|_{\infty} \leq 0.5 \tau_j \big\}$, where $\widehat{\boldsymbol{\xi}}_j  =  \Y_j - \varphi_j(\X \widehat{\V}_{\bullet j}^{0})$ is the residual of the oracle MLE $\widehat{\V}_{\bullet j}^{0}$ for the GLM, with the support $ \{l : \widehat{V}_{lj}^{0} \neq 0 \} = \truesetV$. Consider the data matrix $(\X_{n\times q},\Y_{n\times p})$ and $\Y_j$ refers to the $j$-th column of $\Y$, that is, an $n \times 1$ vector.

Our proof consists of three steps. In {\bf Step 1}, we show by induction that if $\big|\truesetV \cup S_{j}^{[t-1]}\big| \leq 2 \sparsityV$ on $\mathcal{E}_{j}$, then $|\truesetV \cup S_{j}^{[t]}| \leq 2 \sparsityV$, $t=1,\ldots$, so that Assumption~\ref{REcondition} applies. In {\bf Step 2}, we 
estimate the number of iterations to termination $T$. Particularly, we prove that 
$|\text{FP}_j^{[t]}| +  |\text{FN}_j^{[t]}| < 1$ or $|\text{FP}_j^{[t]}|=|\text{FP}_j^{[t]}|=0$ and thus $S_j^{[t]} = S_j^0$, for $t \geq T$. In {\bf Step 3}, we bound 
$\P(\mathcal{E}_{j})$ and show that $1 - \P\big(\cup_{j=1}^p \mathcal{E}_{j}^c\big)$ has a high probability 
tending to one as $n \to \infty$. 

{\bf Step 1:} Suppose $|\truesetV \cup S_{j}^{[t-1]}| \leq 2 \sparsityV$ on $\mathcal{E}_{j}$. By the Taylor's expansion of the gradient $\nabla \loss(\widetilde{\V}_{\bullet j}^{[t]})$ at $\widehat{\V}_{\bullet j}^{0}$,
\begin{align}
\label{eq:eq0}
\nabla \loss(\widetilde{\V}_{\bullet j}^{[t]}) 
 = \nabla \loss(\widehat{\V}_{\bullet j}^{0})  + \nabla^2 \loss(\overline{\V}_{\bullet j}) \big(\widetilde{\V}_{\bullet j}^{[t]} -  \widehat{\V}_{\bullet j}^{0}\big),
\end{align}
where $\overline{\V}_{\bullet j}$ is a vector of intermediate values on the line between $\widehat{\V}_{\bullet j}^{0}$ and $\widetilde{\V}_{\bullet j}^{[t]}$, $\nabla \loss(\widehat{\V}^{0}_{\bullet j})=n^{-1} \sum_{i=1}^n \X_{i\bullet} (-Y_{ij}+ \varphi_j( \X_{i\bullet}^{\top} {\widehat{\V}_{\bullet j}^{0}} ))=-n^{-1} \X^{\top} 
\widehat{\boldsymbol{\xi}}_j$ with $\widehat{\boldsymbol{\xi}}_j= \Y_j - \varphi_j(\X \widehat{\V}_{\bullet j}^{0})$. By the optimality condition of~\eqref{eq:unconstrained} at iteration $t$,
\begin{align}
 0 \leq  \big(\widehat{\V}_{\bullet j}^{0}-\widetilde{\V}_{\bullet j}^{[t]}\big)^{\top}\big( \nabla \loss(\widetilde{\V}_{\bullet j}^{[t]}) +\gamma_j \tau_j \nabla\big\|\big(\widetilde{\V}_{\bullet j}^{[t]}\big)_{\big(S_{j}^{[t-1]}\big)^{c}} \big \|_1\big), 
\label{eq:eq1}
\end{align}
where $\|\cdot\|_1$ denotes the $\ell_1$-norm. On the other hand, by the optimality condition of the oracle estimator $\widehat{\V}_{\bullet j}^{0}$: $\X^{\top}_{\truesetV} (\Y_j - \varphi_j(\X \widehat{\V}_{\bullet j}^{0}))=\X^{\top} \widehat{\boldsymbol{\xi}}_j={\bf 0}$ on $\truesetV$, implying that
$(\Re_{j})_{ S_{j}^{[t-1]} \cap \truesetV }={\bf 0}$, where $\Re_j = \X^{\top} \widehat{\boldsymbol{\xi}}_j / n - \gamma_j \tau_j \nabla\big\|\big(\widetilde{\V}_{\bullet j}^{[t]}\big)_{\big(S_{j}^{[t-1]}\big)^{c}}\big\|_{1}$. Let $\truesetV \Delta S_{j}^{[t-1]} = (\truesetV \backslash S_{j}^{[t-1]}) \cup (S_{j}^{[t-1]} \backslash \truesetV )$,
where $\Delta$ denotes the symmetric difference. 

Hence, combination of \eqref{eq:eq0} and \eqref{eq:eq1} yields that
\begin{align}
	& \big(\widetilde{\V}_{\bullet j}^{[t]} - \widehat{\V}_{\bullet j}^{0}\big)^{\top} \nabla^2 \loss(\overline{\V}_{\bullet j}) \big(\widetilde{\V}_{\bullet j}^{[t]} -  \widehat{\V}_{\bullet j}^{0}\big) 
	\leq \big(\widetilde{\V}_{\bullet j}^{[t]} - \widehat{\V}_{\bullet j}^{0} \big)^{\top}\big(\X^{\top} \widehat{\boldsymbol{\xi}}_j / n - \gamma_j \tau_j \nabla\big\|\big(\widetilde{\V}_{\bullet j}^{[t]}\big)_{\big(S_{j}^{[t-1]}\big)^{c}}\big\|_{1}\big)  \nonumber  \\
	& \leq  \big(\widetilde{\V}_{\bullet j}^{[t]} - \widehat{\V}_{\bullet j}^{0} \big)^{\top}_{\truesetV \Delta S_{j}^{[t-1]}} 
	(\Re_j)_{\truesetV \Delta S_{j}^{[t-1]}} 
	+  \big(\widetilde{\V}_{\bullet j}^{[t]} - \widehat{\V}_{\bullet j}^{0} \big)^{\top}_{\big(\truesetV \cup S_{j}^{[t-1]}\big)^{c}} (\Re_j)_{\big(\truesetV \cup S_{j}^{[t-1]}\big)^c}
   \nonumber \\ 
	& \leq \big\|\big(\widetilde{\V}_{\bullet j}^{[t]}-\widehat{\V}_{\bullet j}^{0}\big)_{\truesetV \Delta S_{j}^{[t-1]}}\big\|_{1}\big(\|\X^{\top} \widehat{\boldsymbol{\xi}}_{j} / n\|_{\infty}+\gamma_j \tau_j\big) \nonumber \\
	& +\big\|\big(\widetilde{\V}_{\bullet j}^{[t]}-\widehat{\V}_{\bullet j}^{0}\big)_{\big(\truesetV \cup S_{j}^{[t-1]}\big)^{c}}\big\|_{1}\big(\|\X^{\top} \widehat{\boldsymbol{\xi}}_{j} / n\|_{\infty}-\gamma_j \tau_j\big), 
	\label{eq:eq2}
\end{align}
where the last inequality holds since $\big(\widetilde{\V}_{\bullet j}^{[t]} - \widehat{\V}_{\bullet j}^{0} \big)^{\top}_{\big(\truesetV \cup S_{j}^{[t-1]}\big)^{c}}  \big(  \nabla\big\|\big(\widetilde{\V}_{\bullet j}^{[t]}\big)_{\big(\truesetV \cup S_{j}^{[t-1]}\big)^{c}}  \big\|_{1}\big) = \big \| \big(\widetilde{\V}_{\bullet j}^{[t]} - \widehat{\V}_{\bullet j}^{0} \big)_{\big(\truesetV \cup S_{j}^{[t-1]}\big)^{c}}  \big\|_1$.
Note that $\big(\widetilde{\V}_{\bullet j}^{[t]} - \widehat{\V}_{\bullet j}^{0}\big)^{\top} \nabla^2 \loss(\overline{\V}_{\bullet j}) \big(\widetilde{\V}_{\bullet j}^{[t]} -  \widehat{\V}_{\bullet j}^{0}\big) \geq 0$ since $\nabla^2 \loss(\overline{\V}_{\bullet j})$ is positive-definite. By \eqref{eq:eq2}, 
$$
\big\|\big(\widetilde{\V}_{\bullet j}^{[t]}-\widehat{\V}_{\bullet j}^{0}\big)_{\big(\truesetV \cup S_{j}^{[t-1]}\big)^{c}}\big\|_{1} \big(\gamma_j \tau_j-\big\|\X^{\top} \widehat{\boldsymbol{\xi}}_{j} / n\big\|_{\infty}\big)
\leq \big\|\big(\widetilde{\V}_{\bullet {j}}^{[t]}-\widehat{\V}_{\bullet j}^{0}\big)_{\truesetV \Delta S_{j}^{[t-1]}}\big\|_{1}
\big(\big\|\X^{\top} \widehat{\boldsymbol{\xi}}_{j} / n\big\|_{\infty}+\gamma_j \tau_j\big).
$$
Note, on event $\mathcal{E}_{j}$, $\big\|\X^{\top} \widehat{\boldsymbol{\xi}}_{j} / n\big\|_{\infty} \leq \gamma_j \tau_j / 2$, and thus 
\begin{align*}
\Big \|\big(\widetilde{\V}_{\bullet j}^{[t]}-\widehat{\V}_{\bullet j}^{0}\big)_{\big(\truesetV \cup S_{j}^{(t-1]}\big)^{c}} \Big\|_1 \leq 3 \Big\|\big(\widetilde{\V}_{\bullet j}^{[t]}-\widehat{\V}_{\bullet j}^{0}\big)_{\truesetV \Delta S_{j}^{[t-1]}}\Big\|_{1} \leq 3\Big\|\big(\widetilde{\V}_{\bullet j}^{[t]}-\widehat{\V}_{\bullet j}^{0}\big)_{\truesetV \cup S_{j}^{[t-1]}} \Big\|_{1}. 
\end{align*}
Note that $\left|\truesetV \cup S_{j}^{[t-1]}\right| \leq 2 \sparsityV$. By Assumption~\ref{REcondition}  and~\eqref{eq:eq2},  
\begin{align}
&\RSC \big\|\widetilde{\V}_{\bullet j}^{[t]}-\widehat{\V}_{\bullet j}^{0}\big\|_{2}^{2} \leq 
\big(\widetilde{\V}_{\bullet j}^{[t]} - \widehat{\V}_{\bullet j}^{0}\big)^{\top} \nabla^2 \loss(\overline{\V}_{\bullet j}) \big(\widetilde{\V}_{\bullet j}^{[t]} -  \widehat{\V}_{\bullet j}^{0}\big) 
\nonumber  \\
&\leq\big(\big\|\X^{\top} \widehat{\boldsymbol{\xi}}_{j} / n\big\|_{\infty}+\gamma_j \tau_j \big)\big\|\big(\widetilde{\V}_{\bullet {j}}^{[t]}-\widehat{\V}_{\bullet {j}}^{0}\big)_{\truesetV \Delta S_{j}^{[t-1]}}\big\|_{1}+\big(\big\|\X^{\top} \widehat{\boldsymbol{\xi}}_{j} / n\big\|_{\infty}-\gamma_j \tau_j \big)  \nonumber \\
& \big\|\big(\widetilde{\V}_{\bullet j}^{[t]}-\widehat{\V}_{\bullet j}^{0}\big)_{\big(\truesetV \cup S_{j}^{[t-1]}\big)^{c}}\big\|_{1}  \nonumber \\ 
&\leq\big(\big\|\X^{\top} \widehat{\boldsymbol{\xi}}_{j} / n\big\|_{\infty}+\gamma_j \tau_j \big)\big\|\big(\widetilde{\V}_{\bullet j}^{[t]}-\widehat{\V}_{\bullet {j}}^{0}\big)_{\truesetV \Delta S_{j}^{[t-1]}}\big\|_{1}, \nonumber \\
& \leq 1.5 \gamma_j \tau_j \sqrt{|\truesetV \Delta S_{j}^{[t-1]}|} \cdot \big\| \widetilde{\V}_{\bullet j}^{[t]}-\widehat{\V}_{\bullet {j}}^{0} \big\|_2,
\label{eq:eq4RSC}
\end{align}
where the last inequality follows from the Cauchy-Schwarz inequality and $\big\|\X^{\top} \widehat{\boldsymbol{\xi}}_{j} / n\big\|_{\infty} \leq 0.5 \gamma_j \tau_j$ on $\mathcal{E}_{j}$. 
Hence, 
\begin{eqnarray}
\big\|\widetilde{\V}_{\bullet j}^{[t]}-\widehat{\V}_{\bullet j}^{0}\big\|_{2}/\tau_j 
\leq (1.5 \gamma_j/\RSC) \sqrt{2 \sparsityV } \leq \sqrt{\sparsityV }, 
\label{eq:inter}
\end{eqnarray}
since $|\truesetV \Delta S_{j}^{[t-1]}| \leq |\truesetV \cup S_{j}^{[t-1]}| \leq 2 \sparsityV$
and $\gamma_j \leq \RSC /6$ by Condition (1) of Theorem~\ref{thm_anc_sele_cons}.
Moreover, $\big\|\widetilde{\V}_{\bullet j}^{[t]}-\widehat{\V}_{\bullet j}^{0}\big\|_{2}^2  \geq |\text{FP}_j^{[t]}| \cdot \tau_j^2$ since  
$| \widetilde{\V}_{l j}^{[t]}-\widehat{\V}_{l j}^{0} | = | \widetilde{\V}_{ lj}^{[t]}| > \tau_j$
for any $l \in \text{FP}_j^{[t]} = S_{j}^{[t]} \backslash \truesetV$.
By \eqref{eq:inter},  
$|\text{FP}_j^{[t]}| \leq \big\|\widetilde{\V}_{\bullet j}^{[t]}-\widehat{\V}_{\bullet j}^{0}\big\|^2_{2}/\tau^2_j \leq  \sparsityV$.
Therefore, $|\truesetV \cup S_{j}^{[t]}| = |\truesetV| + |\text{FP}_j^{[t]}| \leq 2 \sparsityV$.

{\bf Step 2:}
Suppose $ |\text{FP}_j^{[t]}| +  |\text{FN}_j^{[t]}| \geq 1$. Similarly,
\begin{align*}
\big\|\widetilde{\V}_{\bullet j}^{[t]}-\widehat{\V}_{\bullet j}^{0}\big\|_{2}^{2} \geq  (|\text{FP}_j^{[t]}| +  |\text{FN}_j^{[t]}|) (0.5 \tau_j)^2,
\end{align*}
since $| \widetilde{V}_{l j}^{[t]}-\widehat{V}_{ l j}^{0} |  \geq    | \widetilde{V}_{l j}^{[t]}- V_{l j}^{0} | - | \widehat{V}_{l j}^{0}   -  V_{l j}^{0} | \geq \tau_j - 0.5 \tau_j$
for any $l \in \text{FP}_j^{[t]} \cup \text{FN}_j^{[t]}$, by Assumption~\ref{SNRcondition}.
Therefore, 
$\sqrt{|\text{FP}_j^{[t]}| +  |\text{FN}_j^{[t]}|} \leq \big\|\widetilde{\V}_{\bullet j}^{[t]}-\widehat{\V}_{\bullet j}^{0}\big\|_{2}/0.5 \tau_j$. Moreover, 
by~\eqref{eq:eq4RSC} and the Cauchy-Schwarz inequality,
$\RSC\big\|\widetilde{\V}_{\bullet j}^{[t]}-\widehat{\V}_{\bullet j}^{0}\big\|_{2}^{2}
\leq  1.5 \gamma_j \tau_j \big\|\big(\widetilde{\V}_{\bullet j}^{[t]}-\widehat{\V}_{\bullet {j}}^{0}\big)_{\truesetV \Delta S_{j}^{[t-1]}}\big\|_{1} 
\leq 1.5 \gamma_j \tau_j \sqrt{|\truesetV \Delta S_{j}^{[t-1]}|} \cdot \big\| \widetilde{\V}_{\bullet j}^{[t]}-\widehat{\V}_{\bullet {j}}^{0} \big\|_2$. Hence,
$\big\|\widetilde{\V}_{\bullet j}^{[t]}-\widehat{\V}_{\bullet j}^{0}\big\|_{2}/\tau_j 
\leq  (1.5 \gamma_j/\RSC)  \sqrt{|\text{FP}_j^{[t-1]}| +  |\text{FN}_j^{[t-1]}|}$.
By Conditions (1) and (2) of Theorem~\ref{thm_anc_sele_cons}:
\begin{align*}
\sqrt{|\text{FP}_j^{[t]}| +  |\text{FN}_j^{[t]}|} \leq \frac{\big\|\widetilde{\V}_{\bullet j}^{[t]}-\widehat{\V}_{\bullet j}^{0}\big\|_{2}}{0.5 \tau_j} \leq \frac{3 \gamma_j}{\RSC} \sqrt{|\text{FP}_j^{[t-1]}| +  |\text{FN}_j^{[t-1]}|} \leq 0.5 \sqrt{|\text{FP}_j^{[t-1]}| +  |\text{FN}_j^{[t-1]}|}.
\end{align*}
Iterating this process implies that $\sqrt{|\text{FP}_j^{[t]}| +  |\text{FN}_j^{[t]}|} \leq (\frac{1}{2})^t \sqrt{|S_j^0| + |S_j^{[0]} |}$, $t=0,1,\ldots$. If $t \geq  T=1+\big\lceil\log (2 \sparsityV)/\log 4\big\rceil$, then $|\text{FP}_j^{[t]}| +  |\text{FN}_j^{[t]}| < 1$ or $\mathrm{FP}_{j}^{[t]}=\mathrm{FN}_{j}^{[t]}=\emptyset$ on event $\mathcal{E}_{j}$. 
Consequently, $\{ l: \widetilde V_{lj}^{[T]} \neq 0 \} =  \{l: V_{lj}^0 \neq 0\} = \truesetV$.

{\bf Step 3:} To bound $\\P\big(\bigcup_{j=1}^{p} \mathcal{E}_{j}^{c}\big)$, recall that $\mathcal{E}_{j}=\big\{\big\|\X^{\top} \widehat{\boldsymbol \xi}_{j} / n\big\|_{\infty} \leq 0.5 \gamma_j \tau_j \big\}  \cap  \big\{\big\|\widehat{\V}_{\bullet j}^{0}-\V_{\bullet j}^{0}\big\|_{\infty} \leq 0.5 \tau_j \big\}$. Next, we bound the two events in $\mathcal{E}^c_{j}$ separately. For the first event, by the triangular inequality, 
\begin{align}
&\P\big(\big\|\X^{\top} \widehat{\boldsymbol \xi}_{j} / n\big\|_{\infty}>0.5 \gamma_j \tau_j \big)  =\P\big(\big\|\X^{\top} (\Y_j - \varphi_j(\X \widehat{\V}_{\bullet j}^{0}))     / n\big\|_{\infty}>0.5 \gamma_j \tau_j \big) \nonumber \\
& \leq \P\big(\big\|\X^{\top} (\Y_j - \varphi_j(\X  {\V}_{\bullet j}^{0}) )      / n\big\|_{\infty} >0.25 \gamma_j \tau_j \big)  \nonumber \\
& +   \P\big( \| \X^{\top} ( \varphi_j(\X  {\V}_{\bullet j}^{0}) - \varphi_j(\X  \widehat{\V}_{\bullet j}^{0}))/n \|_{\infty} >0.25 \gamma_j \tau_j     \big) \label{eq:eq4}.
\end{align}
By Assumption \ref{interv_boundedness}, $|X_{ik}| \leq c_1$. 
By Assumption \ref{GLM-residual}, $Y_{ij} - \varphi_j({{\V}_{\bullet j}^{0}}^{\top} \X_{i\bullet} )$ is 
sub-exponential with the bound $\subexp$.
Hence, by Bernstein's inequality (Theorem 2.8.2 of \citet{vershynin2018high}), for any given $k=1,\cdots,q$,
\begin{align*}
&\P\left( \left| \sum_{i=1}^n X_{ik} (Y_{ij} - \E [Y_{ij} | \X ] )/n \right|  \geq 0.25 \gamma_j \tau_j   \right)  \leq 2 \exp \left( -  \min\left( \frac{\gamma_{j}^{2} \tau_{j}^{2}n }{32 \subexp^2 c_1^2} , \frac{\gamma_{j} \tau_{j} n}{8 \subexp c_1} \right)    \right).
\end{align*}
Note that $\left\|\X^{\top} (\Y_j - \varphi_j(\X {\V}_{\bullet j}^{0})) / n\right\|_{\infty}
=\max_{k=1}^q |\sum_{i=1}^n X_{ik} (Y_{ij} - \E [Y_{ij} | \X ] )/n|$.
The union bound yields, for the first quantity in~\eqref{eq:eq4}, that
\begin{align}
\label{bound0}
&  \P\big(\big\|\X^{\top} (\Y_j - \varphi_j(\X {\V}_{\bullet j}^{0}))     / n\big\|_{\infty}>0.25 \gamma_j \tau_j \big) \leq 2 q \exp \big(- \min\big( \frac{\gamma_{j}^{2} \tau_{j}^{2} n}{32 \subexp^2 c_1^{2} } , \frac{\gamma_{j} \tau_{j} n}{8 \subexp c_1 } \big) \big), \nonumber \\
   & \leq 2 \exp( - 2 \log n - \log q) = 2 n^{-2} q^{-1}, 
\end{align}
by the choice of $\gamma_j$ and $\tau_j$, that is, $\gamma_{j} \tau_{j} \geq \sqrt{64 \subexp^2 c_{1}^{2}  (\log q + \log n) / n}$. 

For the second quantity in~\eqref{eq:eq4}, we bound $\| \X^{\top} ( \varphi_j(\X  {\V}_{\bullet j}^{0})-\varphi_j(\X \widehat{\V}_{\bullet j}^{0}))/n \|_{\infty}$. Towards this end, note that $V_{kj}^0 = \widehat V_{kj}^0 = 0$ on $k \notin \truesetV$. Therefore, for ${\V}_{\bullet j}^{0}$ and $\widehat{\V}_{\bullet j}^{0}$ constrained on the set $\truesetV$, ${\V}_{\bullet j}^{0}
={\V}_{\truesetV, j}^{0}$ and $\widehat{\V}_{\bullet j}^{0}=\widehat{\V}_{\truesetV, j}^{0}$. Then, 
by Assumption \ref{GLM-residual}, 
\begin{eqnarray}
\label{bound1}
&\| \X^{\top} (\varphi_j(\X  {\V}_{\bullet j}^{0}) - \varphi_j(\X \widehat{\V}_{\bullet j}^{0})) \|_{\infty} \leq \lipsc \| \X^{\top} \X_{\truesetV} ( {\V}_{\truesetV, j}^{0} -   \widehat{\V}_{\truesetV, j}^{0})\|_{\infty},
\end{eqnarray}
for some Lipschitz constant $L_1>0$.
Moreover, by Lemma~\ref{oracle_expression}, for the oracle estimator constrained on $\truesetV$, namely, $\hat{\V}_{\truesetV, j}^{0}$, $\widehat{\V}_{\truesetV, j}^{0}-\V_{\truesetV, j}^{0} = (\X_{\truesetV}^\top \hessianV \X_{\truesetV})^{-1} \X_{\truesetV}^\top (\Y_j - \boldsymbol{\zeta}^0 - \r)$, where $\hessianV$, $\boldsymbol{\zeta}^0$ and $\r$ will be defined in Lemma~\ref{oracle_expression}. Let $\K = \X^{\top} \X_{\truesetV} (\X_{\truesetV}^{\top} \hessianV \X_{\truesetV})^{-1} \X_{\truesetV}^{\top}$. Plugging the above expression into 
\eqref{bound1} yields that 
\begin{align}
\label{bound}
& \P\big( \| \X^\top ( \varphi_j(\X  {\V}_{\bullet j}^{0}) - \varphi_j(\X  \widehat{\V}_{\bullet j}^{0}))/n \|_{\infty} >0.25 \gamma_j \tau_j     \big)  \leq \P\big( \|   \K (\Y_j - \boldsymbol{\zeta}^0 - \r) /n \|_{\infty} >  \frac{\gamma_j \tau_j}{4 \lipsc}     \big) \nonumber \\
& \leq \P \big(  \|   \K (\Y_j - \boldsymbol{\zeta}^0 ) /n \|_{\infty} >  \frac{\gamma_j \tau_j}{4 \lipsc}  - \|   \K \r /n \|_{\infty}      \big).
\end{align}
By Assumption~\ref{interv_boundedness}, there exists a constant $c_3>0$ such that $\| \X^{\top} \X_{\truesetV} (\X_{\truesetV}^{\top} \hessianV \X_{\truesetV})^{-1} \X_{\truesetV}^{\top} \|_{\infty} \leq c_3$ or $\max_{l=1}^q |K_{li}| \leq c_3$. Then,
by Lemmas~\ref{oracle_expression} and ~\ref{lemma_l2_consistency} with the choice of $\tau_j$ and $\gamma_j$,
\begin{align*}
\|\K \r /n \|_{\infty} &=\max_{l=1}^q |\sum_{i=1}^n K_{li} r_i  |/n \leq \max_l \sum_{i=1}^n |K_{li}| |r_i| /n \leq c_3 \sum_{i=1}^n |r_i|/n \\
& \leq c_3 \thirdder (\hat{\V}_{\truesetV, j}^{0}-\V_{\truesetV, j}^{0})^\top  (\X_{\truesetV}^\top \X_{\truesetV} / n) (\hat{\V}_{\truesetV, j}^{0}-\V_{\truesetV, j}^{0}) \\
&\leq c_3 c_{\max} \thirdder  \| \hat{\V}_{\truesetV, j}^{0}-\V_{\truesetV, j}^{0} \|_2^2
\leq c_3 c_{\max} \thirdder \frac{16 \subexp^2 c_1^2}{m^2} \cdot \frac{\sparsityV \log (n \sparsityV)}{n}
\leq  \frac{1}{2} \cdot \frac{\gamma_j \tau_j}{4 \lipsc},
\end{align*}
with probability at least $ 1 - 2 \exp (-\log (\sparsityV) - 2 \log n) = 1 - 2 (\sparsityV)^{-1} n^{-2}$.

Next, in \eqref{bound}, we bound $\K (\Y_j - \boldsymbol{\zeta}^0)/n$.
Note that $\max_{k=1}^q| K_{ki} | \leq c_3$. By Assumption \ref{GLM-residual}, $Y_{ij} - {\zeta}_{ij}^0=(Y_{ij} - \E[Y_{ij}| \X ])$ is sub-exponential with the bound $\subexp$. 
By Theorem 2.8.2 of \citet{vershynin2018high} and a union bound as in \eqref{bound0},
\begin{align}
\label{bound3}
&\P\Big( \| \X^\top ( \varphi_j(\X  {\V}_{\bullet j}^{0}) - \varphi_j(\X  \widehat{\V}_{\bullet j}^{0}))/n \|_{\infty} > 0.25 \gamma_j \tau_j     \Big)  
\leq \P \Big(  \|   \K (\Y_j - \boldsymbol{\zeta}^0 ) /n \|_{\infty} >  0.5 \frac{\gamma_j \tau_j}{4\lipsc}      \Big) \nonumber \\
&\leq  2 q \exp \Big(- \min \big( \frac{\gamma_{j}^2 \tau_{j}^{2} n}{8 \subexp^2 c_3^2 \cdot 16 \lipsc^2},  \frac{\gamma_{j} \tau_{j} n}{16  \subexp c_3 \cdot \lipsc}   \big) \Big) + 2 (\sparsityV)^{-1} n^{-2} \nonumber \\
&\leq 2 q \exp \Big(-\frac{\gamma_{j}^2 \tau_{j}^{2} n}{8 \subexp^2 c_3^2 \cdot 16 \lipsc^2 }\Big) + 2 (\sparsityV)^{-1} n^{-2} .
\end{align}
Combining \eqref{eq:eq4}, \eqref{bound0}, and \eqref{bound3} yields that  
\begin{align*}
& \P\left(\left\|\X^{\top} \widehat{\boldsymbol \xi}_{j} / n\right\|_{\infty}>0.5 \gamma_j \tau_j \right) = \P\left(\left\|\X^{\top} (\Y_j - \varphi_j(\X \widehat{\V}_{\bullet j}^{0}))     / n\right\|_{\infty}>0.5 \gamma_j \tau_j \right) \\
& \leq  2 q \exp \left(-\frac{\gamma_{j}^{2} \tau_{j}^{2} n}{32 \subexp^2 c_1^{2} }\right) + 2 q \exp \left(-\frac{\gamma_{j}^2 \tau_{j}^{2} n}{8 \subexp^2 c_3^2 \cdot 16 \lipsc^2 }\right) + 2 (\sparsityV)^{-1} n^{-2} \\
&\leq 2 n^{-2} q^{-1} + 2 n^{-2} q^{-1} + 2 (\sparsityV)^{-1} n^{-2} \leq 6 n^{-2} q^{-1}.
\end{align*}

Next, we bound the second event $\big\{\big\|\widehat{\V}_{\bullet j}^{0}-\V_{\bullet j}^{0}\big\|_{\infty} \leq 0.5 \tau_j \big\}$ in $\mathcal{E}_{j}^c$. Since $\widehat{\V}_{\bullet j}^{0} = \V_{\bullet j}^{0} = 0$ on $(\truesetV)^c$, it suffices to consider the entries of $\widehat{\V}_{\bullet j}^{0}$ constrained on $\truesetV$, or the oracle estimator $\widehat{\V}_{\truesetV, j}^{0}$. By Lemma~\ref{oracle_expression}, $\widehat{\V}_{\truesetV, j}^{0}-\V_{\truesetV, j}^{0} = \H (\Y_j - \boldsymbol{\zeta}^0 - \r)$, where
$\H = \big(\X_{\truesetV}^{\top} \hessianV \X_{\truesetV}\big)^{-1} \X_{\truesetV}^{\top}=(H_{ki})$.

To bound $ \| \H \r \|_{\infty}$, by Assumption~\ref{interv_boundedness},  $\|(\X_{\truesetV}^{\top} \hessianV \X_{\truesetV} / n)^{-1} \X_{\truesetV}^{\top}\|_{\infty} \leq c_2$ or $\|\H\|_{\infty} \leq c_2 n^{-1}$, for some constant $c_2>0$. By Lemmas~\ref{oracle_expression} and ~\ref{lemma_l2_consistency} with the choice of $\tau_j$, 
\begin{align}
\| \H \r \|_{\infty} & = \max_k |\H_{k \bullet} \r|  = \max_k | \sum_{i=1}^n H_{ki} r_i  | \leq \max_k \sum_{i=1}^n |H_{ki}| |r_i|  \leq c_2 \sum_{i=1}^n |r_i|/n \nonumber \\
& \leq c_2 \thirdder (\widehat{\V}_{\truesetV, j}^{0}-\V_{\truesetV, j}^{0})^{\top}  (\X_{\truesetV}^{\top} \X_{\truesetV} / n) (\widehat{\V}_{\truesetV, j}^{0}-\V_{\truesetV, j}^{0}) \nonumber \\
	& \leq c_2 c_{\max} \thirdder \| \widehat{\V}_{\truesetV, j}^{0}-\V_{\truesetV, j}^{0} \|_2^2
\leq  c_2 c_{\max} \thirdder  \frac{16 \subexp^2 c_1^2}{m^2} \frac{\sparsityV \log (n\sparsityV)}{n} \leq  0.25 \tau_j. 
\label{H-exp}
\end{align}

To bound $\| \H (\Y_j - \boldsymbol{\zeta}^0)\|_{\infty}$,  note that $\max_{k=1}^q| H_{ki} | \leq c_2/n$. By Assumption \ref{GLM-residual}, $Y_{ij} - {\zeta}_{ij}^0=(Y_{ij} - \E[Y_{ij}| \X ])$ is 
sub-exponential with the bound $\subexp$. 
By the triangular inequality, Theorem 2.8.2 of 
\citet{vershynin2018high} and the same argument as in \eqref{bound0}, we obtain that 
\begin{align*}
&	\P\big(\big\|\widehat{\V}_{\bullet j}^{0}-\V_{\bullet j}^{0}\big\|_{\infty}>0.5 \tau_j\big) \leq  
\P\big(\H(\Y_j - \boldsymbol{\zeta}^0) \big\|_{\infty}  >0.5 \tau_j - \big\|\H  \r \big\|_{\infty} \big) \\ 
	&\leq  \P\big(\|\H (\Y_j - \boldsymbol{\zeta}^0)\|_{\infty}  >0.25 \tau_j \big)
	 \leq  \P\big(\|\H  {\boldsymbol  \xi}_j \|_{\infty}  >0.25 \tau_j \big) \\
	& \leq 2 \sparsityV \exp \big(- \min \big( \frac{\tau_{j}^{2} n}{32 \subexp^2 c_{2}^{2}}, \frac{\tau_{j} n}{8 \subexp c_{2}} \big) \big) + 2 (\sparsityV)^{-1} n^{-2}.
\end{align*}
To conclude, on $\mathcal{E}_{j}$, $\widehat S_j \equiv S_j^{[T]}=S_j^0$, which means $\widehat{\V}_{\bullet j} = \widetilde{\V}_{\bullet j}^{[T]} =  \widehat{\V}_{\bullet j}^{0}$. Hence, for $j=1,\ldots,p$, 
\begin{align}
	\label{alg}
	&  \P(\widehat{\V}_{\bullet j} \neq \widehat{\V}_{\bullet j}^{0} ) \leq  \P(\mathcal{E}_{j}^{c}) 
\leq  2 q \exp \big(-\frac{\gamma_{j}^{2} \tau_{j}^{2} n}{32 \subexp^2 c_1^{2} }\big) + 2 q \exp \big(-\frac{\gamma_{j}^2 \tau_{j}^{2} n}{8 \subexp^2 c_3^2 \cdot 16 \lipsc^2}\big)
\nonumber \\
	& + 2 \sparsityV \exp \big(-\frac{\tau_{j}^{2} n}{32 \subexp^2 c_{2}^{2} }\big) + 2 (\sparsityV)^{-1} n^{-2} \leq 8 n^{-2} q^{-1}.
\end{align}

It remains to show that $\widehat{\V}_{\bullet j}^{0}$ is a global minimizer of~\eqref{eq:constrained} with high probability. Towards this end, we will show that Assumptions~\ref{REcondition} and~\ref{SNRcondition} imply the degree of separation condition (3) of \citet{shen2013constrained}. 
To see this, let $g(y_{ij} | \theta,\bx_i) = e^{-\ell(y_{ij}, \theta^\top \bx_i)}$ be a probability density for $y_{ij}$ where we denote $\theta = \V_{\bullet j}$ for notation simplicity. In addition, denote $\theta^0 = (\V_{\truesetV,j},\textbf 0)$ and $\theta_{A_j} = (\textbf 0,\V_{A_j,j})$. By the mean value theorem, there exists $\overline \theta_0$ between $\theta_{A_j}$ and $\theta^0$ such that
\begin{align*}
& \left( g^{1/2}(y_{ij}|\theta_{A_j},\bx_i) - g^{1/2}(y_{ij}|\theta^0,\bx_i) \right)^2 = \left(  \left( \nabla g^{1/2}(y_{ij}|\overline \theta_0,\bx_i)\right)^\top (\theta_{A_j} - \theta^0) \right)^2 \\
&= \left( ( \nabla e^{-\ell(y_{ij}, \overline \theta_0^\top \bx_i)/2} )^\top (\theta_{A_j} - \theta^0) \right)^2 
= \frac{1}{4} e^{-\ell(y_{ij}, \overline \theta_0^\top \bx_i)} \left( \nabla \ell(y_{ij}, \overline \theta_0^\top \bx_i)^\top  (\theta_{A_j} - \theta^0) \right)^2.
\end{align*}
Then, the Hellinger-distance can be written as:
\begin{align*}
& h^2\left(\theta_{A_j}, \theta^0\right)=\frac{1}{4}\left(\int\left( g^{1/2}(y_{ij}|\theta_{A_j},\bx_i) - g^{1/2}(y_{ij}|\theta^0,\bx_i) \right)^2 d \mu(y_{ij})\right) \\
&= \frac{1}{16} \left(\int e^{-\ell(y_{ij}, \overline \theta_0^\top \bx_i)} (\theta_{A_j} - \theta^0)^\top  \nabla \ell(y_{ij}, \overline \theta_0^\top \bx_i) \nabla \ell(y_{ij}, \overline \theta_0^\top \bx_i)^\top (\theta_{A_j} - \theta^0)   d \mu(y_{ij})\right) \\
&= \frac{1}{16}  (\theta_{A_j} - \theta^0) ^\top \left(\int   e^{-\ell(y_{ij}, \overline \theta_0^\top \bx_i)} \cdot \nabla \ell(y_{ij}, \overline \theta_0^\top \bx_i)  \nabla \ell(y_{ij}, \overline \theta_0^\top \bx_i)^\top   d \mu(y_{ij})\right)  (\theta_{A_j} - \theta^0) \\
&= \frac{1}{16}  (\theta_{A_j} - \theta^0) ^\top \E_{\overline \theta_0} \left[\nabla^2 \ell(y'_{ij}, 	\overline \theta_0^\top \bx_i)\right]  (\theta_{A_j} - \theta^0) ,
\end{align*}
where $\E_{\overline \theta_0}$ is the expectation with respect to $Y'_{ij} \sim g(y'_{ij} | \overline \theta_0^\top, \bx_i)$ 
while the last equality follows by the fact that $\E_{\theta} \left[  \nabla \log g(y_{ij} | \theta, \bx_i)  \nabla \log g(y_{ij} | \theta, \bx_i)^\top \right] = - \E_{\theta} \left[ \nabla^2 \log g(y_{ij} | \theta, \bx_i) \right]$ for any $\theta$.

Let $\widetilde \theta = \theta_{A_j} - \theta^0$. Then, $\| \widetilde \theta \|_2^2 \geq |\truesetV \setminus A_j| \| \V_{\truesetV,j} \|_2^2 $. By the definition of $C_{\min}$ of \citet{shen2013constrained} and Assumption~\ref{REcondition},
\begin{align*}
C_{\min} & = \min_{A_j \neq \truesetV, |A_j| \leq \sparsityV} \frac{h^2\left(\theta_{A_j}, \theta^0\right)}{\max(|\truesetV \setminus A_j|, 1)} \\ 
&\geq \min_{A_j \neq \truesetV, |A_j| \leq \sparsityV} | \truesetV \backslash A_j |^{-1} (\theta_{A_j} - \theta^0) ^\top \E\left[\nabla^2 \ell(y_{ij}, \overline \theta_0^\top \bx_i)\right]  (\theta_{A_j} - \theta^0) \\
& \geq  m \| \V_{\truesetV,j} \|_2^2
\geq m (100M c_2)^2 \frac{\log q+\log n}{n} \geq  m (100M c_2)^2 \frac{\log q}{n}, 
\end{align*}
where the last inequality uses Assumptions \ref{REcondition} and \ref{SNRcondition} and the fact 
$\overline \theta_0 = \theta_{A_j} + t(\theta_{A_j} - \theta^0)$, $t \in [0,1]$ so that
$\| \overline \theta_0 \|_0 \leq 2\sparsityV$.
This implies
the degree of separation condition (3) of \citet{shen2013constrained}.
By Theorem 2 there,
$ \mathbb{P}\left(\widehat{\V}_{\bullet j}^{0} \text { is not a global minimizer of }\eqref{eq:constrained} \right)
\leq 3 \exp (-2(\log (q)+\log (n))), 1 \leq j \leq p$, 
implying that 
\begin{align*}
& \mathbb{P}\left(\widehat{\V}_{\bullet j}^{0} \text { is not a global minimizer of }\eqref{eq:constrained} , 1 \leq j \leq p\right) \leq 3 p \left(\exp (-2(\log (q)+\log (n)))\right). 
\end{align*}
Hence, $\widehat{\V}_{\bullet j}$ is a global minimizer of~\eqref{eq:constrained} with probability tending to 1 as $n \to \infty$. Finally, we have shown that ${\widehat S}_j \equiv\{l: \widehat V_{lj} \neq 0\} = S_j^0\equiv\{l: V_{lj}^{0} \neq 0\}$, implying $\{(l,j): \widehat V_{lj} \neq 0\} = \{(l,j): V_{lj}^{0} \neq 0\}$. By Proposition~\ref{prop_peeling}, the estimated $\widehat \supergraph$ via $\widehat{\V}$ reconstructs the true super-graph $\supergraph^0$ correctly.

\begin{lemma}[Expression of the oracle MLE $\widehat{\V}_{\truesetV, j}^{0}$]\label{oracle_expression} 
Let $\widehat{\V}_{\truesetV, j}^{0}$ be the oracle MLE, defined as the minimizer of $\loss(\V_{\truesetV,j} 
| \Y_j, \X_{\truesetV}) =  n^{-1} \sum_{i=1}^{n} \left(
-Y_{i j} ( \bx^{\top}_{i,\truesetV} \V_{\truesetV, j}) + A_j(\bx^{\top}_{i,\truesetV} \V_{\truesetV, j}) \right)$
over $\V_{\truesetV,j}$. Then, 
\begin{align}
\widehat{\V}_{\truesetV, j}^{0}-\V_{\truesetV, j}^{0} = (\X^{\top}_{\truesetV} \hessianV \X_{\truesetV})^{-1} \X^{\top}_{\truesetV} (\Y_j - \boldsymbol{\zeta}^0 - \r), 
\end{align}
where $\boldsymbol{\zeta}^0=(\zeta^0_1,\ldots,\zeta^0_n)$ with $\zeta^0_i  = \varphi_j(\x^{\top}_{i,\truesetV} {\V}_{\truesetV, j}^{0})$, $\hessianV$ is a diagonal matrix with the $i$th diagonal $\hessianV_{ii} = A_j''(\X_{i\bullet}^{\top}{\V_{\bullet j}^0})$ and  $\r=(r_1,\cdots,r_p)$ is the integral form of the reminder for Taylor's expansion satisfying 
\begin{align}
\label{ri}
\sum_{i=1}^n |r_i| \leq \thirdder (\widehat{\V}_{\truesetV, j}^{0}-\V_{\truesetV, j}^{0})^{\top}  (\sum_{i=1}^n \x_{i,\truesetV} \x^{\top}_{i,\truesetV}) (\widehat{\V}_{\truesetV, j}^{0}-\V_{\truesetV, j}^{0}),
\end{align}
for $\thirdder$ defined in Assumption~\ref{GLM-residual}.
\end{lemma}

\noindent {\bf Proof of Lemma \ref{oracle_expression}}.
By the optimality condition for the constrained oracle MLE, for $k \in \truesetV$,
\begin{align}
&\sum_{i=1}^n X_{ik} (y_{ij} - \varphi_j(\x_{i,\truesetV}^{\top} \widehat{\V}_{\truesetV, j}^{0})) = 0, \nonumber \\
&\sum_{i=1}^n X_{ik} (y_{ij} - \varphi_j(\x_{i,\truesetV}^{\top} \V_{\truesetV, j}^{0}) + \varphi_j(\x_{i,\truesetV}^{\top} \V_{\truesetV, j}^{0}) - \varphi_j(\x_{i,\truesetV}^{\top} \widehat{\V}_{\truesetV, j}^{0})) = 0. \label{eq:KKT}
\end{align}
A Taylor series expansion of $\varphi_j(\x_{i,\truesetV}^{\top} \V_{\truesetV, j})$ at $\V_{\truesetV, j}^{0}$ yields that
\begin{align}
\label{eq:taylor}
\varphi_j(\x_{i,\truesetV}^{\top} \widehat{\V}_{\truesetV, j}^{0}) = \varphi_j(\x_{i,\truesetV}^{\top} \V_{\truesetV, j}^{0}) + w_j(\x_{i,\truesetV}^{\top} \V_{\truesetV, j}^{0}) \x_{i,\truesetV}^{\top}  (\widehat{\V}_{\truesetV, j}^{0}-\V_{\truesetV, j}^{0}) + r_i,
\end{align}
where 
{\small \begin{align*}
r_i = \int_{0}^{1} \varphi_j''\big(\x_{i,\truesetV}^{\top} (\widehat{\V}_{\truesetV, j}^{0} + t(\widehat{\V}_{\truesetV, j}^{0} - \V_{\truesetV, j}^{0}) )\big) (1-t)  \,dt \,
\left((\widehat{\V}_{\truesetV, j}^{0}-\V_{\truesetV, j}^{0})^{\top} \x_{i,\truesetV} 
\x_{i,\truesetV}^{\top} (\widehat{\V}_{\truesetV, j}^{0}-\V_{\truesetV, j}^{0})\right),
\end{align*}}is the integral form of the remainder for Taylor's expansion as \citet{li2019tuning} and 
$w_j(\x^{\top} \u) = \varphi_j'(\x^{\top} \u) = A_j''(\x^{\top} \u)$. 
Let $\hessianV$ be a diagonal matrix whose $i$th diagonal $\hessianV_{ii} =  A_j''( {\V^0_{\bullet j}}^{\top} \X_{i\bullet})$. Write~\eqref{eq:KKT} in a matrix form using \eqref{eq:taylor}:
\begin{align}
\X^{\top}_{\truesetV} \hessianV \X_{\truesetV}  (\hat{\V}_{\truesetV, j}^{0}-\V_{\truesetV, j}^{0}) &= \X^{\top}_{\truesetV} (\Y_j - \boldsymbol{\zeta}^0 - \r) \nonumber \\
\hat{\V}_{\truesetV, j}^{0}-\V_{\truesetV, j}^{0} &= ( \X^{\top}_{\truesetV} \hessianV \X_{\truesetV})^{-1} \X^{\top}_{\truesetV} (\Y_j - \boldsymbol{\zeta}^0 - \r),
\label{eq:KKT_matrix}
\end{align}
where $\zeta^0_i = \varphi_j(\x_{i,\truesetV}^{\top} {\V}_{\truesetV, j}^{0})$.
Further, note that $\varphi_j(\x^{\top} \u) = A_j'(\x^{\top} \u)$. Then, in the expression for $r_i$,
$\varphi_j''\big(\x_{i,\truesetV}^{\top} (\widehat{\V}_{\truesetV, j}^{0} + t(\widehat{\V}_{\truesetV, j}^{0} - \V_{\truesetV, j}^{0}) )\big) = g_i(t)$,
where $g_i(t) = \varphi_j''(\eta_i(t)) = A_j'''(\eta_i(t))$ with $\eta_i(t) = \x_{i,\truesetV}^{\top} (\hat{\V}_{\truesetV, j}^{0} + t(\hat{\V}_{\truesetV, j}^{0} - \V_{\truesetV, j}^{0}))$, $t \in (0,1)$. By Assumption~\ref{GLM-residual}, $|g_i(t)| \leq \thirdder$. 
Hence, \eqref{ri} holds.

\begin{lemma}[Rate of convergence under the $\ell_2$-norm]\label{lemma_l2_consistency} 
	Under Assumption~\ref{REcondition} (restricted strong convexity), 
	\begin{align*}
	\| \widehat \V_{\truesetV, j}^{0} - \V_{\truesetV, j}^{0} \|_2 \leq 
 \frac{2}{m} \sqrt{\sparsityV} \big\|\X_{\truesetV}^{\top} (\Y_j - \varphi_j(\X_{\truesetV} {\V}_{\truesetV, j}^{0}))     / n\big\|_{\infty} \leq \frac{4 \subexp c_1}{m} \sqrt{\frac{\sparsityV \log (n\sparsityV)}{n}},
	\end{align*} 
with probability at least $ 1 - 2 \exp (-\log (\sparsityV) - 2 \log n) = 1 - 2 (\sparsityV)^{-1} n^{-2}$.
\end{lemma}

\noindent {\bf Proof of Lemma \ref{lemma_l2_consistency}}.
We follow the proof of \citet{lee2015model} and consider the entries of $\V_{\bullet j}^{0}$ on $\truesetV$. The negative log-likelihood is 
\begin{align*}
\loss(\V_{\truesetV,j} | \Y_j, \X_{\truesetV}) & = n^{-1} \sum_{i=1}^{n} \left( 
-Y_{i j}(\bx^{\top}_{i,\truesetV} \V_{\truesetV, j}) + A_j ( \bx^{\top}_{i,\truesetV} \V_{\truesetV, j})  \right),
\end{align*}
where $\bx_{i,\truesetV}$ is a subvector of $\x_i$ with elements constrained on the indices $\truesetV$. By the definition of the oracle MLE, $\loss(\widehat \V_{\truesetV, j}^{0}) \leq \loss(\V_{\truesetV, j}^{0})$. Taylor's expansion of $\loss(\cdot)$ at ${\V}_{\truesetV, j}^{0}$ yields that
\begin{align*}
0 \geq  \nabla \loss(\V_{\truesetV, j}^{0}) (\widehat \V_{\truesetV, j}^{0} - \V_{\truesetV, j}^{0}) + \frac{1}{2} (\widehat \V_{\truesetV, j}^{0} - \V_{\truesetV, j}^{0})^{\top} \nabla^2\loss(\overline{\V}_{\truesetV, j}^{0}) (\widehat \V_{\truesetV, j}^{0} - \V_{\truesetV, j}^{0}) .
\end{align*}
Let $\Delta = \widehat{\V}_{\bullet j}^{0} - \V_{\bullet j}^{0}$. Clearly, $0 = \| \Delta_{(\truesetV)^c} \|_1 \leq 3 \| \Delta_{\truesetV} \|_1$. Therefore, by the restricted strong convexity condition $\frac{1}{2} (\widehat{\V}_{\bullet j}^{0} - \V_{\bullet j}^{0})^{\top} \nabla^2\loss(\overline{\V}_{\bullet j}^{0})  (\widehat{\V}_{\bullet j}^{0} - \V_{\bullet j}^{0}) \geq \frac{m}{2} \| \widehat{\V}_{\bullet j}^{0} - \V_{\bullet j}^{0} \|_2^2$, which implies $\frac{1}{2} (\widehat{\V}_{\truesetV, j}^{0} - \V_{\truesetV, j}^{0})^{\top} \nabla^2\loss(\overline{\V}_{\truesetV, j}^{0})  (\widehat{\V}_{\truesetV, j}^{0} - \V_{\truesetV, j}^{0}) \geq \frac{m}{2} \| \widehat{\V}_{\truesetV, j}^{0} - \V_{\truesetV, j}^{0} \|_2^2$ as $\widehat{\V}_{(\truesetV)^c, j}^{0} = \V_{(\truesetV)^c, j}^{0} = \mathbf 0$.

By the restricted strong convexity condition, 
\begin{align*}
\nabla \loss(\V_{\truesetV, j}^{0}) (\widehat \V_{\truesetV, j}^{0} - \V_{\truesetV, j}^{0})  + \frac{m}{2} \| \widehat \V_{\truesetV, j}^{0} - \V_{\truesetV, j}^{0} \|_2^2 \leq 0.
\end{align*}
By the H\"older's inequality, 
\begin{align*}
\frac{m}{2} \| \widehat \V_{\truesetV, j}^{0} - \V_{\truesetV, j}^{0} \|_2^2 \leq -\nabla \loss(\V_{\truesetV, j}^{0}) (\widehat \V_{\truesetV, j}^{0} - \V_{\truesetV, j}^{0})   \leq \| \nabla \loss(\V_{\truesetV, j}^{0}) \|_{\infty}  \| \widehat \V_{\truesetV, j}^{0} - \V_{\truesetV, j}^{0} \|_1  .
\end{align*}
By the Cauchy–Schwarz inequality,
\begin{align*}
\frac{m}{2} \| \widehat \V_{\truesetV, j}^{0} - \V_{\truesetV, j}^{0} \|_2^2    &\leq \sqrt{\sparsityV} \| \nabla \loss(\V_{\truesetV, j}^{0}) \|_{\infty}  \| \widehat \V_{\truesetV, j}^{0} - \V_{\truesetV, j}^{0} \|_2  , \\
\|\widehat \V_{\truesetV, j}^{0} - \V_{\truesetV, j}^{0} \|_2    &\leq \frac{2}{m} \sqrt{\sparsityV} \| \nabla \loss(\V_{\truesetV, j}^{0}) \|_{\infty},
\end{align*}
where $\nabla \loss(\V_{\truesetV, j}^{0}) = n^{-1} \X_{\truesetV}^{\top} (\Y_j - \varphi_j(\X_{\truesetV} {\V}_{\truesetV, j}^{0}))$. Therefore,
$$ \P\big(\big\|\X_{\truesetV}^{\top} (\Y_j - \varphi_j(\X_{\truesetV} {\V}_{\truesetV, j}^{0}))     / n\big\|_{\infty}> \epsilon \big) \leq 2 \sparsityV \exp \big(- \min\big( \frac{ n \epsilon^{2}}{2 \subexp^2 c_1^{2} }, \frac{ n \epsilon}{2 \subexp c_1} \big) \big).
$$
Hence,  
$\| \widehat \V_{\truesetV, j}^{0} - \V_{\truesetV, j}^{0} \|_2 \leq
      \frac{2}{m} \sqrt{\sparsityV} \epsilon =
        \frac{4 \subexp c_1}{m} \sqrt{\frac{\sparsityV \log (n \sparsityV)}{n}}$,
with probability $ 1 - 2 \sparsityV \exp \big(- \min\big( \frac{ n \epsilon^{2}}{2 \subexp^2 c_1^{2} }, \frac{ n \epsilon}{2 \subexp c_1} \big) \big) \geq 
	1 - 2 \exp (-\log (\sparsityV) - 2 \log n) = 1 - 2 (\sparsityV)^{-1} n^{-2}$,
	$\| \widehat \V_{\truesetV, j}^{0} - \V_{\truesetV, j}^{0} \|_2 \leq 
	\frac{2}{m} \sqrt{\sparsityV} \epsilon =
	\frac{4 \subexp c_1}{m} \sqrt{\frac{\sparsityV \log (n \sparsityV)}{n}}$,
	where $\epsilon = 2 \subexp c_1 \sqrt{\frac{\log (n \sparsityV)}{n}}$.

\subsection{Proof of Theorem~\ref{thm_confounder_consistency}}
For the $j$th equation, the TLP estimator minimizes:
\begin{align*}
&   (\widehat{\W}_{{\steponesuperset{\text{in}}(j)}, j},\widehat{\U}_{\steponesuperset{\text{an}}(j), j},\widehat{\balpha}_{\steponesuperset{\text{an}}(j), j})  \\ & =\argmin_{\W_{{\steponesuperset{\text{in}}(j)}, j},\U_{\steponesuperset{\text{an}}(j), j},\balpha_{\steponesuperset{\text{an}}(j), j}} \hspace{2mm}    n^{-1} \sum_{i=1}^{n}
-Y_{i j} \left( \W^{\top}_{\steponesuperset{\text{in}}(j), j} \X_{i,\steponesuperset{\text{in}}(j)}  +  \U^{\top}_{\steponesuperset{\text{an}}(j), j} \Y_{i,\steponesuperset{\text{an}}(j)}  +  \balpha^{\top}_{\steponesuperset{\text{an}}(j), j} \widehat \h_{i,\steponesuperset{\text{an}}(j)} \right)  \\ & \quad \quad \quad \quad \quad \quad \quad 
+ A_j\left( \W^{\top}_{\steponesuperset{\text{in}}(j), j} \X_{i,\steponesuperset{\text{in}}(j)}  + \U^{\top}_{\steponesuperset{\text{an}}(j), j} \Y_{i,\steponesuperset{\text{an}}(j)}  +  \balpha^{\top}_{\steponesuperset{\text{an}}(j), j} \widehat \h_{i,\steponesuperset{\text{an}}(j)} \right)  \\
&\text {  subject to } \quad     \sum_{k \in \steponesuperset{\text{an}}(j)} I(U_{kj} \neq 0) \leq K_{j}, \quad \sum_{k \in \steponesuperset{\text{an}}(j)} I(\alpha_{kj} \neq 0) \leq K'_{j},  \quad j=1, \ldots, p.
\end{align*}

In the absence of confounders ($\h_{i,\text{an}(j)} = 0$), if we use standard constrained GLM regression without deconfounding, ($\widehat \h_{i,\steponesuperset{\text{an}}(j)} = 0$),
it is straightforward to show that $\widehat \U_{\steponesuperset{\text{an}}(j),j} \to \U_{\text{an}^{0}(j),j}^{0}$ and $\widehat \W_{\steponesuperset{\text{in}}(j),j} \to \W_{\text{in}^{0}(j),j}^{0}$ by standard high-dimensional statistics results.

We now show the causal graph selection consistency of the TLP estimator in the presence of the confounders. We follow the same proof procedure of Theorem~\ref{thm_anc_sele_cons}. 
Denote the oracle M-estimator $\widehat \btheta^{m l} = (\widehat \W^{ml}_{\steponesuperset{\text{in}}(j),j},\widehat \U^{ml}_{\steponesuperset{\text{an}}(j),j},\widehat \balpha^{ml}_{\steponesuperset{\text{an}}(j), j})) = \argmin \loss(\btheta| \Y_{\steponesuperset{\text{an}}(j)}, \X_{\steponesuperset{\text{in}}(j)}, \widehat \h_{\steponesuperset{\text{an}}(j)} )$ such that 
$\{ k: \widehat U_{kj}^{ml} \neq 0\} = \{ k: U_{kj}^{0} \neq 0\} = \text{pa}^{0}(j)$, $\{ l: \widehat W_{lj}^{ml} \neq 0\}  = \{ l: W_{lj}^{0} \neq 0\} = \text{in}^{0}(j)$ and $\{ k: \widehat \alpha_{kj}^{ml} \neq 0\} = \{ k: \alpha_{kj}^{0} \neq 0\}  $. 
Further, denote $\constrainedset$ as the set of non-zero indices of the concatenated vector $\btheta^{0} = (\W_{\text{in}^{0}(j), j}^{0},\U_{\text{an}^{0}(j), j}^{0},\balpha_{\text{an}^{0}(j), j}^{0})$. Therefore, $\btheta^{0}_{\constrainedset} = (\W_{\text{in}^{0}(j), j}^{0},\U_{\text{pa}^{0}(j), j}^{0},\balpha_{\text{an}^{0}(j), j}^{0})$ and $\widehat \btheta^{ml}_{\constrainedset} = (\widehat \W^{ml}_{\text{in}^{0}(j), j},\widehat \U^{m l}_{\text{pa}^{0}(j), j},\widehat \balpha^{ml}_{\text{an}^{0}(j), j})$; $\text{supp}(\btheta^{0}) = \text{supp}(\widehat \btheta^{m l}) = \constrainedset$. 
By the proof of Theorem~\ref{thm_anc_sele_cons}, it suffices to bound the event $\left\{\left\|\widehat \btheta^{m l}-\btheta^{0}\right\|_{\infty} \leq 0.5 \tau_j \right\}$, or equivalently, $\left\{\left\|\widehat \btheta^{ml}_{\constrainedset} - \btheta^{0}_{\constrainedset}\right\|_{\infty} \leq 0.5 \tau_j \right\}$, as $\widehat U^{ml}_{kj} = U^{0}_{kj} = 0$ on $k \in (\constrainedset)^c$.
Alternatively, by Proposition 1 of \citet{shen2012likelihood},
\begin{align*}
P \left( \widehat{\btheta} \neq \widehat{\btheta}^{m l}\right) \leq \exp \left( -c_2 n C_{\min}(\btheta^0) + 2 \log (p+1) + 3 \right), 
\end{align*}
where $\widehat{\btheta}  = (\widehat{\W}_{{\steponesuperset{\text{in}}(j)}, j},\widehat{\U}_{\steponesuperset{\text{an}}(j), j},\widehat{\balpha}_{\steponesuperset{\text{an}}(j), j})$ is the final TLP estimator at iteration $T$, i.e., $\widehat{\btheta}^{[T]}$.
Therefore, $\left\|\widehat \btheta^{ml} - \btheta^{0} \right\|_{\infty} \leq 0.5 \tau_j $ 
implies that $\left\| \widehat{\btheta}  - \btheta^{0} \right\|_{\infty} \leq 0.5 \tau_j$.

For the root equations, note that the confounders $\h_k$ is independent of the instrumental variable $\X_{\text{in}^0(k)}$. Hence, the confounders do not interfere with the estimation of the coefficient $\W_{\text{in}(k),k}$. By the standard GLM result, $\|  \W^{0}_{\text{in}^{0}(k), k} - \widehat{\W}^{m l}_{\text{in}^{0}(k), k} \|_{\infty} \propto \sqrt{\frac{\log (n  \widetilde s) }{n}} $. We prove the error bound in detail in Lemma~\ref{lemma_error_bound_confounder_root}.

For the child equations, let $\btheta_{\constrainedset} = (\W_{\text{in}^{0}(j), j},\U_{\text{pa}^{0}(j), j},\balpha_{\text{an}^{0}(j), j})$ and  $\widetilde \Z = [\X_{\text{in}^{0}(j)},\Y_{\text{pa}^{0}(j)},\widehat \h_{\text{an}^{0}(j)}]$.
Let $s = \max_{1 \leq j \leq p} \| \U_{\bullet j}^0 \|_0$ and $\widetilde s = \max_{1 \leq j \leq p} \| \W_{\bullet j}^0 \|_0$. The log-likelihood that $\widehat \btheta^{m l}_{\constrainedset} = (\widehat \W^{m l}_{\text{in}^{0}(j), j},\widehat \U^{m l}_{\text{pa}^{0}(j), j},\widehat \balpha^{m l}_{\text{an}^{0}(j), j})$ minimizes is:
\begin{align*}
\loss(\btheta_{\constrainedset}| \widetilde \Z) &= \loss(\W_{\text{in}^{0}(j), j},\U_{\text{pa}^{0}(j), j},\balpha_{\text{an}^{0}(j), j} | \X_{\text{in}^{0}(j)}, \Y_{\text{pa}^{0}(j)}, \widehat \h_{\text{an}^{0}(j)} ) \\
& = \frac{1}{n} \sum_{i=1}^{n}
-Y_{i j} \left( \W^{\top}_{\text{in}^{0}(j), j} \X_{i,\text{in}^{0}(j)}  +  \U^{\top}_{\text{pa}^{0}(j), j} \Y_{i,\text{pa}^{0}(j)}  +  \balpha^{\top}_{\text{an}^{0}(j), j} \widehat \h_{i,\text{an}^{0}(j)}  \right) \\
& + A_j \left( \W^{\top}_{\text{in}^{0}(j), j} \X_{i,\text{in}^{0}(j)}  +  \U^{\top}_{\text{pa}^{0}(j), j} \Y_{i,\text{pa}^{0}(j)} +  \balpha^{\top}_{\text{an}^{0}(j), j} \widehat \h_{i,\text{an}^{0}(j)}  \right).
\end{align*} 
Since $\widehat \btheta^{m l}_{\constrainedset} = (\widehat{\W}^{m l}_{\text{in}^{0}(j), j},\widehat{\U}^{m l}_{\text{pa}^{0}(j), j},\widehat{\balpha}^{m l}_{\text{an}^{0}(j), j})$ minimizes $\loss(\btheta_{\constrainedset} |\Y_{\text{pa}^{0}(j)}, \X_{\text{in}^{0}(j)}, \widehat \h_{\text{an}^{0}(j)} )$, by the KKT condition for the oracle MLE constrained on the true set:
\begin{align}
&\sum_{i=1}^n \widetilde Z_{ik} (y_{ij} - \varphi_j(\widetilde \z_i^{\top} \widehat{\btheta}_{\constrainedset}^{ml} )) = 0, \nonumber \\
&\sum_{i=1}^n \widetilde Z_{ik} (y_{ij} - \varphi_j(\widetilde \z_i^{\top} \btheta_{\constrainedset}^{0}) + \varphi_j(\widetilde \z_i^{\top} \btheta_{\constrainedset}^{0}) - \varphi_j(\widetilde \z_i^{\top} \widehat{\btheta}_{\constrainedset}^{ml})) = 0. \label{eq:KKTthm2}
\end{align}
As in Lemma~\ref{oracle_expression}, applying Taylor series expansion, \eqref{eq:KKTthm2} can be written in matrix form:
\begin{align*}
\widetilde \Z^{\top} \hessianTheta \widetilde \Z  (\btheta_{\constrainedset}^{0} - \widehat{\btheta}_{\constrainedset}^{ml}) = \widetilde \Z^{\top} (\Y_j -  \varphi_j(\widetilde \z_i^{\top} \btheta_{\constrainedset}^{0}) - \r).
\end{align*}
Therefore, $\btheta_{\constrainedset}^{0} - \widehat{\btheta}_{\constrainedset}^{ml} = (\widetilde \Z^{\top} \hessianTheta \widetilde \Z )^{-1} \widetilde \Z^{\top} (\Y_j -  \varphi_j(\widetilde \z_i^{\top} \btheta_{\constrainedset}^{0}) - \r)$. Then we calculate the $\ell_{\infty}$-norm of the estimation error:
\begin{align*}
\| \btheta_{\constrainedset}^{0} - \widehat{\btheta}_{\constrainedset}^{ml} \|_{\infty} &=  \| (\widetilde \Z^{\top} \hessianTheta \widetilde \Z )^{-1} \widetilde \Z^{\top} (\Y_j -  \varphi_j(\widetilde \z_i^{\top} \btheta_{\constrainedset}^{0}) - \r) \|_{\infty} \\
&= \|   (\widetilde \Z^{\top} \hessianTheta \widetilde \Z )^{-1} \widetilde \Z^{\top} (\Y_j - \varphi_j(\z_i^{\top} \btheta_{\constrainedset}^{0}) + \varphi_j(\z_i^{\top} \btheta_{\constrainedset}^{0}) -  \varphi_j(\widetilde \z_i^{\top} \btheta_{\constrainedset}^{0}) - \r) \|_{\infty} \\
& \leq \| \H (\Y_j - \varphi_j(\z_i^{\top} \btheta_{\constrainedset}^{0})) \|_{\infty} +  \| \H (\varphi_j(\z_i^{\top} \btheta_{\constrainedset}^{0}) -  \varphi_j(\widetilde \z_i^{\top} \btheta_{\constrainedset}^{0}) ) \|_{\infty} +  \| \H \r  \|_{\infty},
\end{align*}
where $\H =  (\widetilde \Z^{\top} \hessianTheta \widetilde \Z )^{-1} \widetilde \Z^{\top}$. Denote $\Z = [\X_{\text{in}^{0}(j)},\Y_{\text{pa}^{0}(j)},\h_{\text{an}^{0}(j)}]$ as the true predictor variable. Again, by the bounded domain for interventions condition, there exists $\childconsttwo$ such that $\| n (\widetilde \Z^{\top} \hessianTheta \widetilde \Z )^{-1} \widetilde \Z^{\top} \|_{\infty} \leq \childconsttwo$. Note $\widetilde \Z =[\X_{\text{in}^{0}(j)},\Y_{\text{pa}^{0}(j)},\widehat \h_{\text{an}^{0}(j)}] \in \mathbb R^{2s + \widetilde s}$; $\h_{\text{an}^{0}(j)}$ refers to a submatrix consisting of $\h_k, k \in \text{an}^{0}(j)$ and $\h_{\text{an}^{0}(j)}  \balpha_{\text{an}^{0}(j),j}^{0} = \sum_{k \in \text{an}^{0}(j)} \alpha^{0}_{kj}  \h_{k}$.

Since $\E[\Y_j | \Z] = \varphi_j(   \X_{\text{in}^{0}(j)} {\W^{0}_{\text{in}^{0}(j),j}}+ \Y_{\text{pa}^{0}(j)} {\U^{0}_{\text{pa}^{0}(j),j}}  +   \sum_{k \in \text{an}^{0}(j)} \alpha^{0}_{kj} \h_k) = \varphi_j(\z_i^{\top} \btheta_{\constrainedset}^{0})$, the first term can be bounded by the Bernstein's inequality. That is,
\begin{align*}
&\P( \| \H ( \Y_j - \varphi_j( \z_i^{\top} \btheta_{\constrainedset}^{0}  )) \|_{\infty} > \epsilon) \leq (2s + \widetilde s) \exp \left(- \min\left( \frac{n \epsilon^2 }{2 \subexp^2 \childconsttwo^{2}}, \frac{n \epsilon }{2 \subexp \childconsttwo}\right) \right).
\end{align*}
Setting $\epsilon = 2\subexp \childconsttwo  \sqrt{\frac{\log (n (2s + \widetilde s)) }{n}}$ leads to
\begin{align*}
\| \H (\Y_j - \varphi_j(\z_i^{\top} \btheta_{\constrainedset}^{0})) \|_{\infty} \leq  2\subexp \childconsttwo  \sqrt{\frac{\log (n(2s + \widetilde s)) }{n}} ,
\end{align*}
with probability at least $1- 2 \exp( - 2 \log n - \log (2s + \widetilde s) ) = 1 - 2 n^{-2} {(2s + \widetilde s)}^{-1}$.

Let $\Delta_k = \widehat \h_k - \h_k$ be the estimation error of the confounders. Note that
\begin{align*}
\| \varphi_j(\z_i^{\top} \btheta_{\constrainedset}^{0}) -  \varphi_j(\widetilde \z_i^{\top} \btheta_{\constrainedset}^{0}) \|_{\infty} 
& = \|   (\z_i^{\top} \btheta_{\constrainedset}^{0} -   \widetilde \z_i^{\top} \btheta_{\constrainedset}^{0} )  \odot \varphi_j'(\boldsymbol{ \xi}) \|_{\infty} \\
& =    \| \sumk \alpha^{0}_{kj}   (\h_k - \widehat \h_k)  \odot \varphi_j'(\boldsymbol{ \xi})\|_{\infty} 
\leq  \lipscthmtwo \sumk |\alpha^{0}_{kj}| \cdot \|  \Delta_k \|_{\infty},
\end{align*}
where  we use the fact that $\z_i^{\top} \btheta_{\constrainedset}^{0} = \X_{\text{in}^{0}(j)} {\W^{0}_{\text{in}^{0}(j),j}} + \Y_{\text{pa}^{0}(j)} {\U^{0}_{\text{pa}^{0}(j),j}} +  \sumk \alpha^{0}_{kj}  \h_k$ and $\widetilde \z_i^{\top} \btheta_{\constrainedset}^{0} = \X_{\text{in}^{0}(j)} {\W^{0}_{\text{in}^{0}(j),j}} +  \Y_{\text{pa}^{0}(j)} {\U^{0}_{\text{pa}^{0}(j),j}} +  \sumk \alpha^{0}_{kj}  \widehat \h_k$; the last inequality holds as $|\varphi_j'(z)| \leq \lipscthmtwo$. 
Therefore, $\| \H ( \varphi_j(\z_i^{\top} \btheta_{\constrainedset}^{0}) -  \varphi_j(\widetilde \z_i^{\top} \btheta_{\constrainedset}^{0})) \|_{\infty} \leq n \cdot \frac{\childconsttwo}{n} \cdot \lipscthmtwo \cdot \sumk |\alpha^{0}_{kj}| \cdot \|  \Delta_k \|_{\infty} = \childconsttwo \lipscthmtwo \sumk  |\alpha^{0}_{kj}| \cdot \|  \Delta_k \|_{\infty} $.

Last, for the remainder of Taylor series expansion $\r$, similar to Theorem~\ref{thm_anc_sele_cons} and Lemma~\ref{oracle_expression},
$|\H_{k \bullet} \r|=  | \sum_{i=1}^n H_{ki} r_i  | 
\leq \childconsttwo \sum_{i=1}^n |r_i|/n \leq \childconsttwo \thirdderthmtwo (\btheta_{\constrainedset}^{0} - \widehat{\btheta}_{\constrainedset}^{ml})^{\top}  (\Z^{\top} \Z / n) (\btheta_{\constrainedset}^{0} - \widehat{\btheta}_{\constrainedset}^{ml})  \leq \childconsttwo \cmaxthmtwo \thirdderthmtwo \| \btheta_{\constrainedset}^{0} - \widehat{\btheta}_{\constrainedset}^{ml}  \|_2^2$. Further, by Lemma~\ref{lemma_l2_consistency_confounder_child}, $\| \widehat \btheta_{\constrainedset}^{m l} - \btheta_{\constrainedset}^{0} \|_2   \leq \frac{2}{m} \sqrt{2s + \widetilde s} \cdot \| \nabla \loss(\btheta_{\constrainedset}^{0} |\Y_{\text{pa}^{0}(j)}, \X_{\text{in}^{0}(j)}, \widehat \h_{\text{an}^{0}(j)}) \|_{\infty} \leq  \frac{2}{m} \sqrt{2s + \widetilde s} \left[ \childineqone \sqrt{\frac{\log (n(2s + \widetilde s)) }{n}} +  \childconstone \lipscthmtwo \sumk  |\alpha^{0}_{kj} | \cdot \|  \Delta_k \|_{\infty} \right]$ with $\childineqone = 2 \subexp \childconstone$. Therefore,
\begin{align*}
& |\H_{k \bullet} \r|   \leq \childconsttwo \cmaxthmtwo \thirdderthmtwo \| \btheta_{\constrainedset}^{0} - \widehat{\btheta}_{\constrainedset}^{ml}  \|_2^2  \\
& \leq \childconsttwo \cmaxthmtwo \thirdderthmtwo \left( \frac{2}{m} \sqrt{2s + \widetilde s} \cdot \left( \childineqone \sqrt{\frac{\log (n(2s + \widetilde s)) }{n}} +  \childconstone \lipscthmtwo \sumk |\alpha^{0}_{kj}| \cdot \|  \Delta_k \|_{\infty} \right)    \right)^2 \\
& \leq \childconsttwo \cmaxthmtwo \thirdderthmtwo \left( \left( \frac{2}{m} \sqrt{2s + \widetilde s} \right)^2 \cdot 2 \left(\childineqone^2 {\frac{\log (n(2s + \widetilde s)) }{n}} + \left( \childconstone \lipscthmtwo \sumk |\alpha^{0}_{kj}| \cdot \|  \Delta_k \|_{\infty} \right)^2 \right) \right) \\
&\leq 2 \childconsttwo \cmaxthmtwo \thirdderthmtwo \left(\frac{2}{m} \right)^2  \left(  \childineqone^2 (2s + \widetilde s)  {\frac{\log (n(2s + \widetilde s)) }{n}} +  (2s + \widetilde s)\cdot \left(\childconstone \lipscthmtwo\right)^2 \cdot  s \cdot  \sumk |\alpha^{0}_{kj}|^2 \cdot \|  \Delta_k \|_{\infty}^2  \right) \\
&\leq  \childconsttwo \cmaxthmtwo \thirdderthmtwo \left( \frac{8}{m^2}   \childineqone^2 \cdot \sqrt{\frac{\log (n(2s + \widetilde s)) }{n}} +  \frac{8}{m^2}    \left(\childconstone \lipscthmtwo\right)^2 \sumk |\alpha^{0}_{kj}|^2 \cdot \|  \Delta_k \|_{\infty} \right) ,
\end{align*}
where the last inequality holds true as $n > (2s + \widetilde s)^2 \log ( n(2s + \widetilde s))$ and $(2s + \widetilde s) s \|  \Delta_k \|_{\infty} \leq (2s + \widetilde s) s \sqrt{ \frac{\log (n \widetilde s)}{n}}\leq 1$. Also, we use the property $(\sum_{i=1}^s a_i)^2 \leq s \sum_{i=1}^s a_i^2$. Combining the three terms leads to
\begin{align*}
& \quad \quad \| \widehat \btheta_{\constrainedset}^{ml} - \btheta_{\constrainedset}^{0} \|_{\infty} \leq   \thmchildone \sqrt{\frac{\log (n(2s + \widetilde s)) }{n}} + \sumk ( \childconsttwo \lipscthmtwo |\alpha^{0}_{kj}| + \frac{ 8 \childconsttwo \cmaxthmtwo \thirdderthmtwo \childconstone^2 \lipscthmtwo^2 |\alpha^{0}_{kj}|^2 }{ m^2}  ) \|  \Delta_k \|_{\infty}  \\
&\leq \thmchildone \sqrt{\frac{\log (n(2s + \widetilde s)) }{n}} + \max_k \left( \childconsttwo \lipscthmtwo |\alpha^{0}_{kj}| + \frac{ 8 \childconsttwo \cmaxthmtwo  \thirdderthmtwo \childconstone^2 \lipscthmtwo^2 |\alpha^{0}_{kj}|^2 }{ m^2} \right)  \cdot \sumk \|  \Delta_k \|_{\infty},  
\end{align*}
with probability greater than $
    1 -  4 \exp( - 2 \log n - \log (2s + \widetilde s) ) = 1 -  4 n^{-2} {(2s + \widetilde s)}^{-1}$. Here, $\thmchildone 
= 2\subexp\childconsttwo + \childconsttwo \cmaxthmtwo \thirdderthmtwo (\frac{32}{m^2}\subexp^2\childconstone^2)$. Denote $\widehat A_j$ as the set of non-zero indices of the concatenated vector $\widehat \btheta$. 
Similar to the proof of Theorem~\ref{thm_anc_sele_cons}, if $\tau_j$ is chosen such that $\tau_j \geq 2 \left\|\widehat \btheta^{m l}-\btheta^{0}\right\|_{\infty}$, then $\widehat \btheta = \widehat \btheta^{m l}$ and $\widehat A_j = {\constrainedset}$, that is, $\widehat{\text{pa}}(j) = \text{pa}^{0}(j)$, $\widehat{\text{in}}(j) = \text{in}^{0}(j)$ and $\widehat{\text{an}}(j) =  \text{an}^{0}(j)$. Additionally, note that $\max(\| \widehat \U_{\bullet j} - \U_{\bullet j}^0 \|_{\infty},\| \widehat \W_{\bullet j} - \W_{\bullet j}^0 \|_{\infty}) \leq \| \widehat \btheta_{\constrainedset} - \btheta_{\constrainedset}^{0} \|_{\infty}$.

We have calculated the parameter estimation errors for the $j$th equation. 
Now we calculate the accumulated error for the confounder.
Note that $\h_j =  \varphi_j^{-1}(\E [ \Y_j | \X_{\text{in}^{0}(j)},\Y_{\text{pa}^{0}(j)},\h_j ]) - (\X_{\text{in}^{0}(j)} \W^{0}_{\text{in}^{0}(j),j} +  \Y_{\text{pa}^{0}(j)} \U^{0}_{\text{pa}^{0}(j),j})$. On the other hand, by construction, $\widehat \h_j = \sumk \widehat \alpha_{kj} \widehat \h_k +  \estconf$ where $\estconf = \Y_j - \varphi_j( \X_{\widehat{\text{in}}(j)} \widehat \W_{\widehat{\text{in}}(j), j} + \Y_{\widehat{\text{pa}}(j)} \widehat \U_{\widehat{\text{pa}}(j), j} +  \sum_{k \in \widehat{\text{an}}(j)} \widehat \alpha_{kj} \widehat h_k)$ is the residual estimated from the $j$th equation. 
In practice, we replace $\widehat \h_j$ with $\estconf$ in the algorithm as $\widehat \h_j$ is a linear combination of $\widehat \h_k$ and $\estconf$; including $\widehat \h_k$ and $\widehat \h_j$ in the GLM regression model is equivalent to including $\widehat \h_k$ and $\estconf$.
Note $\widehat \btheta = \widehat \btheta^{m l}$ implies $\widehat \U_{\widehat{\text{pa}}(j), j} = \widehat \U^{ml}_{\text{pa}^{0}(j),j}$, $\widehat \W_{\widehat{\text{in}}(j), j} = \widehat \W^{ml}_{\text{in}^{0}(j),j}$, and $\widehat \alpha_{kj} = \widehat \alpha^{ml}_{kj}$. Similar to the root node case, 
\begin{align*}
\estconf &= \Y_j - \varphi_j( \X_{\text{in}^{0}(j)} \widehat \W^{ml}_{\text{in}^{0}(j),j}  + \Y_{\text{pa}^{0}(j)} \widehat \U^{ml}_{\text{pa}^{0}(j),j}  + \sumk \widehat \alpha^{ml}_{kj} \widehat \h_k) \\
&= \varphi_j(\varphi_j^{-1}(\Y_j)) - \varphi_j( \X_{\text{in}^{0}(j)} \widehat \W^{ml}_{\text{in}^{0}(j),j}  + \Y_{\text{pa}^{0}(j)} \widehat \U^{ml}_{\text{pa}^{0}(j),j}  + \sumk \widehat \alpha^{ml}_{kj} \widehat \h_k),
\end{align*}
and we use $\varphi_j^{-1}(\E [ \Y_j | \X_{\text{in}^{0}(j)},\Y_{\text{pa}^{0}(j)},\h_j ]) - ( \X_{\text{in}^{0}(j)} \widehat \W^{ml}_{\text{in}^{0}(j),j}  + \Y_{\text{pa}^{0}(j)} \widehat \U^{ml}_{\text{pa}^{0}(j),j}  + \sumk \widehat \alpha^{ml}_{kj} \widehat \h_k)$ to approximate $\estconf$. 
In this way,
\begin{align*}
& \quad\; \Delta_j = \widehat \h_j - \h_j = (\sumk \widehat \alpha^{ml}_{kj} \widehat \h_k +  \estconf)  - \h_j  \\
&=   \sumk \widehat \alpha^{ml}_{kj} \widehat \h_k - ( \X_{\text{in}^{0}(j)} \widehat \W^{ml}_{\text{in}^{0}(j),j} +  \Y_{\text{pa}^{0}(j)} \widehat \U^{ml}_{\text{pa}^{0}(j),j} +  \sumk \widehat \alpha^{ml}_{kj} \widehat \h_k) +  ( \X_{\text{in}^{0}(j)} \W^{0}_{\text{in}^{0}(j),j} +  \Y_{\text{pa}^{0}(j)} \U^{0}_{\text{pa}^{0}(j),j} )   \nonumber \\
&= \sumk ( \widehat \alpha^{ml}_{kj} \widehat \h_k - \alpha^{0}_{kj} \widehat \h_k) +     \widetilde \Z (\btheta_{\constrainedset}^{0} - \widehat \btheta_{\constrainedset}^{ml})    ,  
\end{align*}
where the last equality holds as $\widetilde \Z (\btheta_{\constrainedset}^{0} - \widehat \btheta_{\constrainedset}^{ml}) = \X_{\text{in}^{0}(j)} (\W^{0}_{\text{in}^{0}(j),j}- \widehat \W^{ml}_{\text{in}^{0}(j),j}) +  \Y_{\text{pa}^{0}(j)} (\U^{0}_{\text{pa}^{0}(j),j} - \widehat \U^{ml}_{\text{pa}^{0}(j),j})  +  \sumk (\alpha^{0}_{kj} - \widehat \alpha^{ml}_{kj}) \widehat \h_k$. Taking $\ell_{\infty}$-norm of both sides yields
\begin{align*}
& \quad\; \| \Delta_j \|_{\infty} \leq \sumk   \|  \widehat \alpha^{ml}_{kj} \widehat \h_k - \alpha^{0}_{kj} \widehat \h_k \|_{\infty} + \| \widetilde \Z  (\btheta_{\constrainedset}^{0} - \widehat \btheta_{\constrainedset}^{ml}) \|_{\infty}    \\
&\leq   \max_k | \widehat \alpha^{ml}_{kj}  - \alpha^{0}_{kj} | \cdot   \sumk   \| \widehat \h_k \|_{\infty}   +   \|  \widetilde \Z (\widetilde \Z^{\top} \hessianTheta \widetilde \Z )^{-1} \widetilde \Z^{\top} (\Y_j -  \varphi_j(\widetilde \z_i^{\top} \btheta_{\constrainedset}^{0}) - \r) \|_{\infty} \\
&\leq  \| \btheta_{\constrainedset}^{0} - \widehat \btheta_{\constrainedset}^{ml} \|_{\infty} \cdot \sumk  \| \widehat \h_k \|_{\infty} +   \|  \H_2 (\Y_j - \varphi_j(\z_i^{\top} \btheta_{\constrainedset}^{0}) + \varphi_j(\z_i^{\top} \btheta_{\constrainedset}^{0}) -  \varphi_j(\widetilde \z_i^{\top} \btheta_{\constrainedset}^{0}) - \r) \|_{\infty} \\
&\leq \childconstone s \cdot \| \btheta_{\constrainedset}^{0} - \widehat \btheta_{\constrainedset}^{ml} \|_{\infty} +   \| \H_2 (\Y_j - \varphi_j(\z_i^{\top} \btheta_{\constrainedset}^{0})) \|_{\infty} +  \| \H_2 (\varphi_j(\z_i^{\top} \btheta_{\constrainedset}^{0}) -  \varphi_j(\widetilde \z_i^{\top} \btheta_{\constrainedset}^{0}) ) \|_{\infty}\\
& \quad \quad  + \| \H_2 \r  \|_{\infty},
\end{align*}
\sloppy{where $\H_2 = \widetilde \Z (\widetilde \Z^{\top} \hessianTheta \widetilde \Z )^{-1} \widetilde \Z^{\top}$. Again, by Assumption~\ref{interv_boundedness}, there exists $\childconstthree$ such that $\| n \widetilde \Z (\widetilde \Z^{\top} \hessianTheta \widetilde \Z )^{-1} \widetilde \Z^{\top} \|_{\infty} \leq \childconstthree$. Similarly,}
\begin{align*}
|| \Delta_j ||_{\infty} &\leq  \childconstone s \cdot \| \btheta_{\constrainedset}^{0} - \widehat \btheta_{\constrainedset}^{ml} \|_{\infty} +  2\subexp \childconstthree \sqrt{\frac{\log (n(2s + \widetilde s))}{n}} +  \childconstthree \lipscthmtwo \sumk |\alpha^{0}_{kj}| \cdot \|  \Delta_k \|_{\infty}  \\
& \quad +  \childconstthree \cmaxthmtwo \thirdderthmtwo \left(  \frac{32}{m^2} \subexp^2 \childconstone^2 \cdot \sqrt{\frac{\log (n(2s + \widetilde s)) }{n}} +  \frac{ 8 }{m^2}  \childconstone^2 \lipscthmtwo^2 \sumk |\alpha^{0}_{kj}|^2 \cdot \|  \Delta_k \|_{\infty} \right)   \\
&\leq \thmchilddeltatwo \sumk  \| \Delta_k \|_{\infty} + \thmchilddeltaone \sqrt{\frac{\log (n(2s + \widetilde s)) }{n}}  
\leq \thmchilddeltatwo s \cdot \max_k \| \Delta_k \|_{\infty} + \thmchilddeltaone \sqrt{\frac{\log (n(2s + \widetilde s)) }{n}} ,
\end{align*}
\sloppy{where $\thmchilddeltatwo = \max_k \left( \childconstthree \lipscthmtwo |\alpha^{0}_{kj}| + \frac{ 8 \childconstthree \cmaxthmtwo \thirdderthmtwo \childconstone^2 \lipscthmtwo^2 |\alpha^{0}_{kj}|^2 }{ m^2 }   + \childconstone s\left(\childconsttwo \lipscthmtwo |\alpha^{0}_{kj}| + \frac{ 8 \childconsttwo \cmaxthmtwo \thirdderthmtwo \childconstone^2 \lipscthmtwo^2|\alpha^{0}_{kj}|^2 }{ m^2}   \right)  \right)$ and $\thmchilddeltaone =  \left(2 \subexp \childconstthree + \childconstthree \cmaxthmtwo \thirdderthmtwo \frac{32}{m^2} \subexp^2 \childconstone^2 + \childconstone s(2 \subexp \childconsttwo + \childconsttwo \cmaxthmtwo \thirdderthmtwo \frac{32}{m^2}\subexp^2 \childconstone^2) \right)$. The above inequality can be written as:
$\|\Delta_j\|_{\infty} + c \leq \thmchilddeltatwo s \left( \max_k \|\Delta_k\|_{\infty} + c \right)$,
where $c = \frac{ \thmchilddeltaone }{\thmchilddeltatwo s -1}\sqrt{\frac{\log (n(2s + \widetilde s)) }{n}}$.
Therefore,}
\begin{align*}
\|\Delta_j\|_{\infty} + c &\leq  {(\thmchilddeltatwo s)}^{d_j} ( \|\Delta_0\|_{\infty} + c),
\end{align*}    
where $d_j$ denotes the topology depth of the primary variable $\Y_j$ defined as the maximal length of a directed path in the graph from a root variable with depth zero; therefore, $0 \leq d_j \leq d_{\max} \leq p-1$ with $d_{\max}$ the maximal length of a directed path. Rearranging terms yields
\begin{align*}    
\|\Delta_j\|_{\infty} 
&\leq {(\thmchilddeltatwo s)}^{d_j}  \|\Delta_0\|_{\infty} + ( {(\thmchilddeltatwo s)}^{d_j}   - 1 )c                         \\
&= {(\thmchilddeltatwo s)}^{d_j}  \|\Delta_0\|_{\infty} + ( {(\thmchilddeltatwo s)}^{d_j}   - 1 ) \frac{ \thmchilddeltaone }{\thmchilddeltatwo s -1}\sqrt{\frac{\log (n (2s + \widetilde s)) }{n}} .
\end{align*}

In this way, we derive the general form of the accumulated error for $\|\Delta_j\|_{\infty}$ for the multi-layer case. To conclude, if $\tau_j$ satisfies: $\tau_j \geq 2 \left\|\widehat \btheta^{m l}-\btheta^{0}\right\|_{\infty} = C \sqrt{\frac{\log (n (2s + \widetilde s)) }{n}} $, then the deconfounding algorithm reconstructs the causal graph consistently, i.e., $\{(k,j): \widehat U_{kj} \neq 0 \} = \{(k,j): U^0_{kj} \neq 0 \}$, with probability tending to one as $n \to \infty$.

Lemma~\ref{lemma_l2_consistency_confounder_child} bounds the quantity $\| \widehat \btheta_{\constrainedset}^{m l} - \btheta_{\constrainedset}^{0} \|_2$ in child equations.
\begin{lemma}[Rate of convergence under the $\ell_2$-norm for child equations] \label{lemma_l2_consistency_confounder_child} 
	\begin{align*}
	\| \widehat \btheta_{\constrainedset}^{m l} - \btheta_{\constrainedset}^{0} \|_2   &\leq \frac{2}{m} \sqrt{2s + \widetilde s} \cdot \| \nabla \loss(\btheta_{\constrainedset}^{0} |\Y_{\text{pa}^{0}(j)}, \X_{\text{in}^{0}(j)}, \widehat \h_{\text{an}^{0}(j)}) \|_{\infty} \\
	&\leq  \frac{2}{m} \sqrt{2s + \widetilde s} \left[ 2 \subexp \childconstone \sqrt{\frac{\log (n(2s + \widetilde s)) }{n}} +  \childconstone \lipscthmtwo \sumk  |\alpha^{0}_{kj} | \cdot \|  \Delta_k \|_{\infty} \right],
	\end{align*}
	with probability at least $1- 2 \exp( - 2 \log n - \log (2s + \widetilde s) ) = 1 - 2 n^{-2} {(2s + \widetilde s)}^{-1}$.
\end{lemma}
\noindent {\bf Proof of Lemma~\ref{lemma_l2_consistency_confounder_child}}.
Since $\widehat \btheta_{\constrainedset}^{m l} = (\widehat{\W}^{m l}_{\text{in}^{0}(j), j},\widehat{\U}^{m l}_{\text{pa}^{0}(j), j},\widehat{\balpha}^{m l}_{\text{an}^{0}(j), j})$ minimizes $\loss(\btheta_{\constrainedset}|\Y_{\text{pa}^{0}(j)}, \X_{\text{in}^{0}(j)}, \widehat \h_{\text{an}^{0}(j)} )$, 
$\loss(\widehat \btheta_{\constrainedset}^{m l}|\Y_{\text{pa}^{0}(j)}, \X_{\text{in}^{0}(j)}, \widehat \h_{\text{an}^{0}(j)} ) \leq \loss(\btheta_{\constrainedset}^{0} |\Y_{\text{pa}^{0}(j)}, \X_{\text{in}^{0}(j)}, \widehat \h_{\text{an}^{0}(j)} )$.  
As in Lemma~\ref{lemma_l2_consistency}, $\| \widehat \btheta_{\constrainedset}^{m l} - \btheta_{\constrainedset}^{0} \|_2   \leq \frac{2}{m} \sqrt{2s + \widetilde s} \cdot \| \nabla \loss(\btheta_{\constrainedset}^{0} |\Y_{\text{pa}^{0}(j)}, \X_{\text{in}^{0}(j)}, \widehat \h_{\text{an}^{0}(j)}) \|_{\infty}$. Meanwhile,
\begin{align*}
& \quad\; \| \nabla \loss(\btheta_{\constrainedset}^{0} |\Y_{\text{pa}^{0}(j)}, \X_{\text{in}^{0}(j)}, \widehat \h_{\text{an}^{0}(j)}) \|_{\infty} \\
& = n^{-1} \| \widetilde \Z^{\top} ( \Y_j - \varphi_j(   \X_{\text{in}^{0}(j)} {\W^{0}_{\text{in}^{0}(j),j}} +  \Y_{\text{pa}^{0}(j)} {\U^{0}_{\text{pa}^{0}(j),j}} +   \sumk \alpha^{0}_{kj} \widehat \h_k)) \|_{\infty} \\
&= n^{-1} \big \| T_1 + T_2   \|_{\infty} \leq n^{-1} \big \| T_1 \|_{\infty}  + n^{-1} \big \| T_2 \|_{\infty} \\
& \leq n^{-1} \| T_1  \|_{\infty} + n^{-1} \|   \widetilde \Z^{\top} \cdot \left( \sumk \alpha^{0}_{kj}  (\h_k - \widehat \h_k)  \odot \varphi_j'(\boldsymbol{\xi}) \right) \|_{\infty}  \leq n^{-1} \| T_1  \|_{\infty} + \childconstone \lipscthmtwo \sumk |\alpha^{0}_{kj}| \cdot \|  \Delta_k \|_{\infty},    
\end{align*}
\sloppy{where $T_1 = \widetilde \Z^{\top} ( \Y_j - \varphi_j( \X_{\text{in}^{0}(j)} {\W^{0}_{\text{in}^{0}(j),j}} +  \Y_{\text{pa}^{0}(j)} {\U^{0}_{\text{pa}^{0}(j),j}}  +   \sumk \alpha^{0}_{kj}  \h_k))$ and $T_2 = \widetilde \Z^{\top} (\varphi_j( \X_{\text{in}^{0}(j)} {\W^{0}_{\text{in}^{0}(j),j}} + \Y_{\text{pa}^{0}(j)} {\U^{0}_{\text{pa}^{0}(j),j}}  +  \sumk \alpha^{0}_{kj}  \h_k) - \varphi_j(   \X_{\text{in}^{0}(j)} {\W^{0}_{\text{in}^{0}(j),j}} + \Y_{\text{pa}^{0}(j)} {\U^{0}_{\text{pa}^{0}(j),j}}  + \sumk  \alpha^{0}_{kj}  \widehat \h_k))$. The last inequality holds as $|\varphi_j'(z)| = |A_j^{''}(z)| \leq \lipscthmtwo$. 
}

Note that $\| \widetilde \Z  \|_{\infty} \leq \childconstone$ and $\widetilde \Z =[\X_{\text{in}^{0}(j)},\Y_{\text{pa}^{0}(j)},\widehat \h_{\text{an}^{0}(j)}]  \in \mathbb R^{2s + \widetilde s}$. The first term can be bounded by the Bernstein's inequality since $\E[\Y_j | \Z] = \varphi_j(   \X_{\text{in}^{0}(j)} {\W^{0}_{\text{in}^{0}(j),j}} +  \Y_{\text{pa}^{0}(j)} {\U^{0}_{\text{pa}^{0}(j),j}}  +  \sumk \alpha^{0}_{kj}  \h_k) $. That is,
\begin{align*}
&\P( n^{-1} \| T_1  \|_{\infty} > \epsilon) \leq (2s + \widetilde s) \exp \left(-\min\left( \frac{n \epsilon^2 }{2 \subexp^2 \childconstone^{2}}, \frac{n \epsilon }{2 \subexp \childconstone} \right) \right).
\end{align*}
Setting $\epsilon = \childineqone  \sqrt{\frac{\log (n (2s + \widetilde s)) }{n}}$ yields
\begin{align*}
&\P \left( n^{-1} \| T_1  \|_{\infty} \leq  \childineqone  \sqrt{\frac{\log (n (2s + \widetilde s))}{n}} \right) \geq 1 - 2 \exp \left(-\frac{1}{2} (\frac{\childineqone^2}{\subexp^2 \childconstone^2}-2) \log (2s + \widetilde s) - \frac{\childineqone^2}{2 \subexp^2 \childconstone^2} \log n\right).
\end{align*}
In particular, setting $\childineqone = 2 \subexp \childconstone$ yields
\begin{align*}
\| \widehat \btheta_{\constrainedset}^{m l} - \btheta_{\constrainedset}^{0} \|_2   &\leq  \frac{2}{m} \sqrt{2s + \widetilde s} \left[ 2 \subexp \childconstone \sqrt{\frac{\log (n(2s + \widetilde s)) }{n}} +  \childconstone \lipscthmtwo \sumk  |\alpha^{0}_{kj} | \cdot \|  \Delta_k \|_{\infty} \right],
\end{align*}
with probability at least $1- 2 \exp( - 2 \log n - \log (2s + \widetilde s) ) = 1 - 2 n^{-2} {(2s + \widetilde s)}^{-1}$.

Lemma~\ref{lemma_l2_consistency_confounder_root} bounds the quantity $\| \widehat{\W}^{m l}_{\text{in}^{0}(k), k}  - {\W}^{0}_{\text{in}^{0}(k), k} \|_2$ in root equations.
\begin{lemma}[Rate of convergence under the $\ell_2$-norm for root equations] \label{lemma_l2_consistency_confounder_root} 
	\begin{align*}
	\| \widehat{\W}^{m l}_{\text{in}^{0}(k), k}  - {\W}^{0}_{\text{in}^{0}(k), k} \|_2   &\leq  \frac{2}{m} \sqrt{ \widetilde s} \left[ 2 \subexp \rootconstone \sqrt{\frac{\log (n \widetilde s)}{n}} \right],
	\end{align*}
	with probability at least $1- 2 \exp( - 2 \log n - \log \widetilde s) = 1 - 2 n^{-2} {\widetilde s}^{-1}$. 
\end{lemma}

\noindent {\bf Proof of Lemma \ref{lemma_l2_consistency_confounder_root}}.	
Consider the log-likelihood for a root variable $Y_k$: 
\begin{align*}
\loss(\W_{\text{in}^{0}(k), k} |\Y_k, \X_{\text{in}^{0}(k)} ) 
= n^{-1} \sum_{i=1}^{n} 
-Y_{i k} \left( \W^{\top}_{\text{in}^{0}(k), k} \X_{i,\text{in}^{0}(k)}     \right)  + A_k  \left( \W^{\top}_{\text{in}^{0}(k), k} \X_{i,\text{in}^{0}(k)}   \right),
\end{align*}
where the oracle estimator $\widehat{\W}^{m l}_{\text{in}^{0}(k), k}$ is its minimizer with respect to $\W_{\text{in}^{0}(k), k}$.
	
By the definition of $\widehat{\W}^{m l}_{\text{in}^{0}(k), k}$, it follows from Lemma~\ref{lemma_l2_consistency}
that $\| \widehat{\W}^{m l}_{\text{in}^{0}(k), k}  - {\W}^{0}_{\text{in}^{0}(k), k} \|_2   \leq \frac{2}{m} \sqrt{\widetilde s} \cdot \| \nabla \loss( {\W}^{0}_{\text{in}^{0}(k), k} |\Y_k, \X_{\text{in}^{0}(k)}) \|_{\infty}$,
where $\nabla \loss( {\W}^{0}_{\text{in}^{0}(k), k} |\Y_k, \X_{\text{in}^{0}(k)})$ is 
the gradient of $\loss( {\W}^{0}_{\text{in}^{0}(k), k} |\Y_k, \X_{\text{in}^{0}(k)})$.
By the triangular inequality,
{\small{\begin{align}
        & \quad\; \|\nabla \loss( {\W}^{0}_{\text{in}^{0}(k), k} |\Y_k, \X_{\text{in}^{0}(k)}) \|_{\infty} 
        = n^{-1} \| \X_{\text{in}^{0}(k)}^{\top} ( \Y_k - \varphi_k(  {\X_{\text{in}^{0}(k)} \W^{0}_{\text{in}^{0}(k),k} }  )) \|_{\infty} \nonumber \\
        & \leq  n^{-1} \|  \X_{\text{in}^{0}(k)}^{\top}  ( \Y_k - \varphi_k( {\X_{\text{in}^{0}(k)} \W^{0}_{\text{in}^{0}(k),k}} + \h_k )) \|_{\infty}  \nonumber \\
        &  \quad +  n^{-1} \| \X_{\text{in}^{0}(k)}^{\top}  (  \varphi_k( {\X_{\text{in}^{0}(k)} \W^{0}_{\text{in}^{0}(k),k}} + \h_k )  - \varphi_k(  {\X_{\text{in}^{0}(k)} \W^{0}_{\text{in}^{0}(k),k}}  ))   \|_{\infty}  \nonumber \\
        & \equiv G_1 + G_2.
        \label{gradient_root}
        \end{align}}}

Note that $\| \X_{\text{in}^{0}(k)}  \|_{\infty} \leq \rootconstone$ and $\E[\Y_k | \X_{\text{in}^{0}(k)}, \h_k ] =  \varphi_k( {\X_{\text{in}^{0}(k)} \W^{0}_{\text{in}^{0}(k),k}} + \h_k ) $. By the Bernstein's inequality, the first term in~\eqref{gradient_root} is bounded by
\begin{align*}
\P(G_1 > \epsilon) \leq  \widetilde s \exp \left(-\min \left(\frac{n \epsilon^2 }{2 \subexp^2 \rootconstone^{2}}, \frac{n \epsilon }{2 \subexp \rootconstone}\right) \right).
\end{align*}
Setting $\epsilon = 2 \subexp \rootconstone  \sqrt{\frac{\log (n \widetilde s) }{n}}$ leads to
$G_1 \leq 2 \subexp \rootconstone  \sqrt{\frac{\log ( n\widetilde s)}{n}}$  
with probability at least $1- 2 \exp( - 2 \log n - \log \widetilde s) = 1 - 2 n^{-2} {\widetilde s}^{-1}$.

For $G_2$ in~\eqref{gradient_root}, by the Taylor series expansion,
\begin{align*}
G_2
&= n^{-1} \|  \X_{\text{in}^{0}(k)}^{\top} \cdot \left( \varphi_k'({\X_{\text{in}^{0}(k)} \W^{0}_{\text{in}^{0}(k),k}}) \odot \h_k   \right) \|_{\infty} \\
& =  n^{-1} \| (\X_{\text{in}^{0}(k)}^{\top}  \text{diag}(\varphi_k'(  {\X_{\text{in}^{0}(k)} \W^{0}_{\text{in}^{0}(k),k}} ) )) \cdot \h_k \|_{\infty}.
\end{align*}

Note that $\X_{\text{in}^{0}(k)}$ and $\h_k$ are independent.
    Hence, $\E[ \left(\X_{\text{in}^{0}(k)}^{\top}  \text{diag}(\varphi_k'(  {\X_{\text{in}^{0}(k)} \W^{0}_{\text{in}^{0}(k),k}} ) )\right) \cdot \h_k  ] = 0$.
    By the bounded domain for interventions condition, there exists $\rootconfone$ such that $\| \X_{\text{in}^{0}(k)}^{\top}  \text{diag}(\varphi_k'(  {\X_{\text{in}^{0}(k)} \W^{0}_{\text{in}^{0}(k),k}} ) ) \|_{\infty} \leq \rootconfone$. For $j \in \text{in}^{0}(k)$, by the Hoeffding's inequality,
\begin{align*}
\P( n^{-1}\|\X_{j}^{\top}  \text{diag}(\varphi_k'(  {\X_{\text{in}^{0}(k)} \W^{0}_{\text{in}^{0}(k),k}} ) ) \h_k \|_{\infty} > \epsilon) \leq   2\exp \left(- \frac{n \epsilon^2 }{2 \sigma_j^2 \rootconfone^{2}}    \right).
\end{align*}
Applying the union bound and setting $\epsilon = 2 \sigma_j \rootconfone  \sqrt{\frac{\log (n \widetilde s) }{n}}$ yield $G_2 \leq 2 \sigma_j \rootconfone \sqrt{\frac{\log ( n\widetilde s)}{n}}$, 
with probability at least $1- 2 \exp( - 2 \log n - \log \widetilde s) = 1 - 2 n^{-2} {\widetilde s}^{-1}$.
For simplicity, set $G_2  = o(\sqrt{\frac{\log ( n\widetilde s)}{n}})$.

Finally, combining the two terms in~\eqref{gradient_root} yields: 
\begin{align*}
\| \widehat{\W}^{m l}_{\text{in}^{0}(k), k}  - {\W}^{0}_{\text{in}^{0}(k), k} \|_2   & \leq  \frac{2}{m} \sqrt{ \widetilde s} \cdot \|\nabla \loss( {\W}^{0}_{\text{in}^{0}(k), k} |\Y_k, \X_{\text{in}^{0}(k)}) \|_{\infty} 
\leq \frac{2}{m} \sqrt{ \widetilde s} \big( 2 \subexp \rootconstone \sqrt{\frac{\log (n \widetilde s)}{n}} \big).
\end{align*}

Lemma~\ref{lemma_error_bound_confounder_root} derives the estimation bound for $\| \widehat{\W}_{{\steponesuperset{\text{in}}(k)}, k} - \W^{0}_{\text{in}^{0}(k), k} \|_{\infty}$ in root equations.

\begin{lemma}[Rate of convergence under the $\ell_{\infty}$-norm for root equations] \label{lemma_error_bound_confounder_root}
	\begin{align*}
	\|   \widehat{\W}_{{\steponesuperset{\text{in}}(k)}, k} - \W^{0}_{\text{in}^{0}(k), k} \|_{\infty}
	\leq  \left(2 \subexp \rootconsttwo + \rootconsttwo \cmaxthmtwo \thirdderthmtwo \frac{16}{m^2} \subexp^2 \rootconstone^2\right) \sqrt{\frac{\log (n  \widetilde s) }{n}},
	\end{align*}
	with probability at least $1- 4 \exp( - 2 \log n - \log \widetilde s) = 1 - 4 n^{-2} {\widetilde s}^{-1}$. Further, the estimation error for the confounders satisfies:
	$|| \Delta_k ||_{\infty}  \leq \left(2\subexp\rootconstthree + \rootconstthree \cmaxthmtwo \thirdderthmtwo \frac{16}{m^2} \subexp^2 \rootconstone^2\right) \sqrt{\frac{\log (n \widetilde s) }{n}}$.
\end{lemma}

\noindent {\bf Proof of Lemma \ref{lemma_error_bound_confounder_root}}.
Note that by Theorem~\ref{thm_anc_sele_cons},  $\{l: \widehat V_{lk} \neq 0\} = \{l: V_{lk}^{0} \neq 0\}$, implying that $\steponesuperset{\text{in}}(k) = \text{in}^{0}(k)$ in root equations. Therefore, by construction, $\widehat{\W}_{{\steponesuperset{\text{in}}(k)}, k} = \widehat{\W}^{m l}_{\text{in}^{0}(k), k}$, as both are GLM estimators constrained on the same set. It suffices to derive the error bound for the oracle estimator.

To establish the $\ell_{\infty}$-norm of the oracle estimator, as in Lemma~\ref{oracle_expression}, 
we apply the Taylor series expansion of $\varphi_k( {\X_{\text{in}^{0}(k)} \W_{\text{in}^{0}(k),k}}  )$ as:
\begin{align*}
\X_{\text{in}^{0}(k)}^{\top} \M \X_{\text{in}^{0}(k)}  (\W^{0}_{\text{in}^{0}(k), k} - \widehat{\W}^{m l}_{\text{in}^{0}(k), k}) =  \X_{\text{in}^{0}(k)}^{\top} (\Y_k -  \varphi_k(  {\X_{\text{in}^{0}(k)} \W^{0}_{\text{in}^{0}(k),k}} ) 
- \r).
\end{align*}
This implies that $\W^{0}_{\text{in}^{0}(k), k} - \widehat{\W}^{m l}_{\text{in}^{0}(k), k} = \H (\Y_k -  \varphi_k(  {\X_{\text{in}^{0}(k)} \W^{0}_{\text{in}^{0}(k),k}} ) - \r)$, where $\H =  (\X_{\text{in}^{0}(k)}^{\top} \M \X_{\text{in}^{0}(k)} )^{-1} \X_{\text{in}^{0}(k)}^{\top}$. Then,
\begin{align*}
& \| \W^{0}_{\text{in}^{0}(k), k} - \widehat{\W}^{m l}_{\text{in}^{0}(k), k} \|_{\infty}  
= \| \H (\Y_k -   \varphi_k( {\X_{\text{in}^{0}(k)} \W^{0}_{\text{in}^{0}(k),k}} + \h_k ) \\
& \quad +   \varphi_k( {\X_{\text{in}^{0}(k)} \W^{0}_{\text{in}^{0}(k),k}} + \h_k ) - \varphi_k(  {\X_{\text{in}^{0}(k)} \W^{0}_{\text{in}^{0}(k),k}} ) - \r) \|_{\infty}  \\
& \leq \| \H (\Y_k - \varphi_k( {\X_{\text{in}^{0}(k)} \W^{0}_{\text{in}^{0}(k),k}} + \h_k ) ) \|_{\infty} \\
& \quad +  \| \H (\varphi_k( {\X_{\text{in}^{0}(k)} \W^{0}_{\text{in}^{0}(k),k}} + \h_k ) - \varphi_k(  {\X_{\text{in}^{0}(k)} \W^{0}_{\text{in}^{0}(k),k}} ) ) \|_{\infty} +  \| \H \r  \|_{\infty}, 
& \equiv I_1 + I_2 + I_3.
\end{align*}
By Assumption~\ref{interv_boundedness}, there exists $\rootconsttwo$ such that $\| n (\X_{\text{in}^{0}(k)}^{\top} \M \X_{\text{in}^{0}(k)} )^{-1}\X_{\text{in}^{0}(k)}^{\top} \|_{\infty} \leq \rootconsttwo$. Note that $\E[\Y_k | \X_{\text{in}^{0}(k)}, \h_k ] =  \varphi_k( {\X_{\text{in}^{0}(k)} \W^{0}_{\text{in}^{0}(k),k}} + \h_k ) $. Then, by Bernstein's inequality:
\begin{align*}
\P( I_1 > \epsilon) \leq \widetilde s \exp \left(- \min \left(\frac{n \epsilon^2 }{2 \subexp^2 \rootconsttwo^{2}}, \frac{n \epsilon }{2 \subexp \rootconsttwo}\right) \right).
\end{align*}

Setting $\epsilon = 2\subexp\rootconsttwo  \sqrt{\frac{\log (n \widetilde s) }{n}}$ yields
$I_1 \leq  2\subexp \rootconsttwo \sqrt{\frac{\log (n \widetilde s) }{n}}$,
with probability at least $1- 2 \exp( - 2 \log n - \log \widetilde s) = 1 - 2 n^{-2} {\widetilde s}^{-1}$.
On the other hand, as in Lemma~\ref{lemma_l2_consistency_confounder_root}, we have  
\begin{align*}
& I_2 =  \|  \H  \left( \h_k  \odot  \varphi_k'(  {\X_{\text{in}^{0}(k)} \W^{0}_{\text{in}^{0}(k),k}} ) \right) \|_{\infty} \\ 
&=  \|  (\X_{\text{in}^{0}(k)}^{\top} \M \X_{\text{in}^{0}(k)} )^{-1} \X_{\text{in}^{0}(k)}^{\top} \cdot \left( \h_k  \odot  \varphi_k'(  {\X_{\text{in}^{0}(k)} \W^{0}_{\text{in}^{0}(k),k}} )  \right) \|_{\infty}  = o (\sqrt{\frac{\log (n  \widetilde s) }{n}}).
\end{align*}
Finally, as in Theorem~\ref{thm_anc_sele_cons} and Lemma~\ref{oracle_expression}, 
\begin{eqnarray*}
    |\H_{k \bullet} \r| & = &| \sum_{i=1}^n H_{ki} r_i  | \leq \sum_{i=1}^n |H_{ki}| |r_i|  \leq \rootconsttwo \sum_{i=1}^n |r_i|/n \\
    & \leq & \rootconsttwo \thirdderthmtwo (\widehat{\W}^{m l}_{\text{in}^{0}(k), k}  - {\W}^{0}_{\text{in}^{0}(k), k})^{\top}  (\X_{\text{in}^{0}(k)}^{\top} \X_{\text{in}^{0}(k)} / n) (\widehat{\W}^{m l}_{\text{in}^{0}(k), k}  - {\W}^{0}_{\text{in}^{0}(k), k}) \\
    & \leq & \rootconsttwo \cmaxthmtwo \thirdderthmtwo \| \widehat{\W}^{m l}_{\text{in}^{0}(k), k}  - {\W}^{0}_{\text{in}^{0}(k), k} \|_2^2.
\end{eqnarray*}
By Lemma~\ref{lemma_l2_consistency_confounder_root}, $\|\widehat{\W}^{m l}_{\text{in}^{0}(k), k}  - {\W}^{0}_{\text{in}^{0}(k), k} \|_2   \leq  \frac{2}{m} \sqrt{ \widetilde s} \left[ 2 \subexp \rootconstone \sqrt{\frac{\log (n \widetilde s)}{n}} \right]$. Therefore,
\begin{align*}
I_3 = \max_k |\H_{k \bullet} \r|
& \leq \rootconsttwo \cmaxthmtwo \thirdderthmtwo \left( \frac{2}{m} \sqrt{ \widetilde s} \left[ 2 \subexp \rootconstone \sqrt{\frac{\log (n \widetilde s)}{n}}  \right]  \right)^2 
\leq  \rootconsttwo \cmaxthmtwo \thirdderthmtwo \frac{16}{m^2}  \subexp^2 \rootconstone^2  \sqrt{\frac{\log (n \widetilde s)}{n} },
\end{align*}
where the last inequality holds as $n > \widetilde s^2 \log ( n\widetilde s)$. Therefore, combining $I_1$, $I_2$ and $I_3$ yields
\begin{align*}
\|  \W^{0}_{\text{in}^{0}(k), k} - \widehat{\W}^{m l}_{\text{in}^{0}(k), k} \|_{\infty} &\leq  \thmrootone \sqrt{\frac{\log (n  \widetilde s) }{n}} ,
\end{align*}
with probability greater than $
1- 4 \exp( - 2 \log n - \log \widetilde s) = 1 - 4 n^{-2} {\widetilde s}^{-1}$. Here, $\thmrootone 
= 2 \subexp \rootconsttwo + \rootconsttwo \cmaxthmtwo \thirdderthmtwo \frac{16}{m^2} \subexp^2 \rootconstone^2$.  Lastly, recall that $\steponesuperset{\text{in}}(k) = \text{in}^{0}(k)$ and $\widehat{\W}_{{\steponesuperset{\text{in}}(k)}, k} = \widehat{\W}^{m l}_{\text{in}^{0}(k), k}$. Therefore, $\|  \W^{0}_{\text{in}^{0}(k), k} - \widehat{\W}_{{\steponesuperset{\text{in}}(k)}, k} \|_{\infty} \leq  \thmrootone \sqrt{\frac{\log (n  \widetilde s) }{n}}$.

To compute the estimation error of the confounder, note that it follows from~\eqref{eq:true_model} that 
\begin{align*}
\E [ \Y_k | \X_{\text{in}^{0}(k)}, \h_k ] &=  \varphi_k( \X_{\text{in}^{0}(k)} \W_{\text{in}^{0}(k),k}^{0} + \h_k) \\
\h_k &=  \varphi_k^{-1}(\E [ \Y_k | \X_{\text{in}^{0}(k)},\h_k ]) - \X_{\text{in}^{0}(k)} \W_{\text{in}^{0}(k),k}^{0} .
\end{align*}
On the other hand, by construction, $\widehat \h_k =  \Y_k - \varphi_k( \X_{\steponesuperset{\text{in}}(k)}  \widehat \W_{\steponesuperset{\text{in}}(k), k} ) =  \Y_k - \varphi_k( \X_{\text{in}^{0}(k)} \widehat \W^{m l}_{\text{in}^{0}(k),k} ) = \varphi_k(\varphi_k^{-1}(\Y_k)) - \varphi_k( \X_{\text{in}^{0}(k)} \widehat \W^{m l}_{\text{in}^{0}(k),k} ) = \varphi_k'({\boldsymbol \xi}) \odot \left[ \varphi_k^{-1}(\Y_k) - \X_{\text{in}^{0}(k)} \widehat \W^{m l}_{\text{in}^{0}(k),k} \right]$, where the last equality holds by the mean value theorem. 
Now, we use $\varphi_k^{-1}(\E [ \Y_k | \X_{\text{in}^{0}(k)},\h_k ])$ to approximate $\varphi_k^{-1}(\Y_k)$ since $\Y_k = \E [ \Y_k | \X_{\text{in}^{0}(k)},\h_k ] + {\boldsymbol \epsilon}$, where $\epsilon_i$ follows an exponential family distribution. Further, we use $\varphi_k^{-1}(\E [ \Y_k | \X_{\text{in}^{0}(k)},\h_k ]) - \X_{\text{in}^{0}(k)} \widehat \W^{m l}_{\text{in}^{0}(k),k}$ to approximate $\widehat \h_k$ as we estimate its coefficient in subsequent equations. This
reparametrization and approximations permits a comparison of $\widehat \h_k$ and $\h_k$ at the same scale;
see \citet{johnston2008use} for some details about such approximations.
Hence,
\begin{align*}
\Delta_k = \widehat \h_k - \h_k = - \X_{\text{in}^{0}(k)} ( \widehat \W^{m l}_{\text{in}^{0}(k),k} - \W^{0}_{\text{in}^{0}(k),k} ) 
= -  \H_2(\Y_k -  \varphi_k(  {\X_{\text{in}^{0}(k)} \W^{0}_{\text{in}^{0}(k),k}} ) - \r),
\end{align*}
where $\H_2 = \X_{\text{in}^{0}(k)} (\X_{\text{in}^{0}(k)}^{\top} \M \X_{\text{in}^{0}(k)} )^{-1} \X_{\text{in}^{0}(k)}^{\top}$. By the triangular inequality,
\begin{align*}
& \quad\; \| \Delta_k \|_{\infty} \leq \| \H_2 (\Y_k -  \varphi_k(  {\X_{\text{in}^{0}(k)} \W^{0}_{\text{in}^{0}(k),k}} ) - \r) \|_{\infty} \\
& \leq \| \H_2 (\Y_k - \varphi_k( {\X_{\text{in}^{0}(k)} \W^{0}_{\text{in}^{0}(k),k}} + \h_k ) ) \|_{\infty} \\ 
& \quad + \| \H_2 (\varphi_k( {\X_{\text{in}^{0}(k)} \W^{0}_{\text{in}^{0}(k),k}} + \h_k ) - \varphi_k(  {\X_{\text{in}^{0}(k)} \W^{0}_{\text{in}^{0}(k),k}} ) ) \|_{\infty} +   \| \H_2 \r  \|_{\infty}. 
\end{align*}
By the bounded domain for interventions condition, there exists $\rootconstthree$ such that $\| n \X_{\text{in}^{0}(k)} (\X_{\text{in}^{0}(k)}^{\top} \M \X_{\text{in}^{0}(k)} )^{-1} \X_{\text{in}^{0}(k)}^{\top}  \|_{\infty} \leq \rootconstthree$. Similarly,
    $|| \Delta_k ||_{\infty}  \leq \thmroottwo \sqrt{\frac{\log (n \widetilde s) }{n}}$,
    with probability greater than $
    1- 4 \exp( - 2 \log n - \log \widetilde s) = 1 - 4 n^{-2} {\widetilde s}^{-1}$. Here, $\thmroottwo 
    = 2\subexp\rootconstthree + \rootconstthree \cmaxthmtwo \thirdderthmtwo \frac{16}{m^2} \subexp^2 \rootconstone^2$.
This completes the proof.

\end{appendix}

\bibliographystyle{abbrvnat}
\bibliography{main.bib}

\begin{thebibliography}{39}
\providecommand{\natexlab}[1]{#1}
\providecommand{\url}[1]{\texttt{#1}}
\expandafter\ifx\csname urlstyle\endcsname\relax
  \providecommand{\doi}[1]{doi: #1}\else
  \providecommand{\doi}{doi: \begingroup \urlstyle{rm}\Url}\fi

\bibitem[Bello et~al.(2022)Bello, Aragam, and Ravikumar]{bello2022dagma}
K.~Bello, B.~Aragam, and P.~Ravikumar.
\newblock {DAGMA}: Learning {DAG}s via {M}-matrices and a log-determinant acyclicity characterization.
\newblock \emph{arXiv preprint arXiv:2209.08037}, 2022.

\bibitem[Chen and Chen(2008)]{chen2008extended}
J.~Chen and Z.~Chen.
\newblock Extended {B}ayesian information criteria for model selection with large model spaces.
\newblock \emph{Biometrika}, 95\penalty0 (3):\penalty0 759--771, 2008.

\bibitem[Chen et~al.(2019)Chen, Ye, and Wang]{chen2019approximation}
Y.~Chen, Y.~Ye, and M.~Wang.
\newblock Approximation hardness for a class of sparse optimization problems.
\newblock \emph{Journal of Machine Learning Research}, 2019.

\bibitem[Choi et~al.(2004)Choi, Eom, Han, Kim, Han, Choi, Oh, Markelonis, Cho, and Oh]{choi2004phosphorylation}
W.-S. Choi, D.-S. Eom, B.~S. Han, W.~K. Kim, B.~H. Han, E.-J. Choi, T.~H. Oh, G.~J. Markelonis, J.~W. Cho, and Y.~J. Oh.
\newblock Phosphorylation of p38 {MAPK} induced by oxidative stress is linked to activation of both caspase-8-and-9-mediated apoptotic pathways in dopaminergic neurons.
\newblock \emph{Journal of Biological Chemistry}, 279\penalty0 (19):\penalty0 20451--20460, 2004.

\bibitem[Chowdhury et~al.(2022)Chowdhury, Wang, Yu, Huntoon, Karnitz, Kaufmann, Gygi, Birrer, Paulovich, Peng, et~al.]{chowdhury2022dagbagm}
S.~Chowdhury, R.~Wang, Q.~Yu, C.~J. Huntoon, L.~M. Karnitz, S.~H. Kaufmann, S.~P. Gygi, M.~J. Birrer, A.~G. Paulovich, J.~Peng, et~al.
\newblock {DAGBagM}: learning directed acyclic graphs of mixed variables with an application to identify protein biomarkers for treatment response in ovarian cancer.
\newblock \emph{BMC Bioinformatics}, 23\penalty0 (1):\penalty0 1--19, 2022.

\bibitem[Du et~al.(2020)Du, Liu, Zhu, Liu, Wang, and Wu]{du2020activating}
Y.~Du, X.~Liu, X.~Zhu, Y.~Liu, X.~Wang, and X.~Wu.
\newblock Activating transcription factor 6 reduces {A}$\beta$1--42 and restores memory in {A}lzheimer’s disease model mice.
\newblock \emph{International Journal of Neuroscience}, 130\penalty0 (10):\penalty0 1015--1023, 2020.

\bibitem[Efron et~al.(2004)Efron, Hastie, Johnstone, and Tibshirani]{efron2004least}
B.~Efron, T.~Hastie, I.~Johnstone, and R.~Tibshirani.
\newblock Least angle regression.
\newblock \emph{The Annals of Statistics}, 32\penalty0 (2):\penalty0 407--499, 2004.

\bibitem[Hastie et~al.(2015)Hastie, Tibshirani, and Wainwright]{hastie2015statistical}
T.~Hastie, R.~Tibshirani, and M.~Wainwright.
\newblock \emph{Statistical learning with sparsity: the lasso and generalizations}.
\newblock CRC press, 2015.

\bibitem[Hausman(1978)]{hausman1978specification}
J.~A. Hausman.
\newblock Specification tests in econometrics.
\newblock \emph{Econometrica: Journal of the Econometric Society}, pages 1251--1271, 1978.

\bibitem[Hossini et~al.(2015)Hossini, Megges, Prigione, Lichtner, Toliat, Wruck, Schr{\"o}ter, Nuernberg, Kroll, Makrantonaki, et~al.]{hossini2015induced}
A.~M. Hossini, M.~Megges, A.~Prigione, B.~Lichtner, M.~R. Toliat, W.~Wruck, F.~Schr{\"o}ter, P.~Nuernberg, H.~Kroll, E.~Makrantonaki, et~al.
\newblock Induced pluripotent stem cell-derived neuronal cells from a sporadic {A}lzheimer’s disease donor as a model for investigating {AD}-associated gene regulatory networks.
\newblock \emph{BMC Genomics}, 16:\penalty0 1--22, 2015.

\bibitem[Johnston et~al.(2008)Johnston, Gustafson, Levy, and Grootendorst]{johnston2008use}
K.~Johnston, P.~Gustafson, A.~Levy, and P.~Grootendorst.
\newblock Use of instrumental variables in the analysis of generalized linear models in the presence of unmeasured confounding with applications to epidemiological research.
\newblock \emph{Statistics in Medicine}, 27\penalty0 (9):\penalty0 1539--1556, 2008.

\bibitem[Kanehisa et~al.(2002)]{kanehisa2002kegg}
M.~Kanehisa et~al.
\newblock The {KEGG} database.
\newblock In \emph{Novartis Foundation Symposium}, pages 91--100. Wiley Online Library, 2002.

\bibitem[Kang et~al.(2016)Kang, Zhang, Cai, and Small]{kang2016instrumental}
H.~Kang, A.~Zhang, T.~T. Cai, and D.~S. Small.
\newblock Instrumental variables estimation with some invalid instruments and its application to mendelian randomization.
\newblock \emph{Journal of the American Statistical Association}, 111\penalty0 (513):\penalty0 132--144, 2016.

\bibitem[Knudson et~al.(2021)Knudson, Benson, Geyer, and Jones]{knudson2021likelihood}
C.~Knudson, S.~Benson, C.~Geyer, and G.~Jones.
\newblock Likelihood-based inference for generalized linear mixed models: Inference with the {R} package glmm.
\newblock \emph{Stat}, 10\penalty0 (1):\penalty0 e339, 2021.

\bibitem[Lee et~al.(2015)Lee, Sun, and Taylor]{lee2015model}
J.~D. Lee, Y.~Sun, and J.~E. Taylor.
\newblock On model selection consistency of regularized {M}-estimators.
\newblock \emph{Electronic Journal of Statistics}, 9\penalty0 (1):\penalty0 608--642, 2015.

\bibitem[Li et~al.(2023)Li, Shen, and Pan]{li2023inference}
C.~Li, X.~Shen, and W.~Pan.
\newblock Inference for a large directed acyclic graph with unspecified interventions.
\newblock \emph{Journal of Machine Learning Research}, 24\penalty0 (73):\penalty0 1--48, 2023.

\bibitem[Li and Lederer(2019)]{li2019tuning}
W.~Li and J.~Lederer.
\newblock Tuning parameter calibration for $\ell_1$-regularized logistic regression.
\newblock \emph{Journal of Statistical Planning and Inference}, 202:\penalty0 80--98, 2019.

\bibitem[Negahban et~al.(2012)Negahban, Ravikumar, Wainwright, and Yu]{negahban2012unified}
S.~N. Negahban, P.~Ravikumar, M.~J. Wainwright, and B.~Yu.
\newblock A unified framework for high-dimensional analysis of {M}-estimators with decomposable regularizers.
\newblock \emph{Statistical Science}, 27\penalty0 (4):\penalty0 538--557, 2012.

\bibitem[Nelder and Wedderburn(1972)]{nelder1972generalized}
J.~A. Nelder and R.~W. Wedderburn.
\newblock Generalized linear models.
\newblock \emph{Journal of the Royal Statistical Society: Series A (General)}, 135\penalty0 (3):\penalty0 370--384, 1972.

\bibitem[Nelsen(2007)]{nelsen2007introduction}
R.~B. Nelsen.
\newblock \emph{An Introduction to Copulas}.
\newblock Lecture Notes in Statistics. Springer, 2nd edition, 2007.

\bibitem[Park and Raskutti(2017)]{park2017learning}
G.~Park and G.~Raskutti.
\newblock Learning quadratic variance function ({QVF}) {DAG} models via overdispersion scoring ({ODS}).
\newblock \emph{Journal of Machine Learning Research}, 18:\penalty0 224--1, 2017.

\bibitem[Pearl(2000)]{pearl2000models}
J.~Pearl.
\newblock Models, reasoning and inference.
\newblock \emph{Cambridge University Press}, 19\penalty0 (2):\penalty0 3, 2000.

\bibitem[Sharma et~al.(2021)Sharma, Chunduri, Gopu, Shatrowsky, Crusio, and Delprato]{sharma2021common}
A.~Sharma, A.~Chunduri, A.~Gopu, C.~Shatrowsky, W.~E. Crusio, and A.~Delprato.
\newblock Common genetic signatures of {A}lzheimer’s disease in {D}own {S}yndrome.
\newblock \emph{F1000Research}, 9:\penalty0 1299, 2021.

\bibitem[Shen et~al.(2012)Shen, Pan, and Zhu]{shen2012likelihood}
X.~Shen, W.~Pan, and Y.~Zhu.
\newblock Likelihood-based selection and sharp parameter estimation.
\newblock \emph{Journal of the American Statistical Association}, 107\penalty0 (497):\penalty0 223--232, 2012.

\bibitem[Shen et~al.(2013)Shen, Pan, Zhu, and Zhou]{shen2013constrained}
X.~Shen, W.~Pan, Y.~Zhu, and H.~Zhou.
\newblock On constrained and regularized high-dimensional regression.
\newblock \emph{Annals of the Institute of Statistical Mathematics}, 65\penalty0 (5):\penalty0 807--832, 2013.

\bibitem[Spirtes et~al.(2000)Spirtes, Glymour, and Scheines]{spirtes2000causation}
P.~Spirtes, C.~Glymour, and R.~Scheines.
\newblock \emph{Causation, prediction, and search}.
\newblock The MIT Press, 2000.

\bibitem[Terza et~al.(2008)Terza, Basu, and Rathouz]{terza2008two}
J.~V. Terza, A.~Basu, and P.~J. Rathouz.
\newblock Two-stage residual inclusion estimation: addressing endogeneity in health econometric modeling.
\newblock \emph{Journal of Health Economics}, 27\penalty0 (3):\penalty0 531--543, 2008.

\bibitem[Theil(1992)]{theil1992estimation}
H.~Theil.
\newblock Estimation and simultaneous correlation in complete equation systems.
\newblock \emph{Henri Theil’s Contributions to Economics and Econometrics: Econometric Theory and Methodology}, pages 65--107, 1992.

\bibitem[Tsamardinos et~al.(2006)Tsamardinos, Brown, and Aliferis]{tsamardinos2006max}
I.~Tsamardinos, L.~E. Brown, and C.~F. Aliferis.
\newblock The max-min hill-climbing {B}ayesian network structure learning algorithm.
\newblock \emph{Machine Learning}, 65\penalty0 (1):\penalty0 31--78, 2006.

\bibitem[van~de Geer and B{\"u}hlmann()]{10.1214/09-EJS506}
S.~A. van~de Geer and P.~B{\"u}hlmann.
\newblock {On the conditions used to prove oracle results for the Lasso}.
\newblock \emph{Electronic Journal of Statistics}, 3:\penalty0 1360--1392.

\bibitem[Vershynin(2018)]{vershynin2018high}
R.~Vershynin.
\newblock \emph{High-dimensional probability: An introduction with applications in data science}, volume~47.
\newblock Cambridge University Press, 2018.

\bibitem[Windmeijer et~al.(2019)Windmeijer, Farbmacher, Davies, and Davey~Smith]{windmeijer2019use}
F.~Windmeijer, H.~Farbmacher, N.~Davies, and G.~Davey~Smith.
\newblock On the use of the lasso for instrumental variables estimation with some invalid instruments.
\newblock \emph{Journal of the American Statistical Association}, 114\penalty0 (527):\penalty0 1339--1350, 2019.

\bibitem[Yang et~al.(2015)Yang, Smith, and Liu]{yang2015graphical}
J.~Yang, J.~Smith, and E.~Liu.
\newblock Graphical models and inference.
\newblock \emph{Journal of Statistical Modeling}, 42\penalty0 (1):\penalty0 10--32, 2015.

\bibitem[Ying et~al.(2019)Ying, Xu, and Murphy]{ying2019two}
A.~Ying, R.~Xu, and J.~Murphy.
\newblock Two-stage residual inclusion for survival data and competing risks — {A}n instrumental variable approach with application to {SEER}-{M}edicare linked data.
\newblock \emph{Statistics in Medicine}, 38\penalty0 (10):\penalty0 1775--1801, 2019.

\bibitem[Yuan et~al.(2019)Yuan, Shen, Pan, and Wang]{yuan2019constrained}
Y.~Yuan, X.~Shen, W.~Pan, and Z.~Wang.
\newblock Constrained likelihood for reconstructing a directed acyclic {G}aussian graph.
\newblock \emph{Biometrika}, 106\penalty0 (1):\penalty0 109--125, 2019.

\bibitem[Zhang(2017)]{zhang2017restricted}
H.~Zhang.
\newblock The restricted strong convexity revisited: analysis of equivalence to error bound and quadratic growth.
\newblock \emph{Optimization Letters}, 11\penalty0 (4):\penalty0 817--833, 2017.

\bibitem[Zhang et~al.(2022)Zhang, Ma, Liu, Du, Zhu, Liu, and Wu]{zhang2022activating}
J.-Y. Zhang, S.~Ma, X.~Liu, Y.~Du, X.~Zhu, Y.~Liu, and X.~Wu.
\newblock Activating transcription factor 6 regulates cystathionine to increase autophagy and restore memory in {A}lzheimer's disease model mice.
\newblock \emph{Biochemical and Biophysical Research Communications}, 615:\penalty0 109--115, 2022.

\bibitem[Zhao et~al.(2018)Zhao, Liu, and Zhang]{zhao2018pathwise}
T.~Zhao, H.~Liu, and T.~Zhang.
\newblock Pathwise coordinate optimization for sparse learning: Algorithm and theory.
\newblock \emph{The Annals of Statistics}, 46\penalty0 (1):\penalty0 180--218, 2018.

\bibitem[Zheng et~al.(2018)Zheng, Aragam, Ravikumar, and Xing]{zheng2018dags}
X.~Zheng, B.~Aragam, P.~Ravikumar, and E.~P. Xing.
\newblock {DAGs} with {NO TEARS}: continuous optimization for structure learning.
\newblock In \emph{Proceedings of the 32nd International Conference on Neural Information Processing Systems}, pages 9492--9503, 2018.

\end{thebibliography}

\end{document}